\documentclass[preprintnumbers,showpacs,amsmath,amssymb,floatfix,10pt,prd,twocolumn,superscriptaddress,nofootinbib]{revtex4-1}
\usepackage{graphicx}
\usepackage{latexsym}
\usepackage{epsfig}
\usepackage{amssymb}
\usepackage{subfig}
\usepackage{float}
\usepackage{hyperref}
\hypersetup{
	colorlinks=true,
	linkcolor=blue,
	filecolor=magenta,
	urlcolor=cyan,
}
\usepackage{mathtools}

\begin{document}
\title{Compact stellar model in Tolman space-time in presence of pressure anisotropy}

\author{Piyali Bhar}
\email{piyalibhar90@gmail.com
 } \affiliation{Department of Mathematics, Government General Degree College, Singur, Hooghly, West Bengal 712 409,
India}

\author{Shyam Das}
\email{dasshyam321@gmail.com
 }  \affiliation{Department of Physics, P. D. Women's College, Jalpaiguri, West Bengal 735101,
India}

\author{Bikram Keshari Parida}
\email{parida.bikram90.bkp@gmail.com
 }  \affiliation{Department of Physics, Pondicherry University, Kalapet, Puducherry 605014, India}

\begin{abstract}\noindent
In this paper, we develop a new relativistic compact stellar model for a spherically symmetric anisotropic matter distribution. The model has been obtained through generating a new class of solutions by invoking the Tolman {\em ansatz} for one of the metric potentials $g_{rr}$ and a physically reasonable selective profile of radial pressure. We have matched our obtained interior solution to the Schwarzschild exterior spacetime over the bounding surface of the compact star. These matching conditions together with the condition of vanishing the radial pressure across the boundary of the star have been utilized to determine the model parameters. We have shown that the central pressure of the star depends on the parameter $p_0$. We have estimated the range of $p_0$ by using the recent data of compact stars 4U 1608-52 and Vela X-1. The effect of $p_0$ on different physical parameters e.g., pressure anisotropy, the subliminal velocity of sound, relativistic adiabatic index etc. have also been discussed. The developed model of the compact star is elaborately discussed both analytically and graphically to justify that it satisfies all the criteria demanded a realistic star. From our analysis, we have shown that the effect of anisotropy becomes small for higher values of $p_0$. The mass-radius (M-R) relationship which indicates the maximum mass admissible for observed pulsars for a given surface density has also been investigated in our model. Moreover, the variation of radius and mass with central density has been shown which allow us to estimate central density for a given radius (or mass) of a compact star.
\end{abstract}

\maketitle

\section{Introduction}
Compact stars are the ultra-high dense objects exist in nature where a huge mass gets compacted in a small region producing the mass density beyond that of the nuclear density \cite{Weber1,Glendenning,Shapiro}. A compact star is born through a non equilibrium process of the gravitational collapse refers collectively to a neutron star, made out of neutron-rich nuclear matter or, a hybrid star made of nuclear matter with a quark matter core or, strange star exclusively composed of quark matter. White dwarfs which are less dense than neutron stars and black holes may be included in this compact star category. Compact stars are unique objects that manifest themselves across a wide range of multi-messenger signals like electromagnetic radiation from radio to gamma-rays, cosmic rays, neutrinos, and gravitational waves. It has been observed that compact stars may have huge magnetic fields, especially at the surface. Their extreme density, gravity and magnetic fields make them exceptional astrophysical laboratories for exploring the fundamental theories and interactions of elementary particles and testing general relativity at extreme conditions. Compact stars are not only extreme concerning their density but some of them also possess rotation in the millisecond regime. In fact, the first compact star has been discovered as pulsars, by observing pulsating radio signals, for the first time in 1967 \cite{Hewish}.

The general theory of relativity (GR) attracts a lot of interest in the fields of astrophysics, cosmology and gravitational-wave astronomy and it potentially leads to major breakthroughs since its introduction in 1915. Adoption of GR by the scientific community has increased dramatically in understanding the physics of compact stellar objects over the years. Following the discovery of quasars in the 1960s, and other very high energy phenomena in the universe such as gamma-ray bursts gravitation theory and relativistic astrophysics have gone through extensive developments in recent decades and the study of a compact star has got tremendous momentum. In relativistic astrophysics, studies of compact stars remain a field of active research since it can be used as a testbed for general relativity as well as particle behavior in the extreme conditions. Understanding the behavior and properties of stellar structure in the strong-field regime remains one of the fundamental questions.

Einstein field Equation is the cornerstone of General Relativity (GR). Exact solutions of the field equations are essential for describing the physical behavior of an astrophysical compact object. Unfortunately, the high nonlinearity associated with the field equations makes the underlying calculations complicated with mathematical difficulties. There are very few exact interior solutions of the field equations satisfying the required general physical conditions inside the star. Delgaty and Lake \cite{Delgaty} have found 127 solutions out of which only 16 qualify the test to meets all the conditions required for describing a physically realistic system. The study of general relativistic compact stellar objects via finding exact solutions of the field equations compatible with observational data has remained one of the major research areas in relativistic astrophysics. Observational data available with the advent of gravitational wave astronomy (LIGO, Virgo, KAGRA, LISA), as well as to high-angular-resolution observations of black hole vicinity (EHT, VLTI/GRAVITY) stimulate many researchers to contribute immensely to the theoretical development of the physics of compact objects.

For a compact star, its mass and radius are considered as the basic properties. One can connect these with the microscopic properties of nuclear and quark matter. This link is made by the equation of state (EOS) of the matter phase(s) inside the star which relates the pressure with the energy density of dense matter. For the study of compact stars, one requires proper understanding of the equation of state (EOS) corresponding to the material compositions of the star. The equation of state used to study the compact star, in particular, for an estimation of the accurate size, maximum mass and other physical features. With the knowledge of the EOS with proper boundary conditions, physical properties (such as mass-radius relationships) of the star can be analyzed by integrating the equation of hydrostatic equilibrium known as Tolman-Oppenheimer-Volkoff \cite{OV} equations. In particular, a stiff equation of state produces a large maximum mass.

In literature, assuming the different form of an equation of state (EOS) researchers generated many exact solutions to the field equations for describing a realistic compact object. For example, a linear EOS have been found to develop stellar structure by many researchers \cite{MAK,Sharma1,Thiru1,Maharaj,Thiru2,Govender1}. Exact solutions to Einstein field equations for an anisotropic sphere admitting a quadratic type of EOS \cite{Feroze, Maharaj2, Sunzu}, polytropic type \cite{Thiru3,Mafa,Ngubelanga,Isayev,Nasim} or Van der Waals type EOS \cite{Thiru4} have also been found in the literature.

One of the main difficulty is the uncertainty estimation in the EOS, which is core to the modeling of compact stellar objects and to understand the physical behavior of a compact object. Such limitation on EOS, encourage many researchers to adopt various ad-hoc approaches for a wide variety scenario of astrophysical systems. Amongst many alternative techniques, one either assumes the geometry or the fall-off behavior of density or pressure of the matter source. Towards this direction the geometrical approach suggested by Vaidya and Tikekar \cite{VT}and Tikekar \cite{Tikekar} having compact three-spheroidal geometry or Finch-Skea {\em ansatz} \cite{Finch} where the associated space-time is paraboloidal are very useful for describing a super-dense compact object. Such alternative methods are useful for the development of stellar models as shown by many investigators such as \cite{MaharajL,Mukherjee,GuptaJ,TT,Jotania,Sharma02S, JohnSD, ThomasVO05, ThomasVO07, TikVO, SharmaBS, PandyaVO, SharmaSDS,PatelSS, MuradS13, Murad2S13,FatemaHM13,Fatema2HM13,MuradS15,ThirukkaneshSD06,MaharajS06,KomathirajSD,IvanovBV02,Tikekar84, Sharma06SS, Tik90, Maharaj96,Tik98, Mukherjee97, Tikekar99VO, Chattopadhyay10, Chattopadhyay12}. Several simple but effective approaches to solve the field equations have been also used where the two metric potentials $g_{tt}$ and $g_{rr}$ are in general linked through an equation. The dependency of the metric potentials describes as Karmakar embedding class one method, conformally flat geometry and conformal motion respectively.

The first interior solution \cite{Karl} corresponding to a spherically symmetric stellar object was given by Karl Schwarzschild by imposing the isotropic condition on the Einstein equations. The matter under consideration was treated to be perfect fluid, which has an equal radial ($p_r$) and tangential ($p_t$) pressures. Later, Lemaitre in $1933$ \cite{Lemaitre} developed a constant density anisotropic stellar model supported with tangential pressure only. This model was generalized for variable density by Florides \cite{Florides}. In $1972$, Ruderman \cite{Ruderman} and Canuto \cite{Canuto} pointed out that at very high-density nuclear matter may become anisotropic in general. From the theoretical point of view, it is important to include the pressure anisotropy in the energy-momentum tensor describing the matter distribution of the system for describing the relativistic stellar structure. Bowers and Liang \cite{Bowers} generalized the equations of hydrostatic equilibrium to include the effects of local anisotropy on relativistic fluid spheres. Their work suggests that the anisotropy effects on the maximum equilibrium mass and surface redshift of a compact star. Interestingly, it was shown by Ivanov \cite{BVIvanov} that by considering a compact system to be anisotropic the  effects of shear, electromagnetic field, etc. can be automatically taken into account. 
The different factors that have been identified for the justification of the existence of anisotropy within a stellar interior such as the presence of a solid core \cite{Kippenhahn}, phase transitions \cite{Sokolov}, a type III superfluid, a pion condensation \cite{Sawyer} or the presence of the electrical field \cite{Usov}, slow rotation \cite{HerreraSantos}. Strong magnetic fields can also generate an anisotropic pressure inside a self-gravitating body as pointed out by Weber \cite{Weber}. It has been shown that the mixture of two gases can be represented by effective anisotropic fluid models \cite{Letelier,Bayin}. On the galactic scale, the existence of the anisotropy has been pointed out \cite{Binney}. Several anisotropic models have been investigated by Maurya and Gupta \cite{Maurya14}, Maurya et al. \cite{Maurya15}, Pandya et al. \cite{Pandya15}, Bhar et al. \cite{Bhar14}, Murad \cite{Murad13}, Maharaj and Mafa Takisa \cite{Maharaj12}, Mafa Takisa et al. \cite{Mafa14a,Mafa14b}, Sunzu et al. \cite{Sunzu14as,Sunzu14bs}. General algorithms for generating static anisotropic solutions was also found by Lake\cite{KLake09}. Some recent work may be found in Refs. \cite{Thiru8,Thiru6,Kumar,Thiru7,Maurya1,THarko,Harko1,Harko2,Harko3,Mak02,MakD02}. A complete review on anisotropic fluid spheres can be found by the work of Herrera and Santos \cite{Herrera97San}.

As far as the dynamical evolution of a self-gravitating system is concerned the anisotropic effects on the evolution was investigated by Herrera and co-workers \cite{HerreraP98,Herrera04T,Herrera11I}, Chan et al. \cite{Chan03J}, Govinder et al. \cite{Govinder98ks}, Herrera and Santos \cite{Herrera02O}, Chan et al. \cite{Chan94R}, Naidu et al. \cite{Naidu06NF} and Rajah and Maharaj \cite{RajahSD}. Conformally flat solutions corresponding to anisotropic compact self-gravitating objects were developed by Herrera et al. \cite{HerreraAdi,HerreraCon}. Raposo et al. also \cite{raposo} studied on the dynamical properties of anisotropic
self-gravitating fluids in a covariant framework. 

Inspired by the previous work done by several researchers, in the present paper we develop a model of a compact star by assuming a physically reasonable choice for the radial pressure. Our paper has been organized as follows. In sect.~\ref{2x} Einstein field equations have been described. Sect.~\ref{3x} deals with the solutions of the field equation. In the next section (sect.~\ref{4x}) we have matched our interior solution to the exterior Schwarzschild spacetime at the boundary. The physical attributes and the stability condition of the model are respectively given in sect.~\ref{5x} and \ref{6x}. Some discussions and concluding remarks have been given in sect.~\ref{7x}.

\section{Interior Spacetime and Einstein field Equations}\label{2x}
It is well known that the energy-momentum tensor for an anisotropic model of the compact object can be described by,
\begin{equation}\label{tmu}
T_{\chi}^{\xi}=(\rho+p_t)u^{\xi}u_{\chi}+p_t g_{\chi}^{\xi}+(p_r-p_t)\eta^{\xi}\eta_{\chi},
\end{equation}
where $\eta^{i}$ is the space-like vector and the vector $u_i$ represent fluid $4$-velocity and  which is orthogonal to $ \eta^{i}$ with $ -u^{i}u_{j} =\eta^{i}\eta_j = 1 $ and $u^{i}\eta_j= 0$. $\rho,\,p_r$ and $p_t$ are the matter density, radial pressure and
tangential pressure respectively.\\
In curvature co-ordinate $(t,\,r,\,\theta,\,\phi)$ for a static spherically symmetry configuration in (3+1)-dimension, the interior spacetime is described by the line element,
\begin{equation}
ds_{-}^{2}=-e^{\nu(r)}dt^{2}+e^{\lambda(r)}dr^{2}+r^{2}(d\theta^{2}+\sin^{2}\theta d\phi^{2}),
\end{equation}
where $\lambda$ and $\nu$ being metric potentials and are the functions of the radial coordinate `r'.\\
Using the Einstein field equations $G_{\mu \nu}=\kappa T_{\mu \nu}$, we get the following set of equations
\begin{eqnarray}
\kappa c^2\rho&=&\frac{\lambda'}{r}e^{-\lambda}+\frac{1}{r^{2}}(1-e^{-\lambda}),\label{f1}\\
\kappa p_r&=& \frac{1}{r^{2}}(e^{-\lambda}-1)+\frac{\nu'}{r}e^{-\lambda},\label{f2} \\
\kappa p_t&=&\frac{1}{4}e^{-\lambda}\left[2\nu''+\nu'^2-\lambda'\nu'+\frac{2}{r}(\nu'-\lambda')\right], \label{f3}
\end{eqnarray}
where $\kappa=\frac{8\pi G}{c^4}$, $G_{\mu\nu}$ being Einstein tensor, $G$ is the universal gravitational constant and $c$ is the speed of the light. In the above equations `prime' denotes differentiation with respect to radial co-ordinate $r$.
In this respect we want to mention that the mass function $m(r)$ inside the radius `r' can be obtained from the relation $e^{-\lambda}=1-\frac{2m(r)}{r}$ which can directly follow from eq.~ (\ref{f1}).
\begin{figure}[htbp]
    \centering
        \includegraphics[scale=.3]{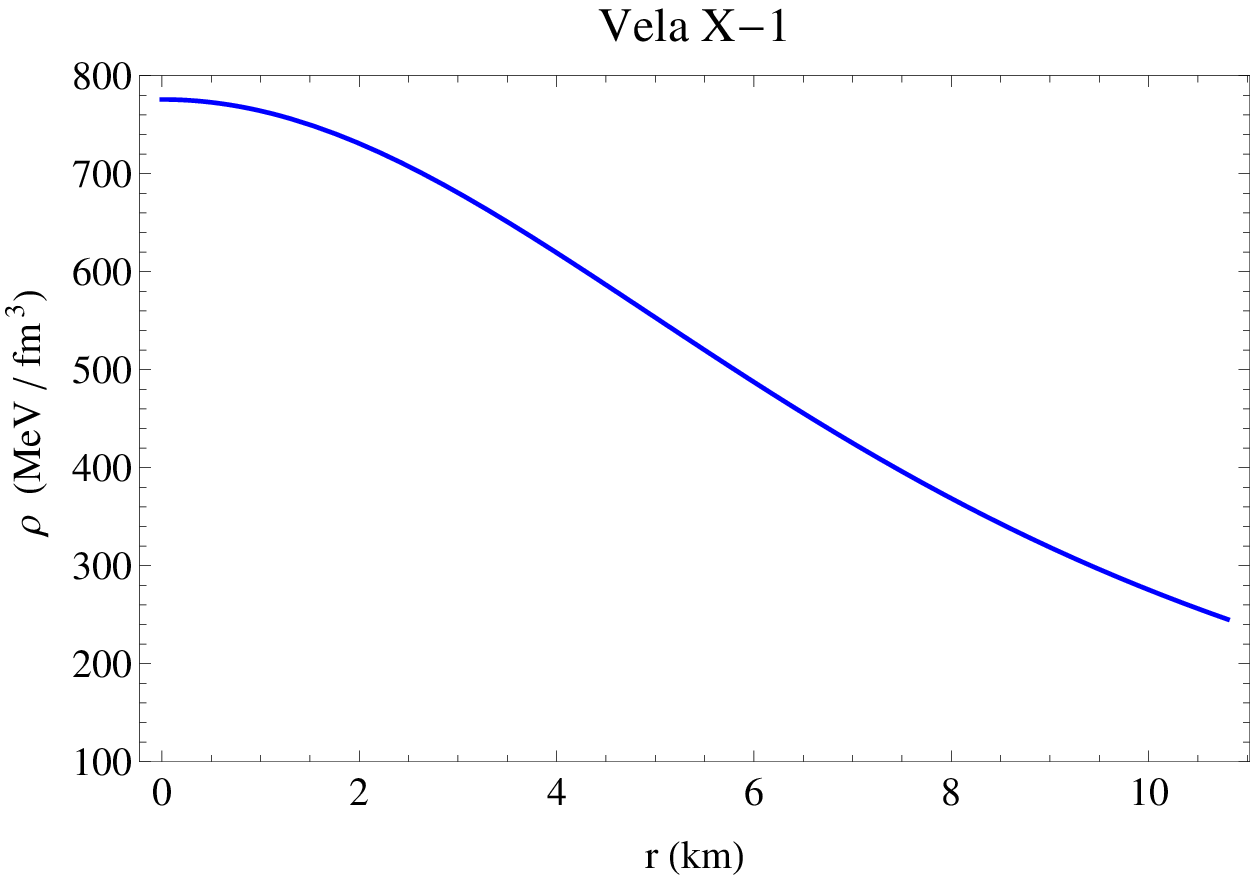}
        \includegraphics[scale=.3]{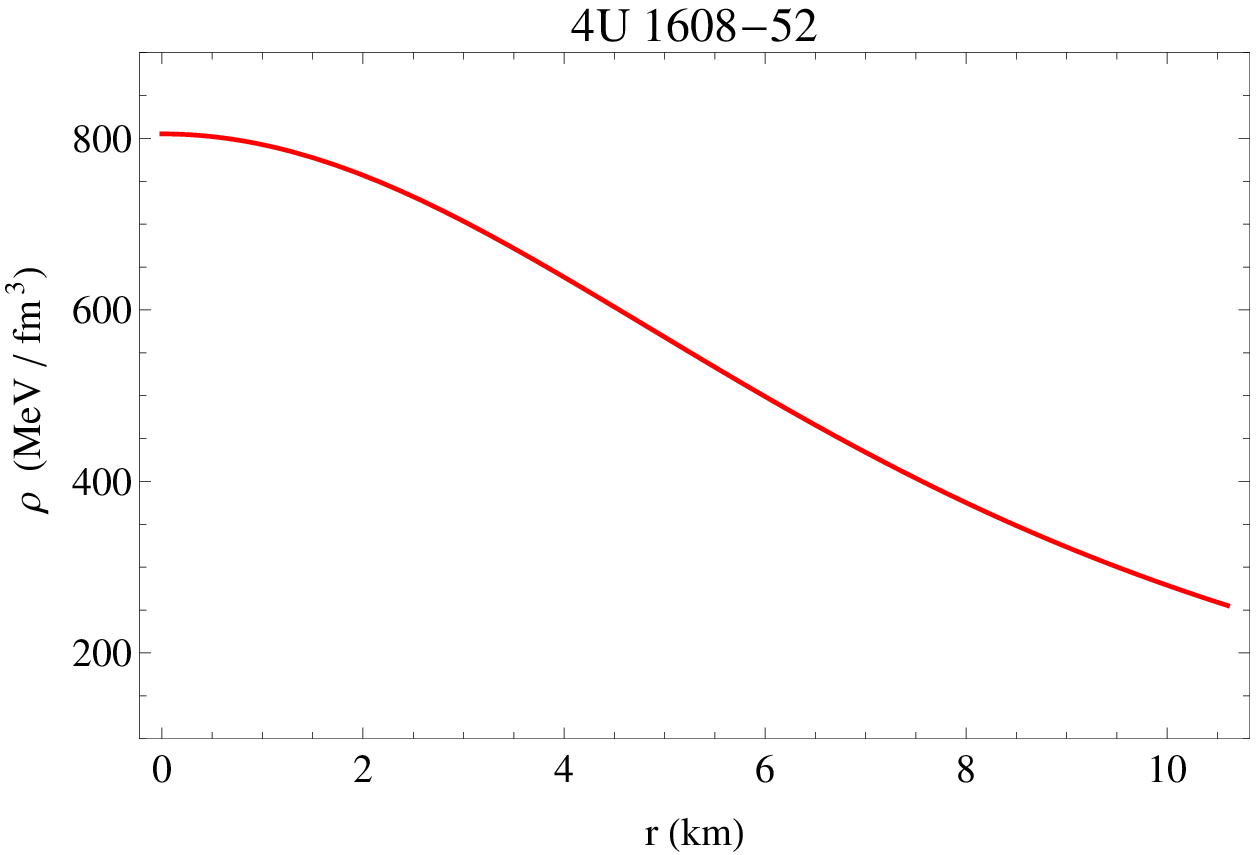}
       \caption{Energy density $\rho$ is plotted against $r$ inside the stellar interior for a possible modelling of the compact stars Vela X-1 and 4U 1608-52.\label{rho}}
\end{figure}

\begin{figure}[htbp]
    \centering
\includegraphics[scale=.3]{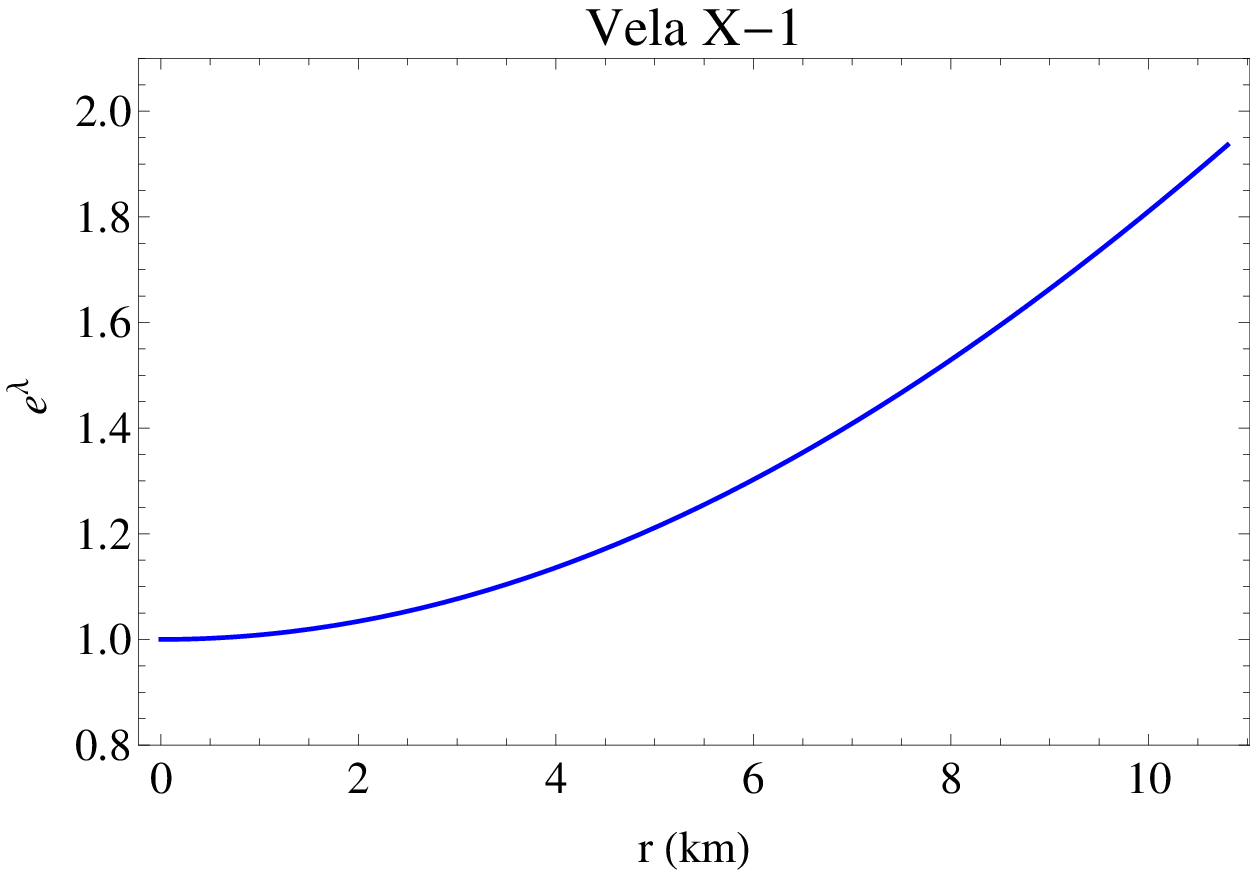}
        \includegraphics[scale=.35]{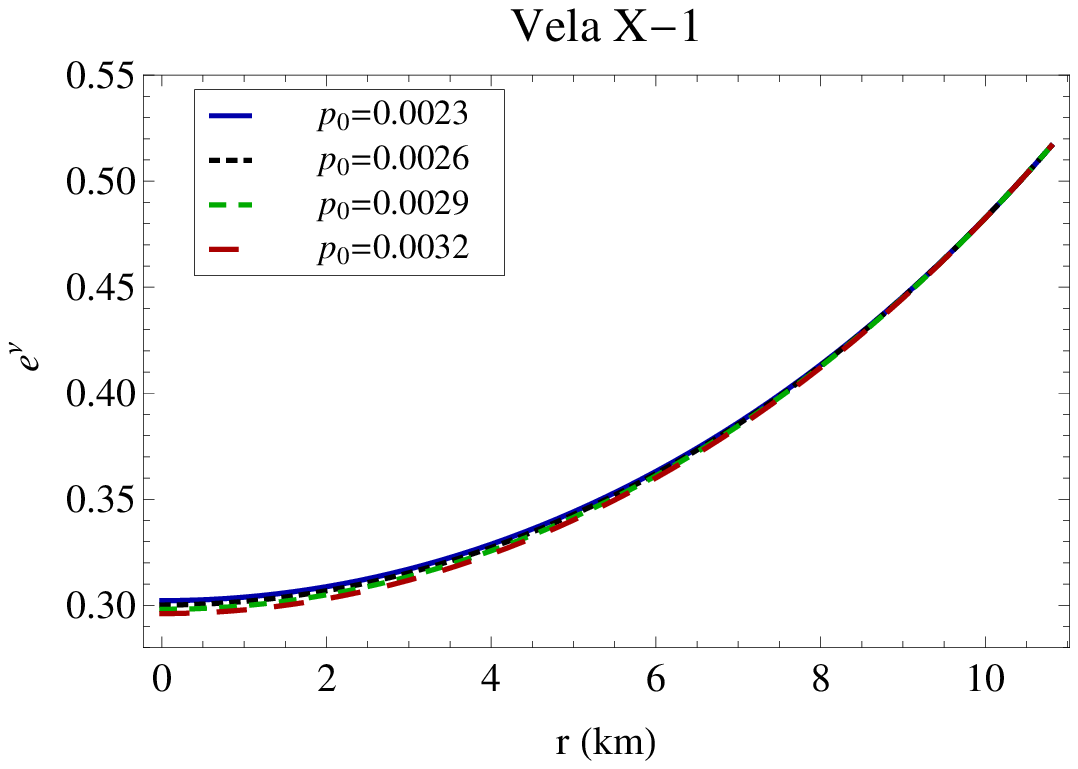}
        \includegraphics[scale=.3]{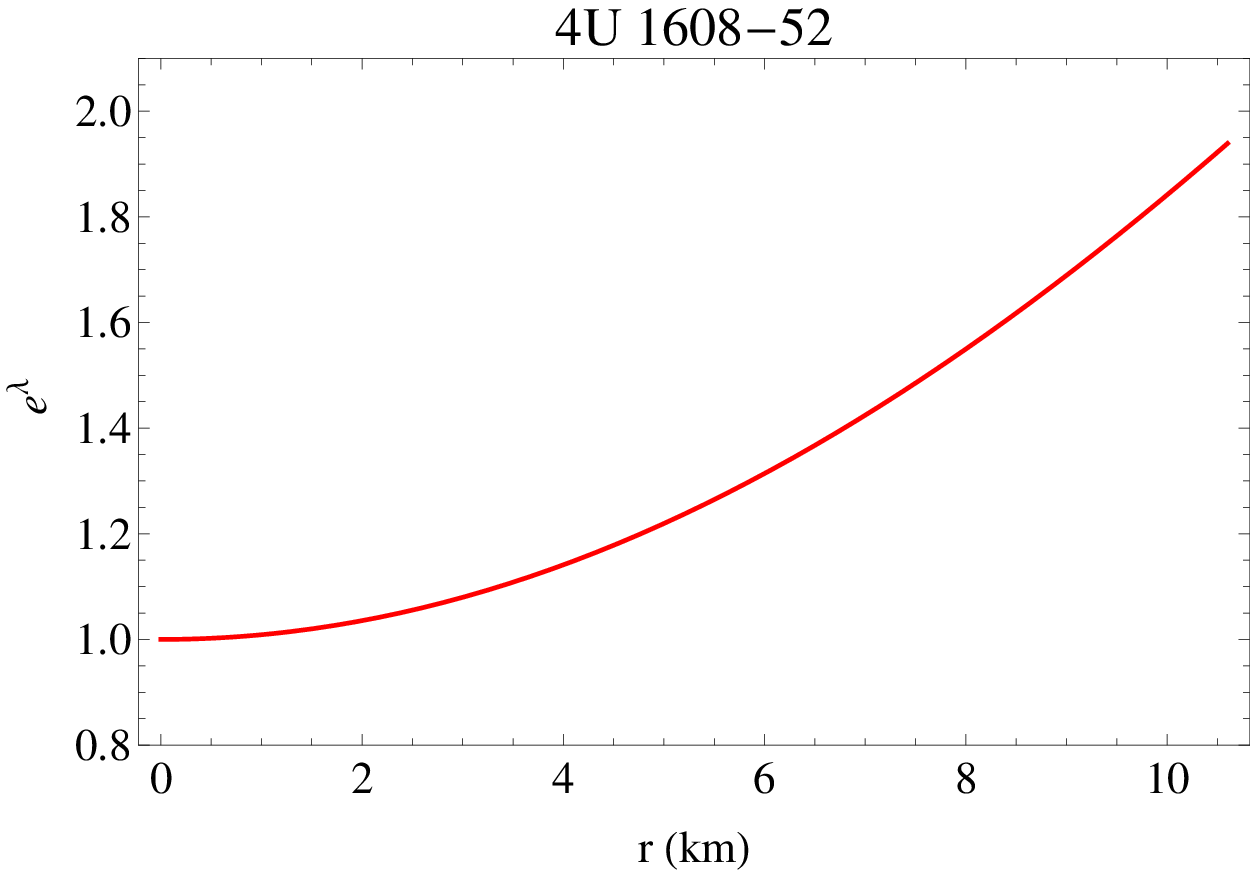}
        \includegraphics[scale=.35]{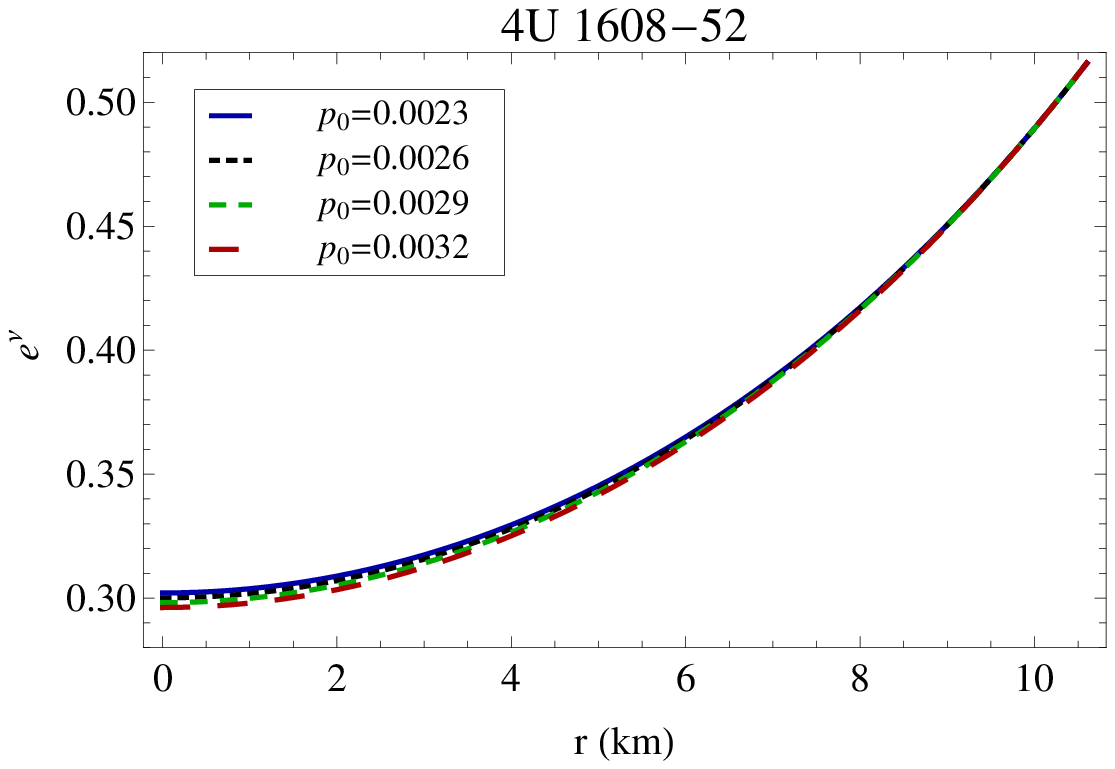}
        \caption{The metric potentials $e^{\nu}$ and $e^{\lambda}$ are  plotted against $r$ inside the stellar interior for a possible modelling of the compact stars Vela X-1 and 4U 1608-52.\label{metric}}
\end{figure}

\section{Choice of metric potential and new anisotropic solution}\label{3x}
It is well known that the solutions of the field equation which could describe a compact stellar structure that satisfies all the physically reasonable conditions are a very difficult job. To reduce the difficulties we apply a geometrical approach by prescribing a suitable form of the metric function. To solve the above set of Eqs.~(\ref{f1})-(\ref{f3}), let us take the metric potential $g_{rr}$ in the following form:
\begin{equation}\label{lam}
e^{\lambda}=1+ar^2+br^4,
\end{equation}
where `$a$' and `$b$' are constant parameters having the units km$^{-2}$ and km$^{-4}$ respectively. This metric potential was proposed by Tolman\cite{tolman1} for describing a compact stellar object. This metric potential is free from central singularity and monotonic increasing function of $r$. This type of ansatz were used earlier by several authors to model compact star in both GTR and modified gravity \cite{x1,x2}.  \\
Substituting the expression for the metric potential  $e^{\lambda}$ from (\ref{lam}) into (\ref{f1}) and assuming $G=1=c$, the matter density can be obtained as,
\begin{equation}
8\pi\rho=\frac{3 a + (a^2 + 5 b) r^2 + 2 a b r^4 + b^2 r^6}{(1 + a r^2 + b r^4)^2},\label{Rhoo}
\end{equation}

\begin{figure}[htbp]
        \includegraphics[scale=.3]{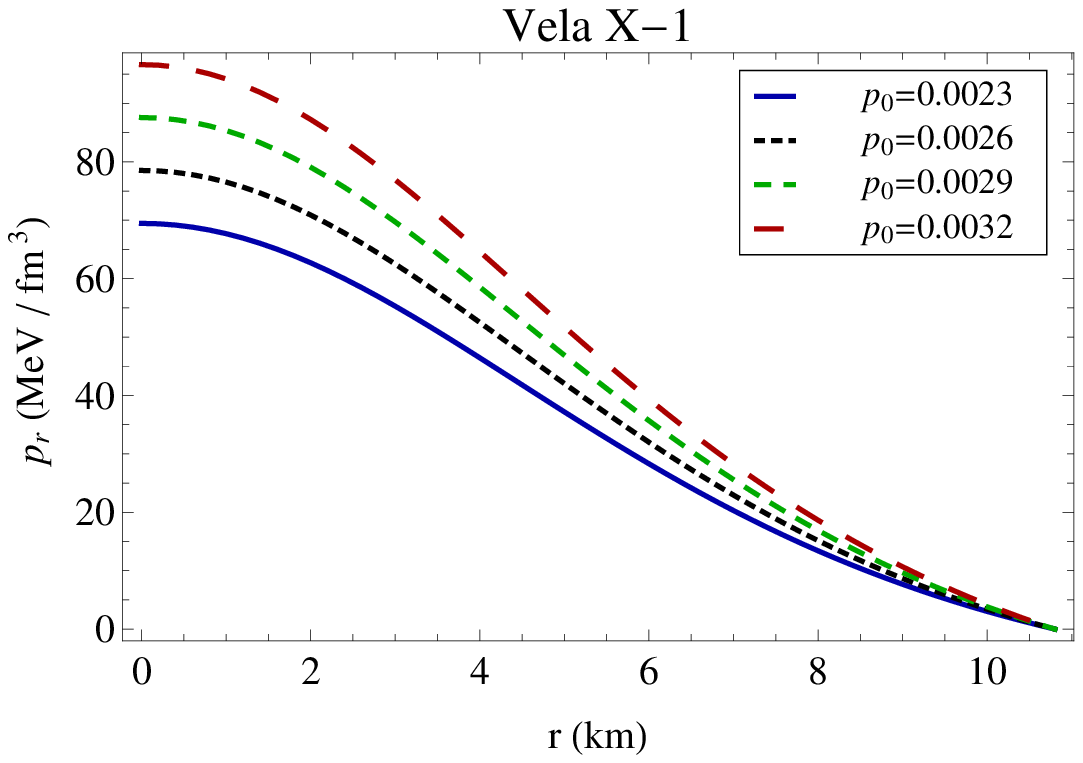}
        \includegraphics[scale=.3]{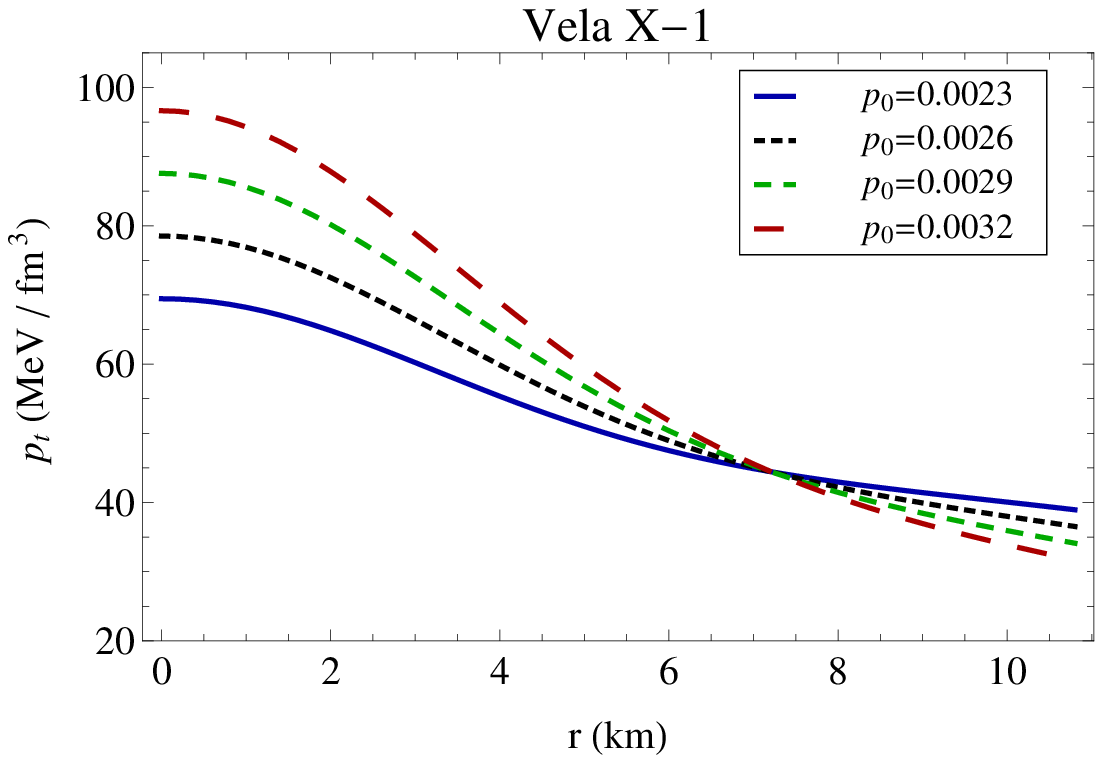}
        \includegraphics[scale=.3]{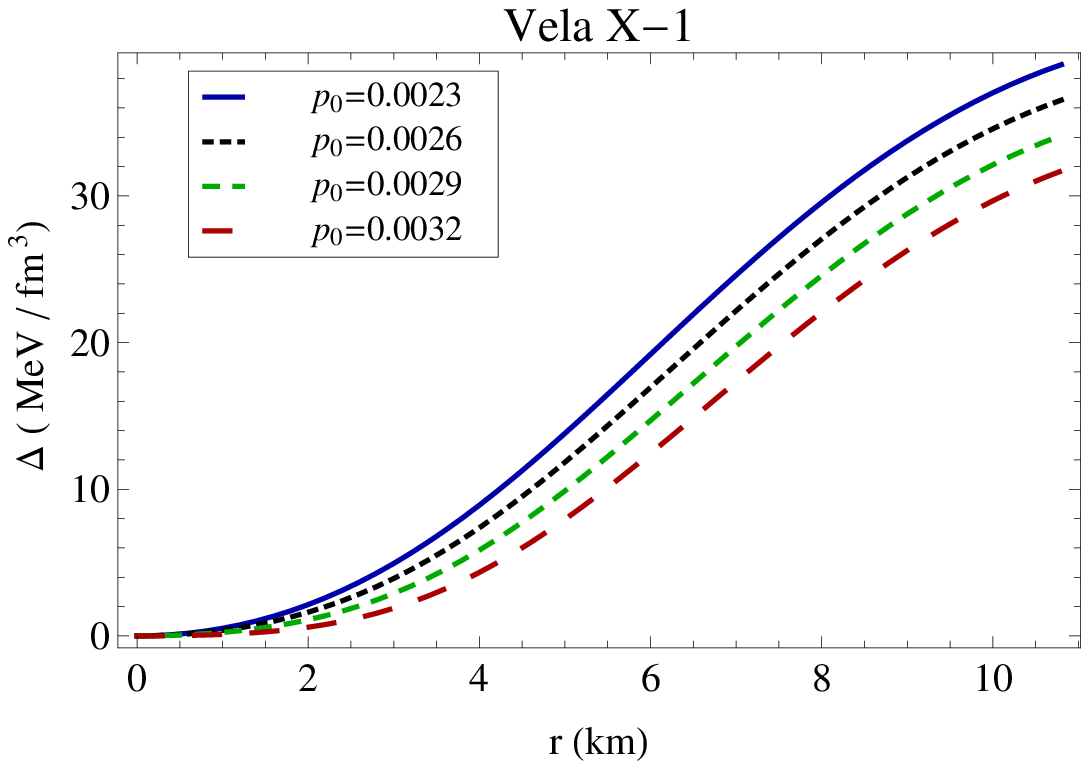}
        \includegraphics[scale=.3]{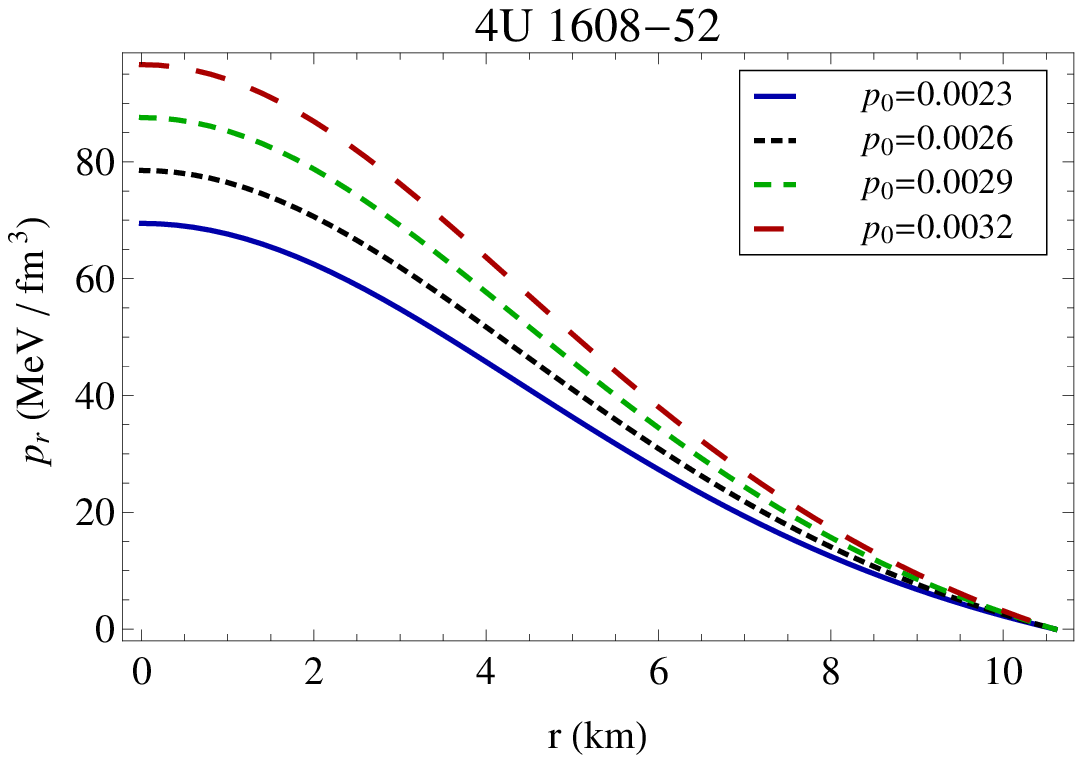}
        \includegraphics[scale=.3]{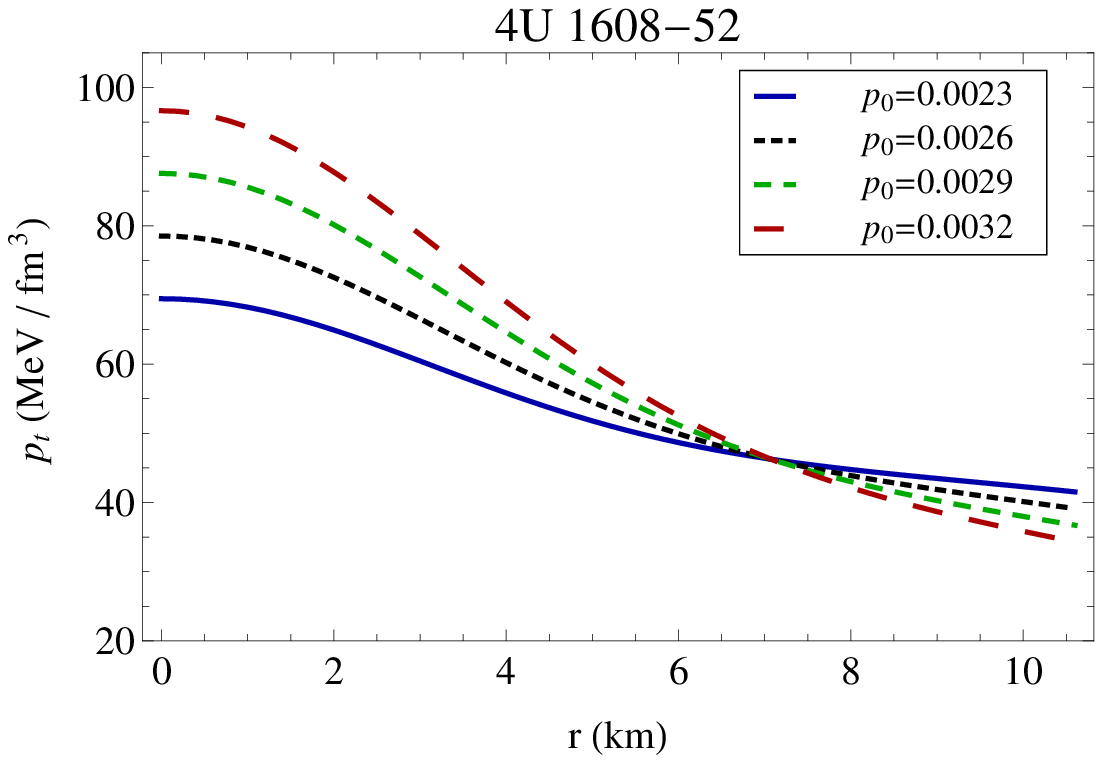}
        \includegraphics[scale=.3]{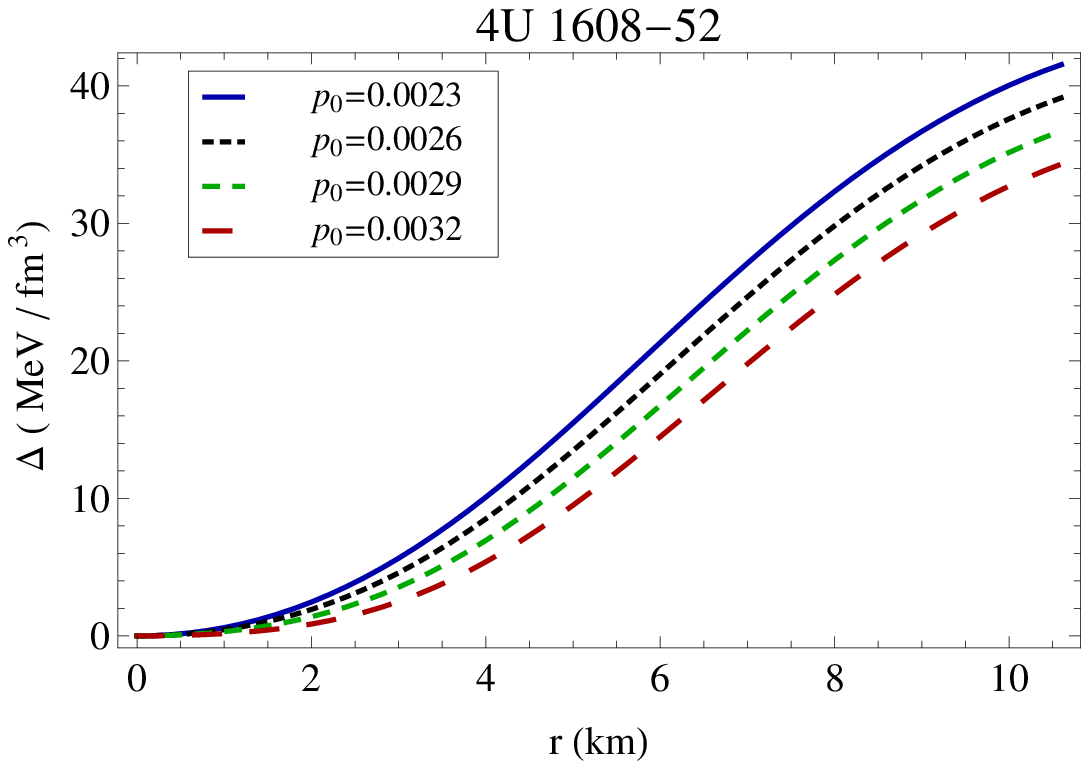}
       \caption{Radial pressure $p_r$, transverse pressure $p_t$ and anisotropic factor $\Delta$ are plotted against $r$ inside the stellar interior for a possible modelling of the compact stars Vela X-1 and 4U 1608-52  for $p_0=0.0023$, $p_0=0.0026$, $p_0=0.0029$ and $p_0=0.0032$ by considering different values of `a' and `b' mentioned in Table~II. \label{pr100}}
\end{figure}

To find the other metric potential $e^{\nu}$ 
let us assume the radial pressure in the form
\begin{eqnarray}
8\pi p_r=\frac{p_0(1-ar^{2})}{(1+ar^2+br^4)^{2}},\label{pr}
\end{eqnarray}

where $p_0$ is a non-negative constant. The expression of $p_r$ is reasonable due to the fact that it is monotonic decreasing function of '$r$' and vanishes at $r=\frac{1}{\sqrt{a}}$. Therefore, the radius of the star is obtained as $R=\frac{1}{\sqrt{a}}$, which is a finite quantity. Moreover, it possess a finite value of central pressure equal to $p_0/8\pi$ for all $p_0>0$. The same type of choice of pressure was earlier used by Sharma and Ratanpal \cite{SharmaRatan} for describing a static spherically symmetric anisotropic stellar configuration.

Substituting the expression for the radial pressure $p_r$ mentioned above and the metric potential $e^{\lambda}$ into (~\ref{f2}) we get,
\begin{eqnarray} \label{dnu}
\frac{d\nu}{d r}  & = & \frac{p_0 \, r(1-a r^{2})}{1+a r^2+b r^4} + r(a+b r^2),
\end{eqnarray}
On integrating equation (\ref{dnu}) we get,
\begin{eqnarray}
\nu & = & \frac{1}{4}\Big[2 a r^2 + b r^4 -\frac{
   2 p_0(a^2 + 2 b)}{b \sqrt{a^2-4b}} \tan^{-1}\Big(\frac{a + 2 b r^2}{\sqrt{a^2-4b}}\Big)\nonumber\\&&
    -\frac{a p_0}{b} \log(1 + a r^2 + b r^4)\Big]+B,
\end{eqnarray}
where $B$ is the constant of integration, which will be determined from the boundary conditions.
Employing the expression of the metric potentials in Eq.~(\ref{f3}) we get the expression of the transverse pressure $p_t$ as,
\begin{eqnarray}
8\pi p_t & =& \frac{1}{4 \Psi^3}\big[p_0^2 r^2 (1-a r^2)^2 +
 r^2 (a + b r^2) \Psi \xi_1 \nonumber\\&& -
 2 p_0 (-2 + 4 a r^2 +  +\xi_2
    )\big],\label{Pt}
\end{eqnarray}
where $\xi_1,\,\xi_2$ defined as,
\begin{eqnarray*}
\xi_1 &=3 a + (a^2 + 5 b) r^2 +
    2 a b r^4 + b^2 r^6 ,\\
\xi_2 &=(a^2 + 3 b) r^4+a (a^2 - 3 b) r^6 + (2 a^2 - b) b r^8 + a b^2 r^{10},\\
\Psi &= 1+ a r^2+b r^4.
\end{eqnarray*}
The anisotropic factor $\Delta=p_t-p_r$ is given by,
\begin{eqnarray}
8\pi \Delta & =& \frac{r^2}{4 \Psi^3}\big[A_1 +
    A_2 r^2 + A_3 r^4 +
   A_4 r^6 +
   A_5 r^8\nonumber\\&&+ 4 a b^3 r^{10} + b^4 r^{12}\big],
\end{eqnarray}
where $A_i$'s are constants given by,
\begin{eqnarray*}
A_1&=&3 a^2 - 8 a p_0 + p_0^2,\\
A_2&=&2 (2 a^3 + 4 a b + a^2 p_0 - 5 b p_0 - a p_0^2),\\
A_3&=&a^4 + 5 b^2 -
      2 a^3 p_0 + 10 a b p_0 + a^2 (14 b + p_0^2),\\
      A_4&=&2 b (2 a^3 + 8 a b - 2 a^2 p_0 + b p_0),\\
      A_5&=&2 b^2 (3 (a^2 + b) - a p_0).
\end{eqnarray*}
The anisotropy factor $\Delta$ measures the anisotropy of the system and $\frac{2\Delta}{r}$ is termed as the anisotropic force which will be repulsive in nature if $p_t>p_r$ and attractive if the inequalities is in reverse direction.
In the coming sections we shall check the physical features of our proposed model.

\section{Exterior spacetime and matching condition}\label{4x}
To fix the model parameters $a,\,b$ and $D$ we match our interior spacetime to the exterior Schwarzschild spacetime at the boundary $r=R$, where $R$ is the radius of the star.
For our present case the interior and exterior line element is given by,
\begin{eqnarray}
ds^2_{-}&=& -\exp\Big\{\frac{1}{4}\Big\{2 a r^2 + b r^4 +\frac{
   2 p_0(a^2 + 2 b)}{b \sqrt{a^2-4b}} \times  \nonumber\\&&\tan^{-1}\Big(\frac{a + 2 b r^2}{\sqrt{a^2-4b}}\Big)
   -\frac{ap_0}{b} \log(1 + a r^2 + b r^4)\Big\}\nonumber\\&&+B\Big\}dt^2+(1 + ar^2 + br^4)dr^2\nonumber\\&&+r^2(d\theta^2+\sin^2\theta d\phi^2),
   \\
    ds^2_{+}&=&-\left(1-\frac{2M}{r}\right)dt^2+\left(1-\frac{2M}{r}\right)^{-1}dr^2\nonumber\\&&+r^2(d\theta^2+\sin^2\theta d\phi^2).
\end{eqnarray}
Now using the matching condition at the boundary of the star, one can get two fundamental form. The first fundamental from consists in the continuity of the metric potential across the boundary $r=R$. Explicitly
\begin{equation}\label{eqm1}
e^{\lambda^{-}}|_{r=R}=e^{\lambda^{+}}|_{r=R} \quad \mbox{and} \quad   e^{\nu^{-}}|_{r=R}=e^{\nu^{+}}|_{r=R}.
\end{equation}
The second fundamental form gives,
\begin{equation}\label{eqm2}
p_{r}(R)=0,
\end{equation}
which determines the size of the compact object. The second fundamental form tells us that the size of a compact star can not be arbitrarily large, i.e., it is finite. \\
Therefore, from first fundamental form we obtain
\begin{eqnarray}
\ln\left(1-\frac{2M}{R}\right)&=&\frac{1}{4}\Big\{2 a R^2 + b R^4 +\frac{
   2 p_0(a^2 + 2 b)}{b \sqrt{a^2-4b}}\times \nonumber\\&& \tan^{-1}\Big(\frac{a + 2 b R^2}{\sqrt{a^2-4b}}\Big)
    \nonumber\\&&-\frac{a p_0}{b} \log(1 + a R^2 + b R^4)\Big\}+B,\nonumber\\ \label{eqm3}\\
    \left(1-\frac{2M}{R}\right)^{-1}&=&1+a R^2+b R^4.\label{eqm4}
\end{eqnarray}
Using the expression of radial pressure (\ref{pr}) from the Eq.~(\ref{eqm2}) we have,
\begin{eqnarray}\label{eqm5}
R^2=\frac{1}{a}.
\end{eqnarray}
Now solving the Eqs.~(\ref{eqm3})-(\ref{eqm5}) we get,
\begin{eqnarray}
a&=&\frac{1}{R^2},\\
b&=&\frac{1}{R^4}\left[\left(1-\frac{2M}{R}\right)^{-1}-2\right],\\
B&=&\ln\left(1-\frac{2M}{R}\right)-\frac{1}{4}\Big\{2 a R^2 + b R^4 +\frac{
   2 p_0(a^2 + 2 b)}{b \sqrt{a^2-4b}}\times \nonumber\\&& \tan^{-1}\Big(\frac{a + 2 b R^2}{\sqrt{a^2-4b}}\Big)
    -\frac{ap_0}{b} \log(1 + a R^2 + b R^4)\Big\}.\nonumber\\
\end{eqnarray}
The above set of equations determines the constants $a$, $b$ and $B$ in terms of mass, radius of the compact object. So we can calculate the model parameters $a$, $b$ and $B$ numerically for a particular compact star from its estimated mass and radius data. For our present work we have estimated the mass and radius for the two compact stars Vela X-1\cite{vela}  and 4U 1608-52 \cite{4u}  which has been presented in Table~I and the values of the constants $a,\,b$ are obtained in Table~II where as the numerical values of $B$ is obtained in Table~III. One interesting thing we can note that here $a,\,b$ do not depends on $p_0$ but $B$ does.

\begin{figure}[htbp]
        \includegraphics[scale=.3]{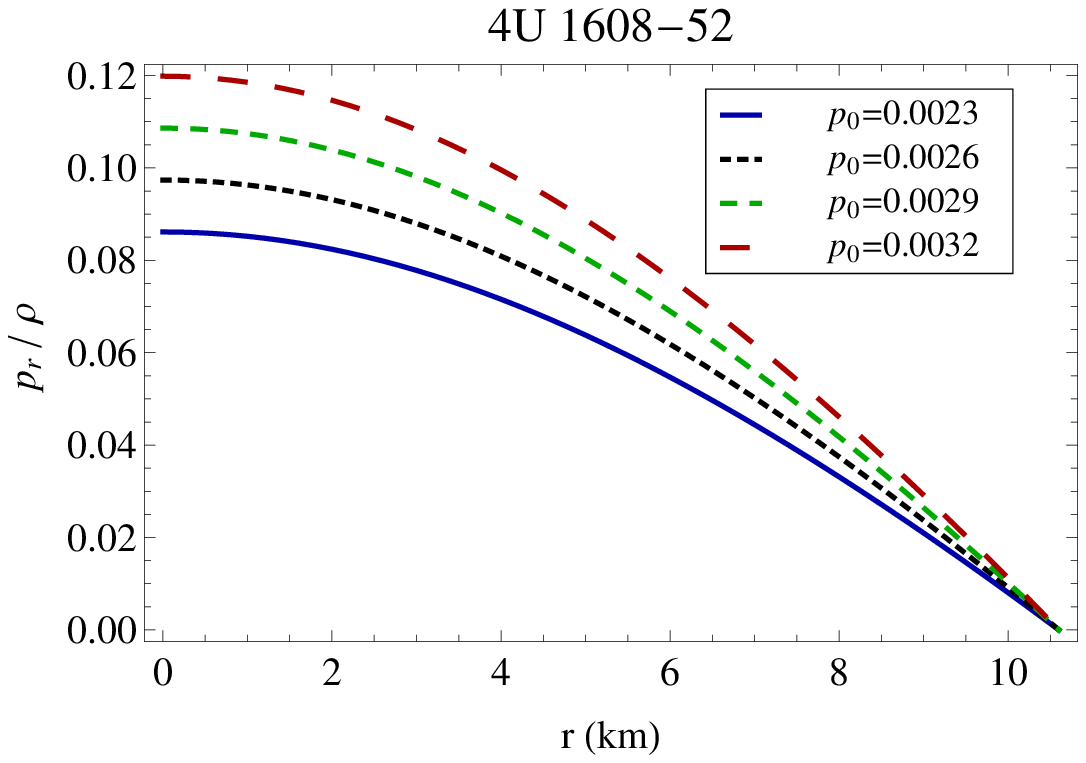}
        \includegraphics[scale=.3]{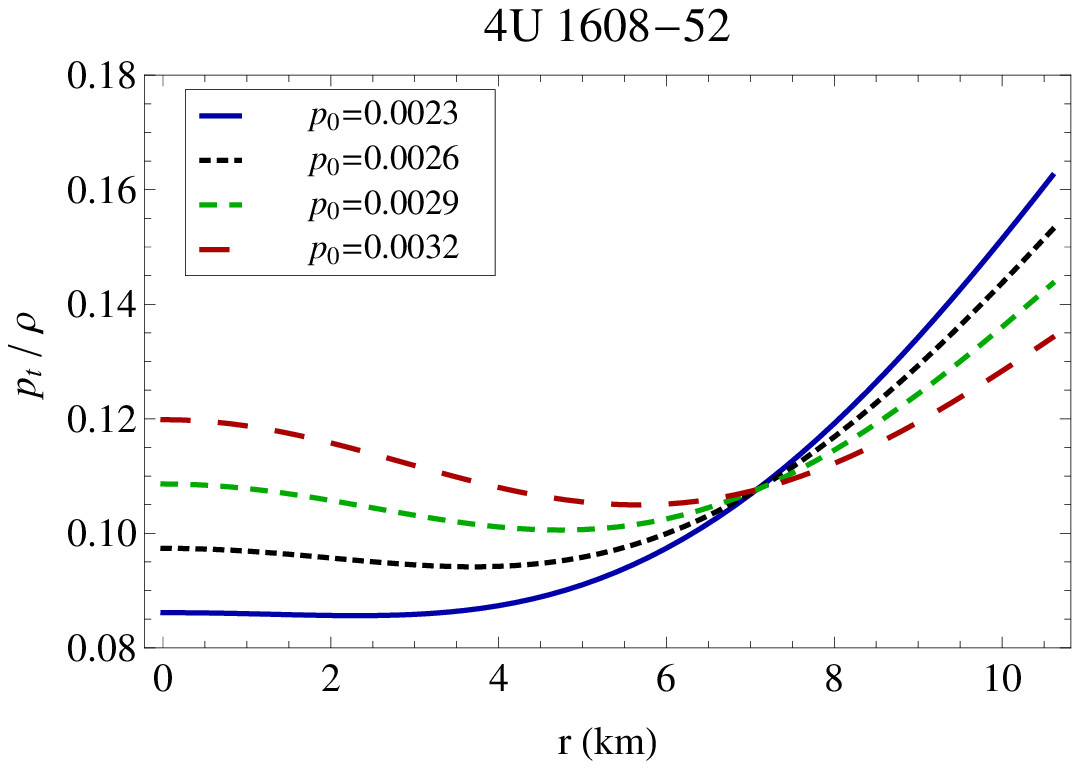}
        \includegraphics[scale=.3]{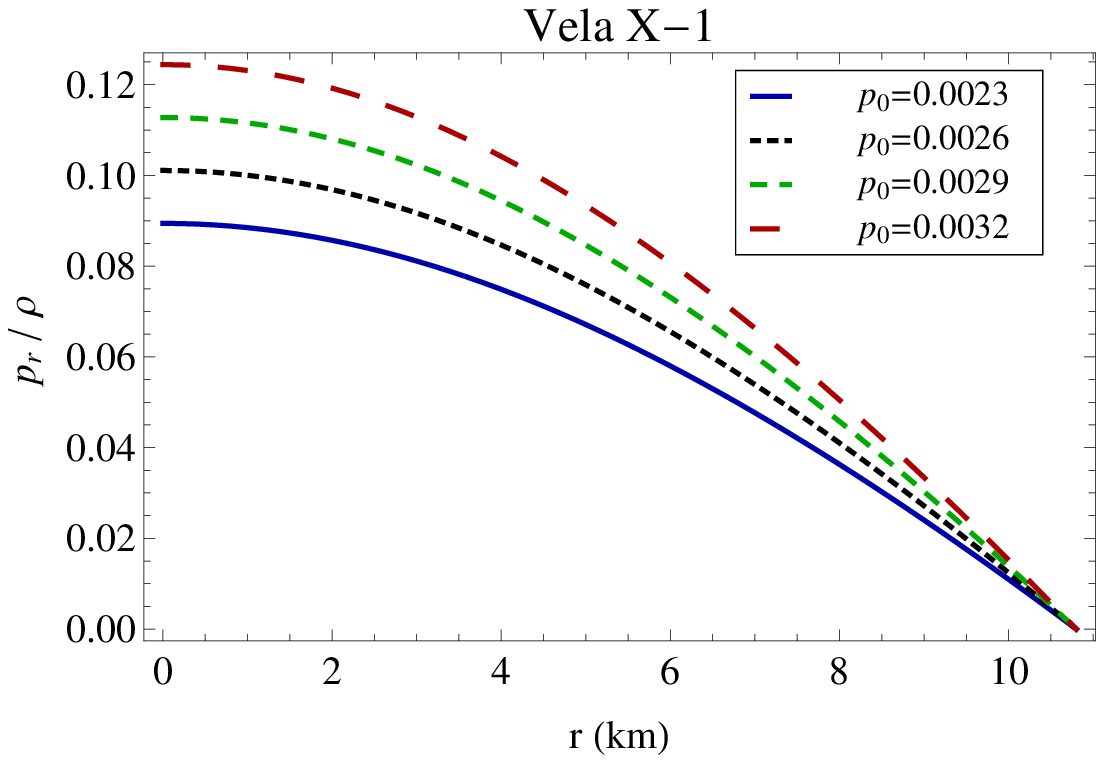}
        \includegraphics[scale=.3]{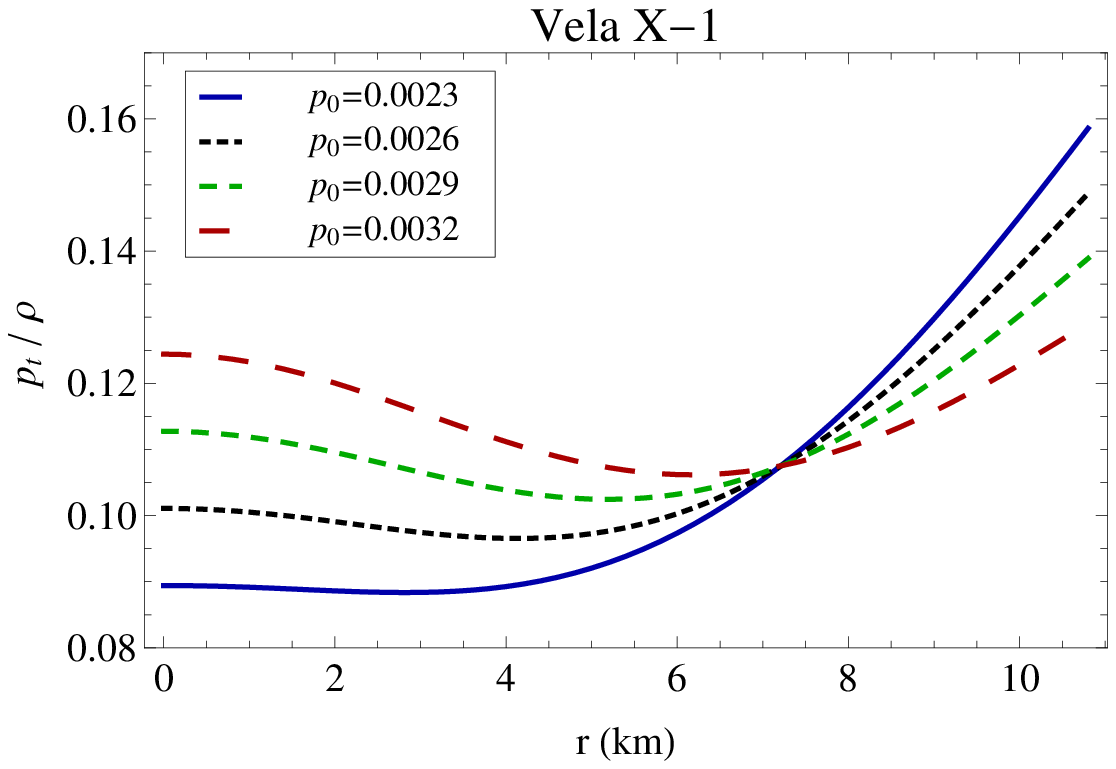}
       \caption{The equation of state parameters are plotted against radial parameter $r$ for a possible modelling of the compact star Vela X-1 and 4U 1608-52 for different values of $p_0$ mentioned in the figures.\label{e8}}
\end{figure}
\section{Analysis of the solution}\label{5x}
\subsection{Nature of pressure and density}
The density and pressure at the core of the star is obtained as,
\[\rho_c=\frac{3a}{\kappa};~~~p_c=\frac{p_0}{\kappa}\].
Now it is well known that for a physically acceptable model, both the central density and central pressure are non-negative, which gives the following two inequalities:
\begin{equation} a>0;~~~~~  p_0>0.\end{equation}
It is an important task to find a physically reasonable bound for the central density $p_0$ which we shall discuss in the coming sections.\par
The density and pressure gradients are obtained as,
\begin{eqnarray*}
  \kappa\frac{d\rho}{dr} &=& -\frac{2 r}{\Psi^3} (5 a^2 - 5 b + B_1 r^2 + B_2 r^4 +
   3 a b^2 r^6 + b^3 r^8),\nonumber\\
   \kappa\frac{dp_r}{dr}&=&-\frac{2 p_0}{\Psi^2} \Big(2 b r^3 + a r (2-b r^4)\Big), \\
   \kappa\frac{dp_t}{dr}&=&\frac{r}{2 \Psi^4}\times\Big[B_3 +
   2 B_4 r^2 - B_5 r^4 -
   4 B_6 r^6 + B_7 r^8 \nonumber\\&&+
   4 b B_8 r^{10} +
   B_9 r^{12} + B_{10} r^{14} +
   B_{11} r^{16}\Big],
\end{eqnarray*}
where $B_i's$ are constants given by,
\begin{eqnarray*}
B_1&=&a (a^2 + 13 b),\\
B_2&=&3 b (a^2 + 4 b),\\
B_3&=&3 a^2 - 16 a p_0 + p_0^2,\\
B_4&=&a^3 + 8 a b + 3 a^2 p_0 - 14 b p_0 - 2 a p_0^2, \\
B_5&=&a^4 -19 a^2 b - 15 b^2 + a (17 a^2 - 58 b) p_0 \\&&+2 (a^2 + b) p_0^2,\\
B_6&=&-8 a b^2 + (a^4 + 10 a^2 b - 9 b^2) p_0 \\&&+
      a (-a^2 + b) p_0^2,\\
B_7&=&b (a^4 + 4 a^2 b + 15 b^2) -
      a (a^4 + 8 a^2 b + 49 b^2) p_0 \nonumber\\&&+ (a^4 + 6 a^2 b -
         3 b^2) p_0^2,\\
B_8&=& a b (a^2 + b) - (a^4 + a^2 b + 2 b^2) p_0 + a b p_0^2,\\
B_9&=&b^2 (b - a p_0) \big(b + a (6 a + p_0)\big),\\
B_{10}&=&4 a b^3 (b - a p_0),\\
B_{11}&=&b^4 (b - a p_0).
\end{eqnarray*}
Now at the center of the star,
\begin{eqnarray}
\kappa \rho''&=&-10 (a^2 - b)<0,\label{x1}\\
\kappa p_r''&=&-4 a p_0<0,\label{x2}\\
\kappa p_t''&=&\frac{3 a^2 - 16 a p_0 + p_0^2}{2}<0.\label{x3}
\end{eqnarray}
From Eq.~(\ref{x1}), we have, \begin{eqnarray}\label{100}
b<a^2,
\end{eqnarray}
and Eq.~(\ref{x2}) is automatically satisfied. Eq.~(\ref{x3}) implies,
\begin{equation}\label{y1}
p_0 \in (0.189 a, 15.81a).
\end{equation}
The equation of state parameters $\omega_r$ and $\omega_t$ are given by,
\begin{eqnarray*}
\omega_r=\frac{p_r}{\rho}&=&\frac{p_0 (1-a r^2) \Psi}{
\xi_1},\\
\omega_t=\frac{p_t}{\rho}&=&\frac{1}{4\Psi \xi_1}\Big[p_0^2 r^2 (1-a r^2)^2 \Psi +
    +\varphi
   \Big],
\end{eqnarray*}
Where,\begin{eqnarray*}\varphi&=&p_0(4 - a^4 r^8 - 3 a^3 \xi_3 -
      a^2 \xi_4 -
      a r^2 \xi_5 +
      b r^4 \xi_6)\\&&+r^2 (a + b r^2) \Psi \xi_1,\\
\xi_3&=&r^6 (2 + b r^4),\\
\xi_4&=&r^4 + 11 b r^8 + 3 b^2 r^{12},\\
\xi_5&=&4 + b r^4 (-2 + b r^4) (6 + b r^4),\\
\xi_6&=&-2 + b r^4 (7 + b r^4).
\end{eqnarray*}

\begin{table*}\label{table1}
\caption{The observed and estimated mass and radius for a possible modelling of the compact star 4U 1608-52 \cite{4u} and Vela X-1 \cite{vela}.}
\begin{tabular*}{\textwidth}{@{\extracolsep{\fill}}lrrrrrrrrl@{}}
\hline
Compact Star &\multicolumn{1}{c}{$M/M_{\odot}$ }&\multicolumn{1}{c}{R(km)}&\multicolumn{1}{c}{$M/M_{\odot}$ }&\multicolumn{1}{c}{R(km)}\\
&(observed)&(observed)&(estimated)&(estimated)\\
\hline
4U 1608-52&$1.74\pm 0.14$ &$10.811\pm 0.197$&1.74&10.6\\
Vela X-1&$1.77\pm0.08$&$10.852\pm 0.108$&1.77&10.8\\
\hline
\end{tabular*}
\end{table*}

\begin{table*}\label{table2}
\caption{The numerical values of $a,\,b$ central density $(\rho_c)$ and surface density $(\rho_s)$ are obtained for a possible modelling of the compact star 4U 1608-52 \cite{4u} and Vela X-1 \cite{vela}.}
\begin{tabular*}{\textwidth}{@{\extracolsep{\fill}}lrrrrrrrrl@{}}
\hline
{Compact Star}&\multicolumn{1}{c}{$a$}&\multicolumn{1}{c}{$b$}&\multicolumn{1}{c}{$\rho_c$}&\multicolumn{1}{c}{$\rho_s$}\\
units&km$^{-2}$&km$^{-4}$&gm.cm$^{-3}$&gm.cm$^{-3}$\\
\hline
4U 1608-52&$0.00889996$&$-4.8392\times10^{-6}$&$1.43346\times10^{15}$&$4.54525\times10^{14}$\\
Vela X-1&$0.00857339$&$-4.70387\times10^{-6}$&$1.38086\times10^{15}$&$4.36711\times10^{14}$\\
\hline
\end{tabular*}
\end{table*}

\subsection{Mass function and redshift}
Introducing the relation between the mass function $m(r)$ and the metric potential $e^{\lambda}$, $e^{-\lambda}=1-\frac{2m}{r}$, the expression for mass function can be obtained as,
\begin{equation}
m(r)= \frac{r^3(a+br^2)}{ 2(1+ar^2+br^4)}.
\end{equation}
Using the formulae of the compactness factor $u(r)$ and surface redshift ($z_s$)
\begin{eqnarray*}
  u(r) &=& \frac{m}{r}, \\
  1+z_s(R) &=& \frac{1}{\sqrt{1-2u(R)}},
\end{eqnarray*}
the expression of these two quantities in our present model are obtained as,
 \begin{eqnarray*}
 u(r)&=& \frac{r^2(a+br^2)}{2(1+ar^2+br^4)},\\
 z_s&=&\sqrt{1+a R^2+b R^4}-1,
 \end{eqnarray*}
 where `$R$' being the radius of the star.
 The gravitational redshift is obtained as,
 \begin{eqnarray*}
   Z &=& \frac{1}{\sqrt{e^{\nu}}}-1, \\
   &=&\bigg[\exp\Big\{2ar^2+\frac{1}{4}\Big[\frac{
   2 p_0(a^2 + 2 b)}{b \sqrt{a^2-4b}} \tan^{-1}\Big(\frac{a + 2 b r^2}{\sqrt{a^2-4b}}\Big)\nonumber\\&&
    + b r^4-\frac{ap_0}{b} \log(1 + a r^2 + b r^4)\Big]+B\Big\}\bigg]^{-\frac{1}{2}}-1.
 \end{eqnarray*}
 Now the central value of the gravitational redshift is obtained as,
 \begin{eqnarray*}
 Z_0 = \frac{1}{\sqrt{\bigg[\exp\Big\{\frac{1}{4}\Big[\frac{
   2 p_0(a^2 + 2 b)}{b \sqrt{a^2-4b}} \tan^{-1}\Big(\frac{a}{\sqrt{a^2-4b}}\Big)
    \Big]+B\Big\}\bigg]}}-1.\end{eqnarray*}
    The gradient of the gravitational redshift is obtained as,
 \begin{eqnarray*}
 \frac{dZ}{dr} &=& -\frac{1}{2\sqrt{e^{\nu}}}\Big[r\frac{p_0(1-ar^{2})}{1+ar^2+br^4}+r(a+br^2)\Big].
 \end{eqnarray*}
 Now at the center of the star $\frac{dZ}{dr}=0$, and \[\left(\frac{d^2Z}{dr^2}\right)_{r=0}=-\frac{(a + p_0)}{
 2 \sqrt{e^{
  B + \frac{(a^2 + 2 b) p_0 \tan^{-1}\left(\frac{a}{\sqrt{a^2 - 4 b}}\right)}{
   2 b \sqrt{a^2 - 4 b}}}}}<0.\] It verifies that the gravitational redshift has maximum value at the center of the star.

\subsection{Mass-radius relationship}
\begin{figure}[htbp]
    \centering
        \includegraphics[scale=.45]{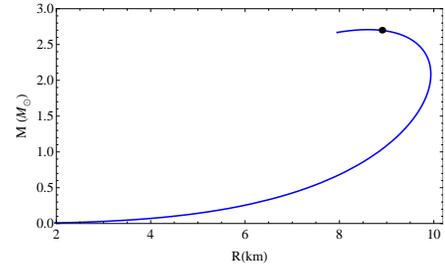}
       \caption{Mass-radius relationship.\label{mr}}
\end{figure}

We have generated the mass-radius ($M - R$) relationship for our developed model as shown in Fig.~\ref{mr}. The mass-radius relationship obtained for an assumed surface density $8.5 \times 10^{14}$ gm/cc. The chosen surface density is roughly close to that considered by Sharma and Maharaj \cite{Sharma1}. The maximum mass allowed in this model is found to be $2.7 M_\odot$ with the corresponding radius of value $8.9$ km. This limit on maximum mass for a neutron star is approximately $3.2 M_\odot$ \cite{Ruffini}.

\subsection{Radius-Central density and mass-central density relation}
\begin{figure}[htbp]
    \centering
        \includegraphics[scale=.45]{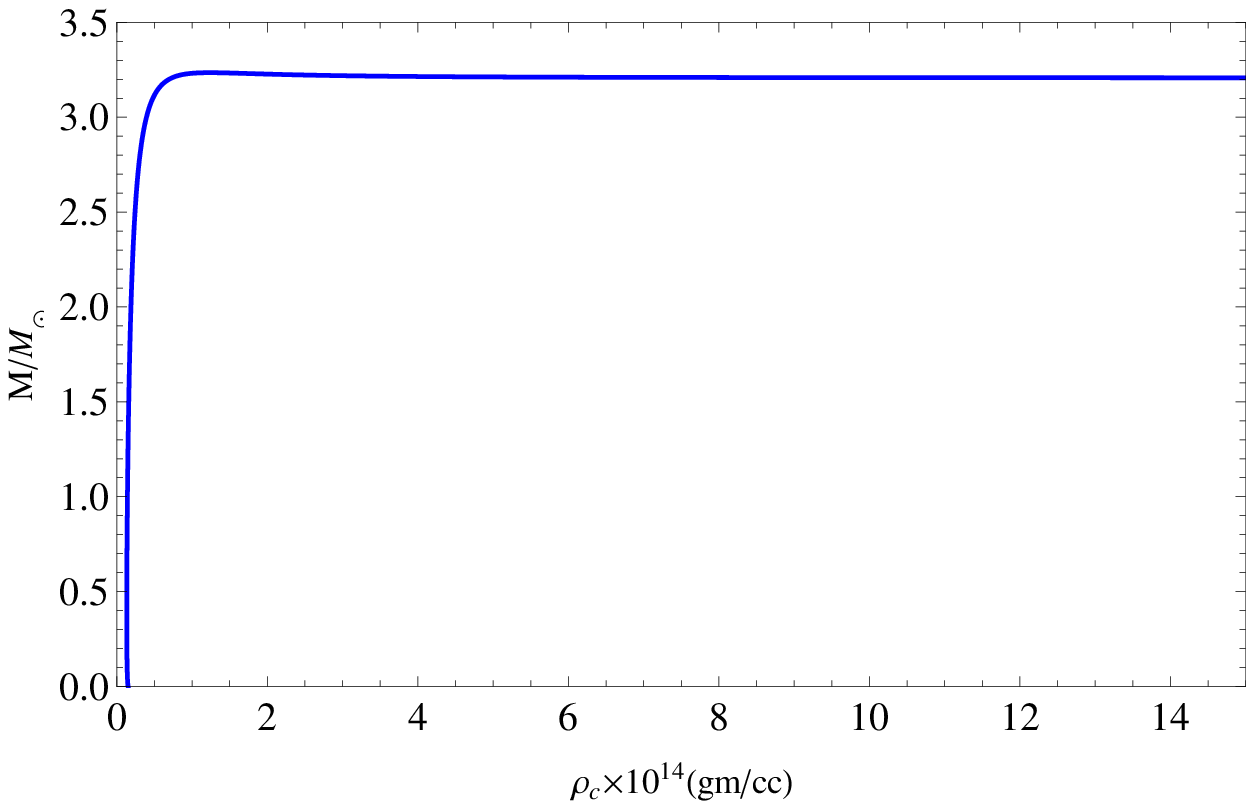}
        \includegraphics[scale=.45]{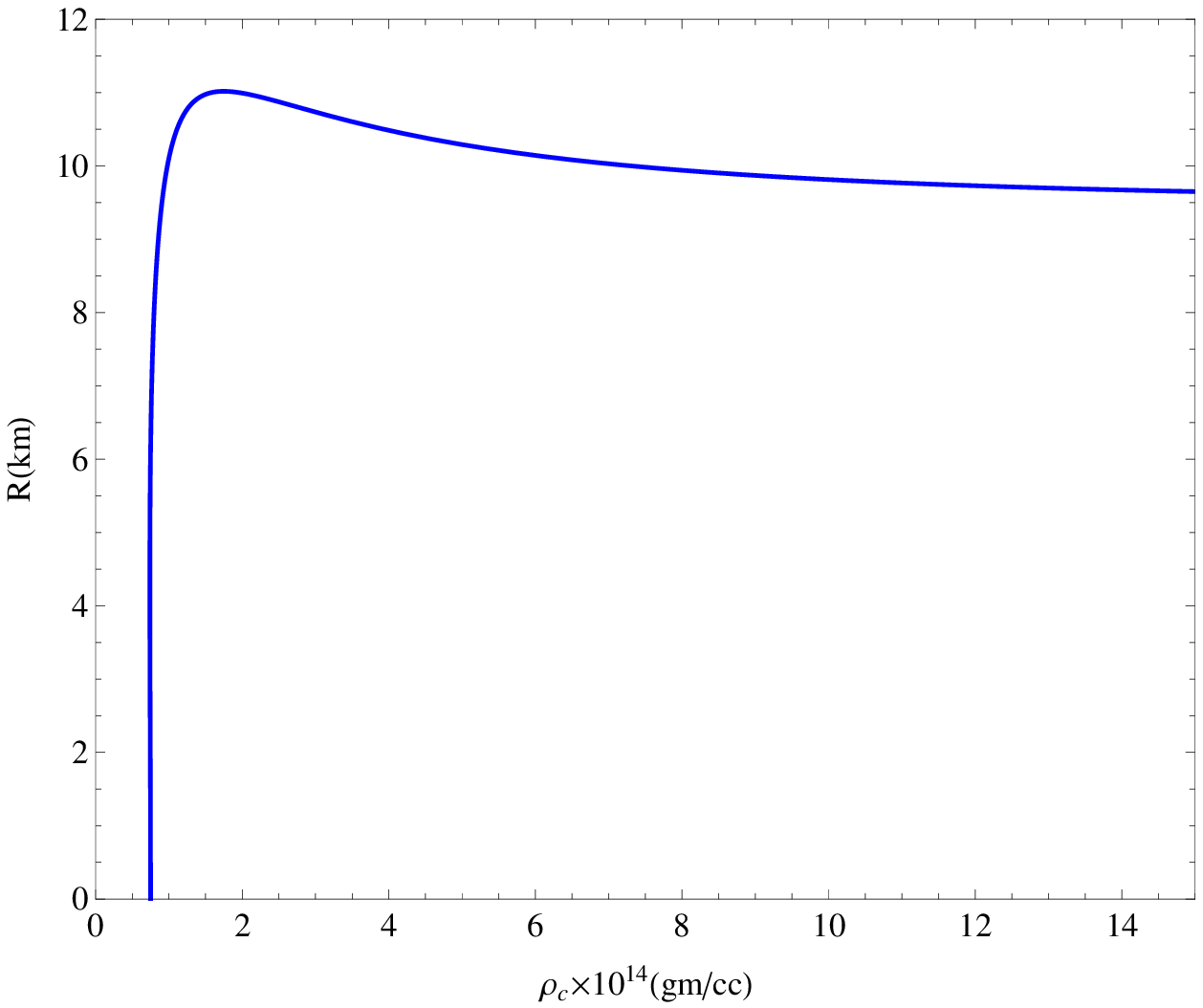}
       \caption{Variation of radius and mass with central density.\label{cendenrm}}
\end{figure}

For compact objects we have plotted the variation of the radius and mass with the central density in Fig.~\ref{cendenrm}. This plot allows us to determine the central density of compact star once the radius or the mass of the corresponding star is known. Similar type of study can be found in the work of Deb et al. \cite{Deb18}.

\subsection{Energy Conditions}
It is well known that for a compact star model, the energy conditions should be satisfied and in this subsection we are in a position to study about it. For an anisotropic compact star, all the energy conditions like, Weak Energy Condition (WEC), Null Energy Condition (NEC), Strong Energy Condition (SEC)  and dominant energy conditions (DEC) are satisfied if and only if the following inequalities hold simultaneously for every points inside the stellar configuration.

\begin{figure}[htbp]
        \includegraphics[scale=.3]{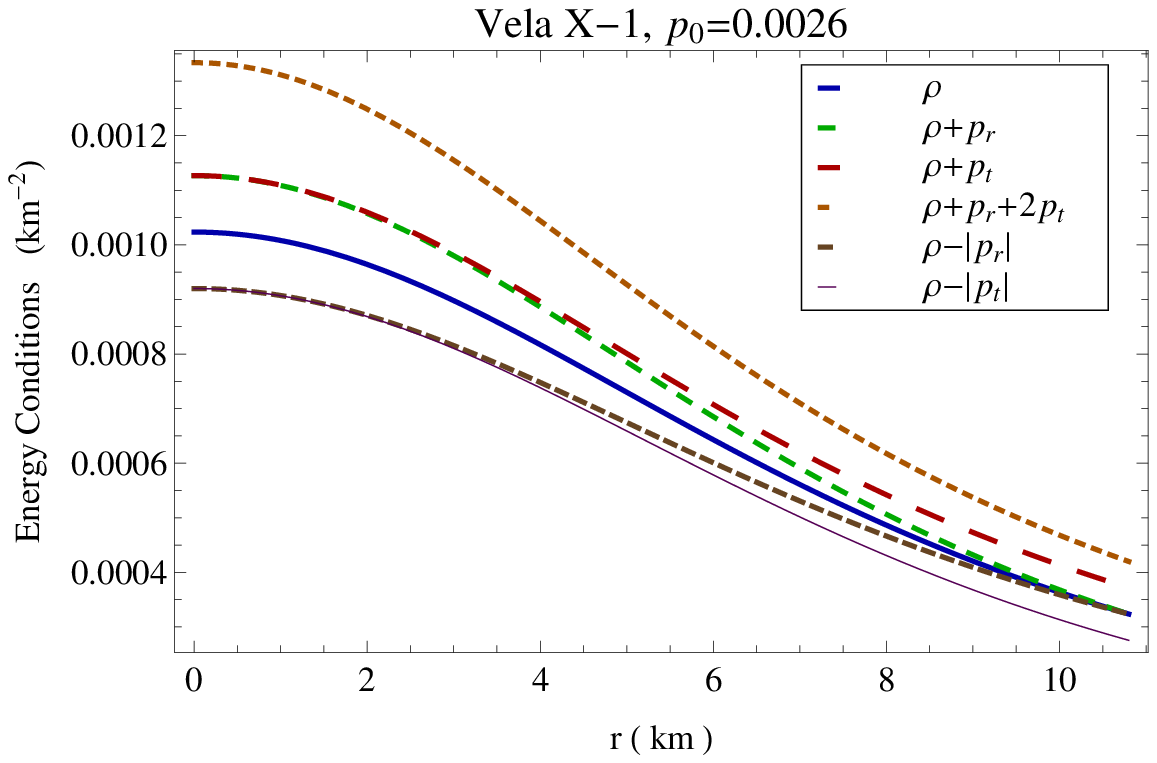}
        \includegraphics[scale=.3]{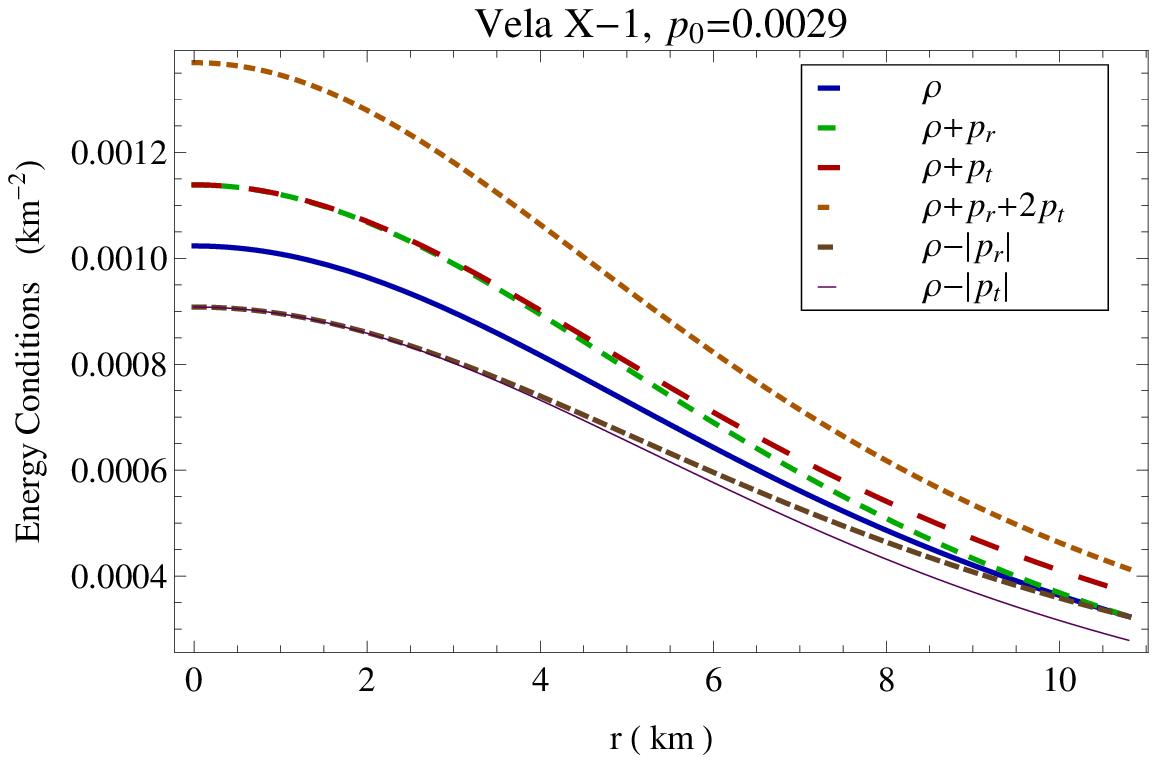}
        \includegraphics[scale=.3]{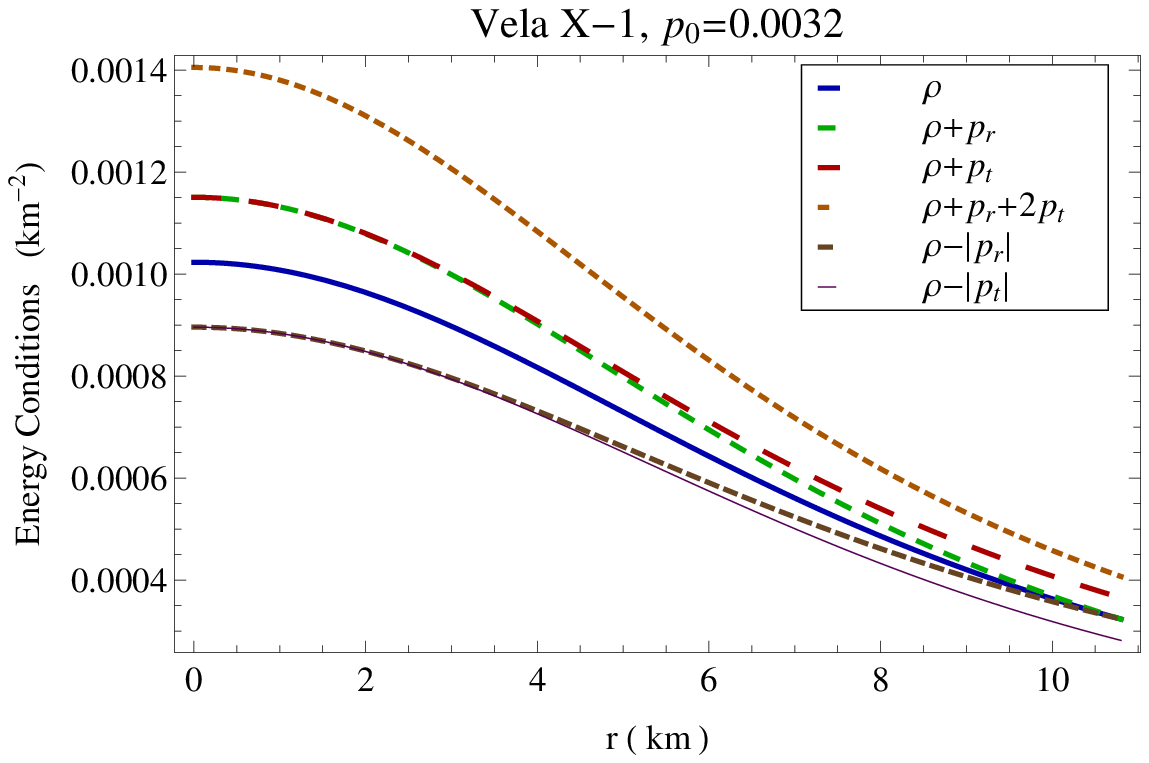}
        \includegraphics[scale=.3]{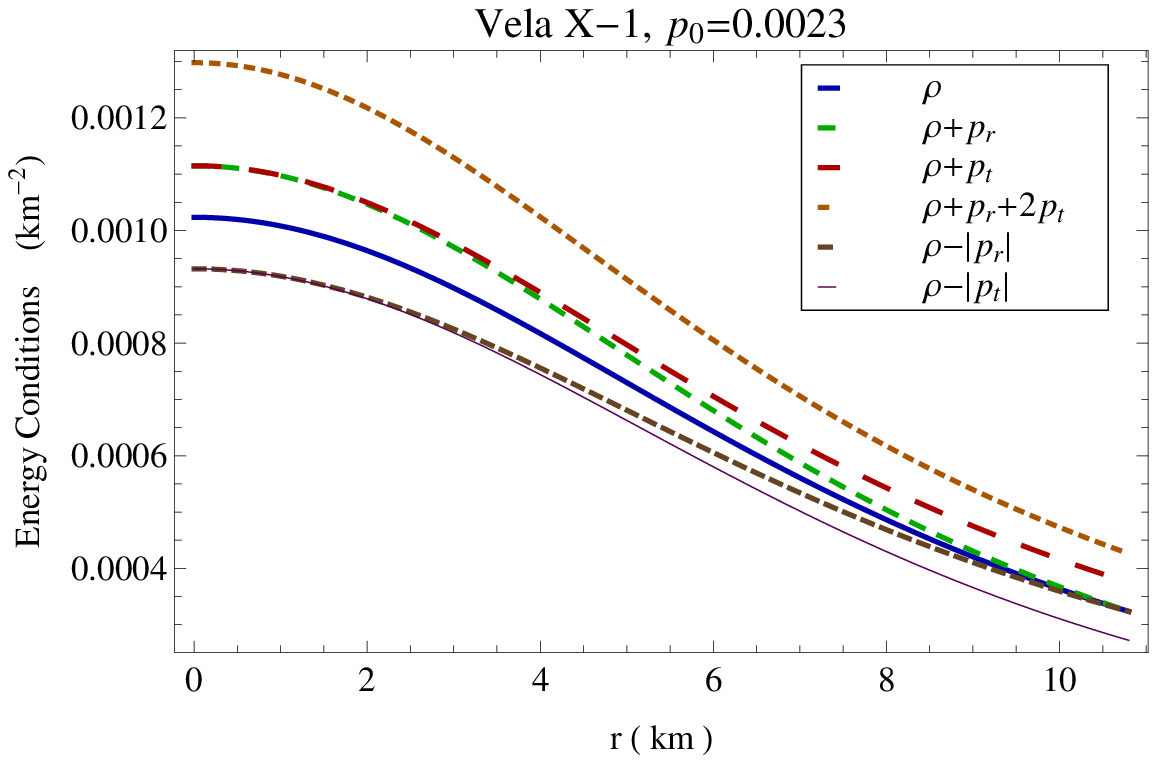}
        \includegraphics[scale=.3]{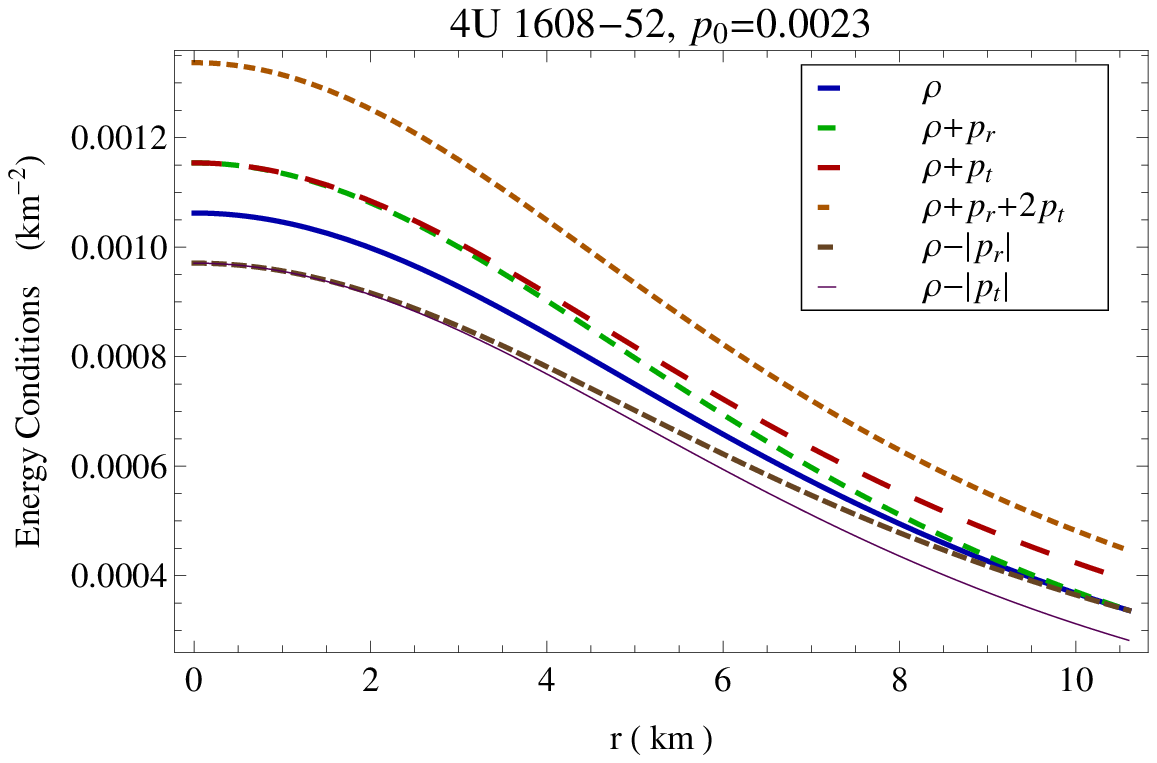}
        \includegraphics[scale=.3]{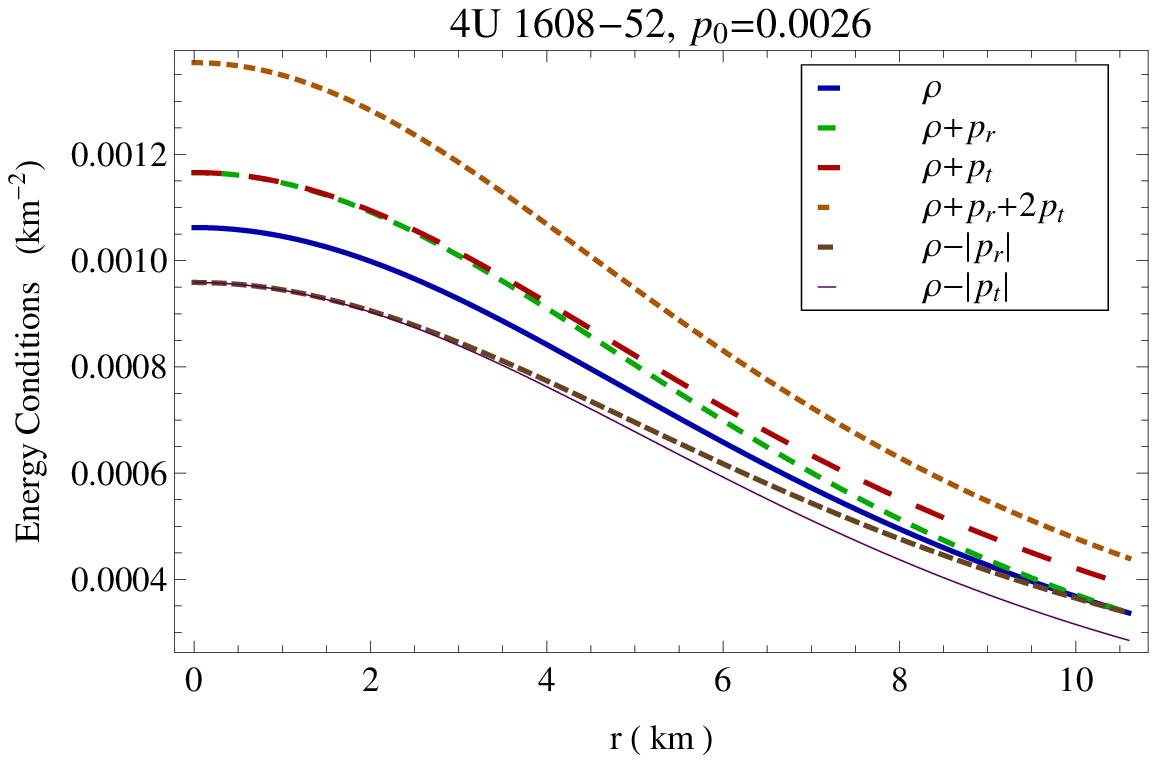}
        \includegraphics[scale=.3]{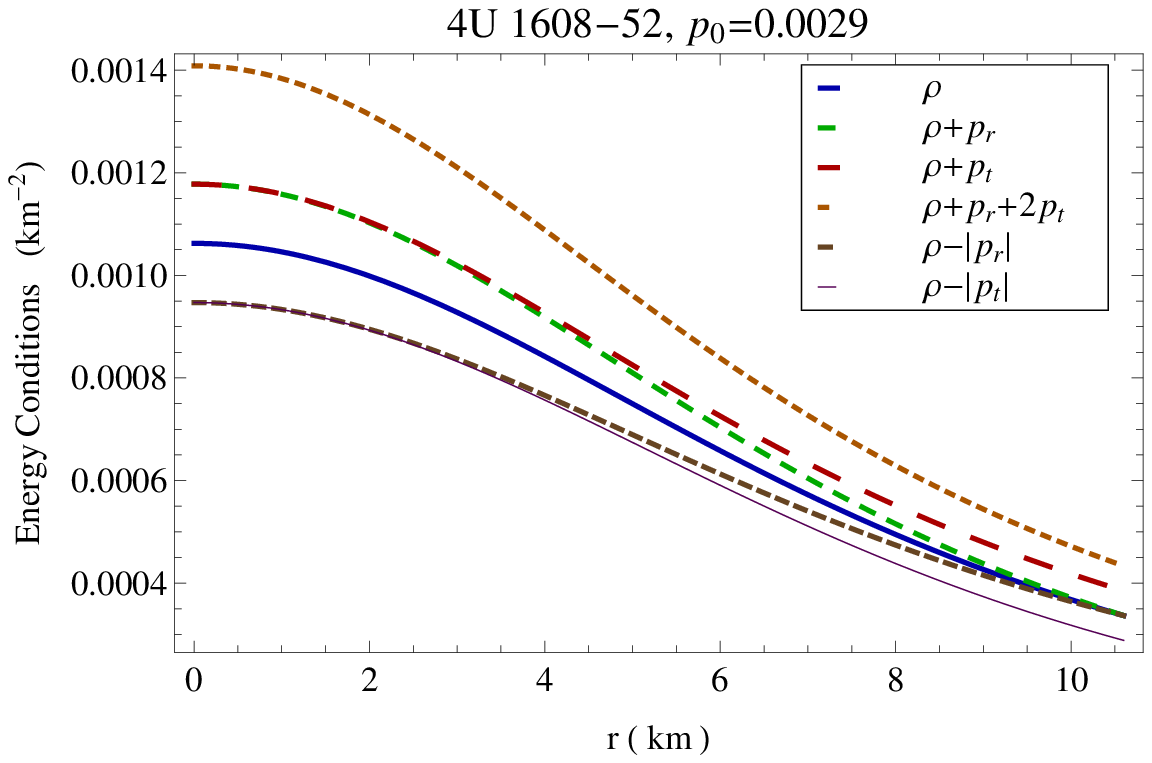}
        \includegraphics[scale=.3]{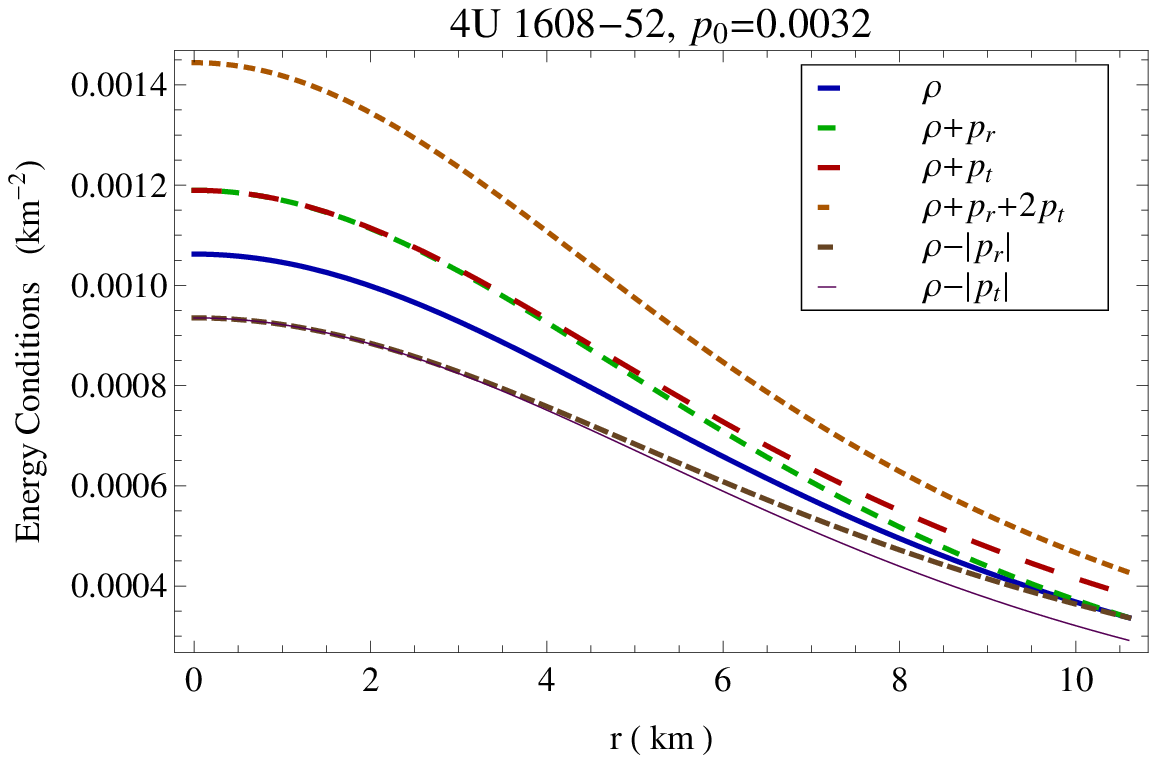}
       \caption{NEC,~WEC,~SEC and DEC are plotted against the radius $r$ for a possible modelling of the compact star Vela X-1 and 4U 1608-52 for different values of $p_0$ mentioned in the figures.\label{ec1}}
\end{figure}

\begin{eqnarray}\label{1}
WEC &:& T_{\mu \nu}t^\mu t^\nu \ge 0~\mbox{or}~\rho \geq  0,~\rho+p_i \ge 0  \label{2k}\\
NEC &:& T_{\mu \nu}l^\mu l^\nu \ge 0~\mbox{or}~ \rho+p_i \geq  0\\ \label{3}
DEC &:& T_{\mu \nu}t^\mu t^\nu \ge 0 ~\mbox{or}~ \rho \ge |p_i|  \\ \label{4}
SEC &:& T_{\mu \nu}t^\mu t^\nu - \frac{1}{2} T^\lambda_\lambda t^\sigma t_\sigma \ge 0 ~\mbox{or}~ \rho+\sum_i p_i \ge 0,\label{4k}\nonumber\\
\end{eqnarray}
where $i$ takes the value $r$ and $t$ for radial and transverse pressure. $~t^\mu$ and $l^\mu$ are time-like vector and null vector respectively and $T^{\mu \nu}t_\mu $ is nonspace-like vector. To check all the inequality stated above we have drawn the profiles of ~l.h.s of (\ref{2k})-(\ref{4k}) in fig~\ref{ec1} in the interior of the compact star PSR J 1614-2230 and 4U1608-52. The figure shows that all the energy conditions are well satisfied by our present model of compact star.

\section{Stability analysis}\label{6x}
In this section we want to check the stability of the present model with the help of (i) causality condition, (ii) relativistic adiabatic index (iii) Harrison-Zeldovich-Novikov's stability condition and (iv) tov equation.
\subsection{Causality Condition}
For a compact star model, the radial and transverse velocity of sound is obtained as,
$$V_r^2=\frac{dp_r}{d\rho}, \hspace{1cm} V_t^2=\frac{dp_t}{d\rho}.$$
\begin{figure}[htbp]
        \includegraphics[scale=.3]{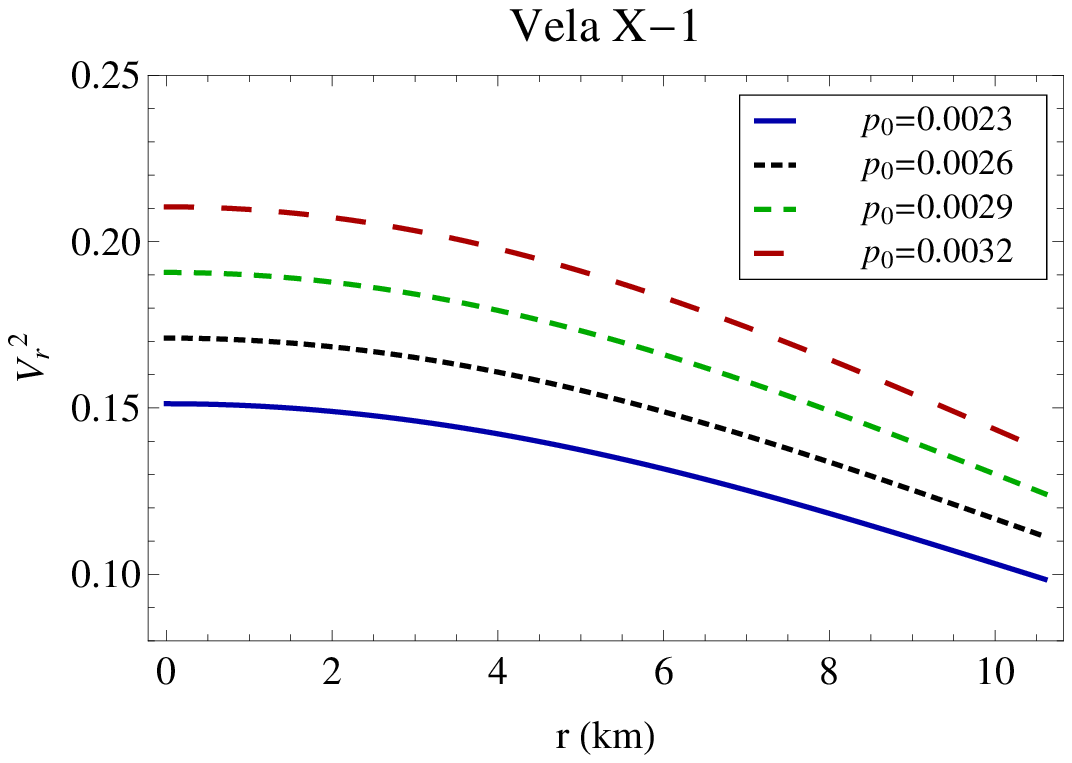}
        \includegraphics[scale=.3]{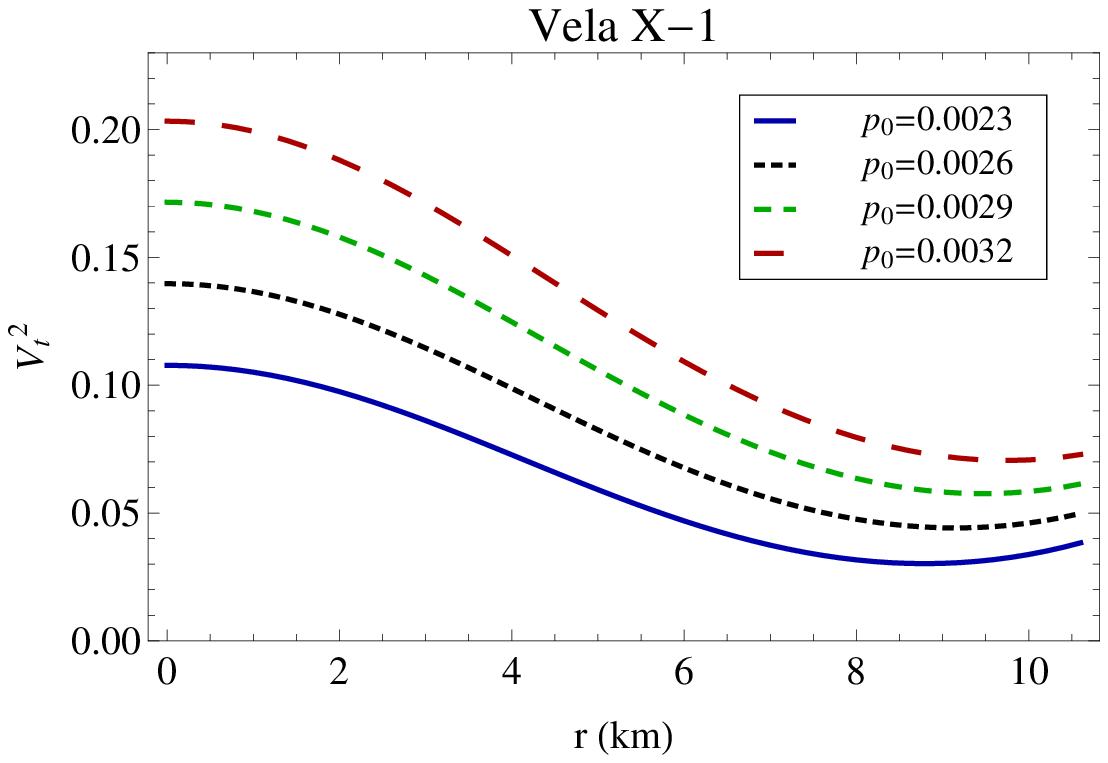}
        \includegraphics[scale=.3]{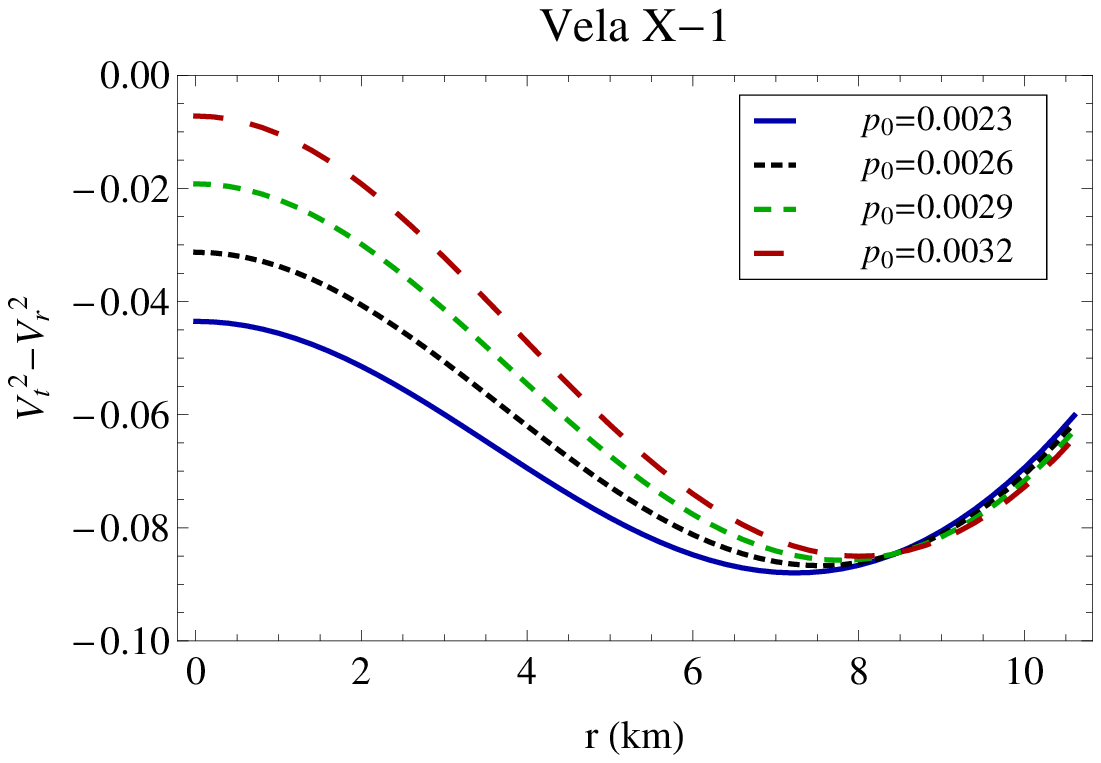}
        \includegraphics[scale=.3]{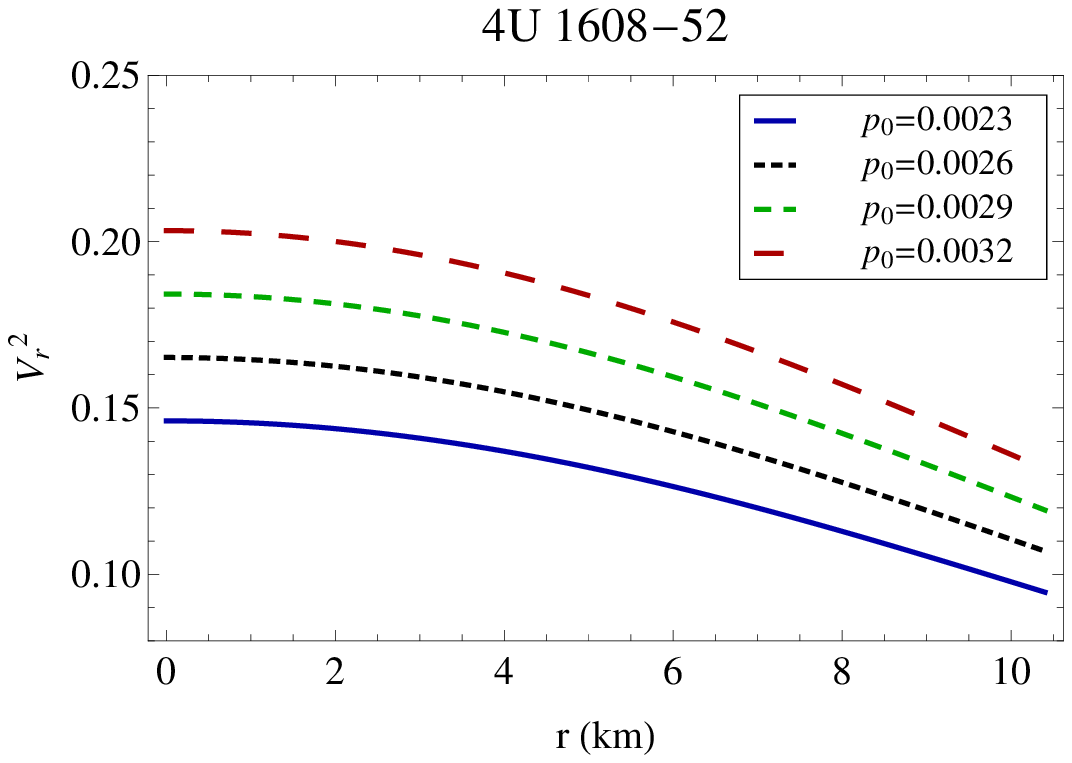}
        \includegraphics[scale=.3]{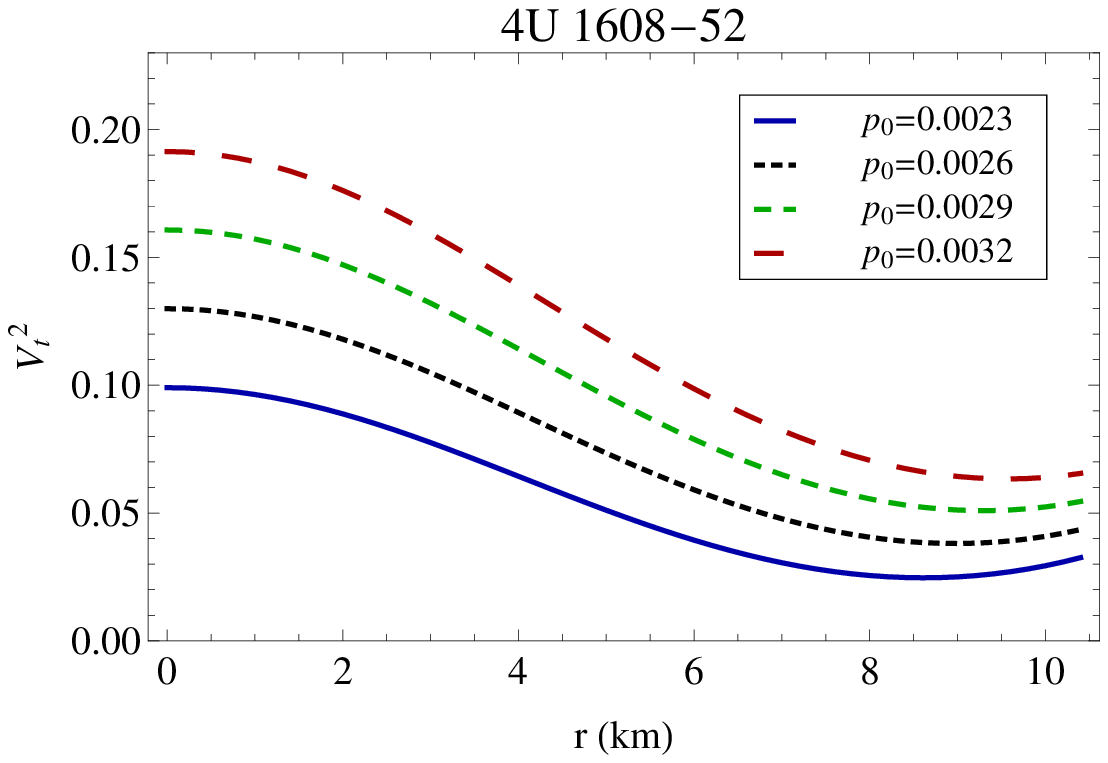}
        \includegraphics[scale=.3]{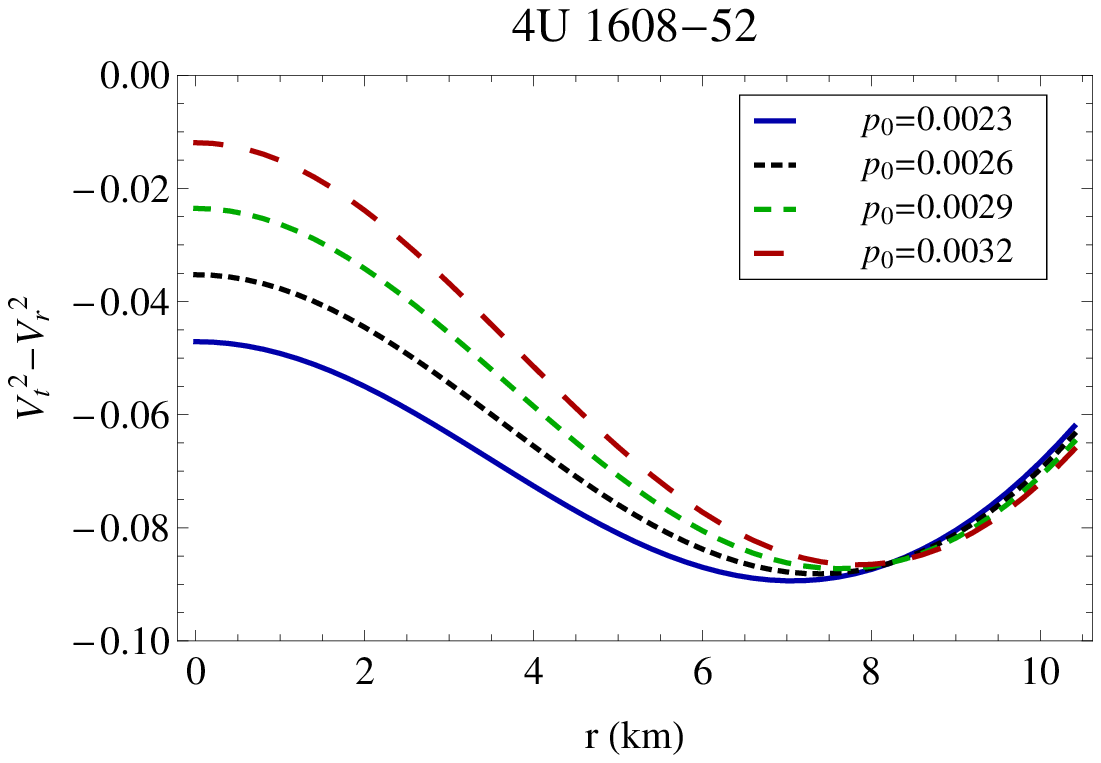}
       \caption{$V_r^2,\,~V_t^2$ and $V_t^2-V_r^2$ are plotted against $r$ for $p_0=0.0023$, $p_0=0.0026$, $p_0=0.0029$ and $p_0=0.0032$
respectively for different values of `a' and `b' mentioned in Table~II \label{v100} for a possible modelling of the compact stars  Vela X-1 and 4U 1608-52.}
\end{figure}
Using the above formulae, for our present model, the radial and transverse velocity of sound are calculated as,
\begin{eqnarray}
V_r^2&=&\frac{p_0 \Psi} {
\zeta}\Big(2 b r^2 + a (2 - b r^4)\Big),\\
V_t^2&=&\frac{1}{4 \Psi \zeta}\Big[C_3 -
   2 C_4 r^2 + C_5 r^4 +
   4 C_6 r^6 + C_7 r^8  \nonumber\\&&-
  C_8 r^{10}+
   C_9 r^{12} +
  C_{10} r^{14} +
   C_{11} r^{16}\Big],
\end{eqnarray}
where \begin{eqnarray*}\zeta=5 a^2 - 5 b +
     C_1 r^2 + C_2 r^4 + 3 a b^2 r^6 +
     b^3 r^8,\end{eqnarray*}
     and $C_i's$ are constants given by,
\begin{eqnarray*}
C_1&=&a (a^2 + 13 b),\\
C_2&=&3 b (a^2 + 4 b),\\
C_3&=&-3 a^2 + 16 a p_0 - p_0^2,\\
C_4&=&a^3 + 8 a b + 3 a^2 p_0 - 14 b p_0 - 2 a p_0^2,\\
C_5&=&a^4 -
      19 a^2 b - 15 b^2 + a (17 a^2 - 58 b) p_0 \\&&+
      2 (a^2 + b) p_0^2,\\
      C_6&=&-8 a b^2 + (a^4 + 10 a^2 b - 9 b^2) p_0 +
      a (-a^2 + b) p_0^2,\\
      C_7&=&a^5 p_0 + 8 a^3 b p_0 + 49 a b^2 p_0 +
      3 b^2 (-5 b + p_0^2) \nonumber\\&&- a^4 (b + p_0^2) -
      2 a^2 b (2 b + 3 p_0^2),\\
      C_8&=& 4 b \big(a b (a^2 + b) - (a^4 + a^2 b + 2 b^2) p_0 + a b p_0^2\big),\\
      C_9&=&b^2 (-b + a p_0) \big(b + a (6 a + p_0)\big),\\
      C_{10}&=& 4 a b^3 (-b + a p_0),\\
      C_{11}&=&b^4 (-b + a p_0).
\end{eqnarray*}
Now at the point $r=0$, the square of radial and transverse velocity of sound are obtained as,
\begin{eqnarray*}
V_r^2=\frac{3 a p_0}{5 (a^2 - b)},~~
V_t^2=-\frac{3 a^2 - 20 a p_0 + p_0^2}{20 (a^2 - b)}.
\end{eqnarray*}
Moreover, at the center of the star,
\begin{eqnarray}
  V_t^2-V_r^2 &=& -\frac{3 a^2 - 8 a p_0 + p_0^2}{20 (a^2 - b)}.
\end{eqnarray}

Now by using Le Chatelier's principle we have the speed of sound must be positive inside the stellar interior, i.e., $0<\frac{p_r'}{\rho'},\frac{p_t'}{\rho'}$. We also know that for a physically acceptable model, the velocity of the sound (both radial and transverse) should be less than the speed of the light i.e., both $\frac{p_r'}{\rho'},\frac{p_t'}{\rho'}<1$ which is known as the causality condition. Here  $\rho'$ and $p_r'$ represent differentiation with respect to r. Combining the above two inequalities we have, $0\leq V_r^2,~V_t^2 \leq 1.$\par
To find a potentially stable region, in 1992, Herrera proposed a method which is known as ``cracking method''. This method tells us that a stellar model will be potentially stable if the square of radial velocity of sound exceeds the square of transverse velocity of sound everywhere within the stellar model, otherwise the stellar model will be potentially unstable i.e.,
\begin{eqnarray*}
  V_t^2-V_r^2\left\{
               \begin{array}{ll}
                 <0 ~\text{for}~ 0 \leq r \leq R~ \Rightarrow~& \hbox{potentially stable} \\
                 >0~\text{for}~ 0 \leq r \leq R~\Rightarrow~ &\hbox{unstable}.
               \end{array}
             \right.
\end{eqnarray*}

Now at the center of the star, $V_t^2-V_r^2~<0$ gives,
\begin{eqnarray}\label{101}
-\frac{3 a^2 - 8 a p_0 + p_0^2}{20 (a^2 - b)}<0
\end{eqnarray}
Using, Eq.~(\ref{100}) from Eq.~(\ref{101}), we further obtain the following inequality:
\begin{eqnarray*}3 a^2 - 8 a p_0 + p_0^2>0.
\end{eqnarray*}
The above inequality gives,
\begin{eqnarray}\label{102}
  p_0 \not\in (0.3944a,~7.6055a).
\end{eqnarray}
Combining (\ref{y1}) and (\ref{102}), we further get,
\begin{eqnarray}
  p_0 \in (0.18975a,~0.3944a)\cup(7.6055a,~15.8102a). \label{centp}
\end{eqnarray}
Hence we have obtained a range for $p_0$ that could describe a physically reasonable stellar structure.
\subsection{Relativistic Adiabatic index}
The adiabatic index determines the stability of a compact object and for an anisotropic stellar configuration it is defined as,
\begin{eqnarray}
\Gamma_r =\frac{\rho(r)+p_r(r) }{ p_r(r)} \, \frac{dp_r(r)}{d\rho(r)}.
\end{eqnarray}

\begin{figure}[htbp]
        \includegraphics[scale=.45]{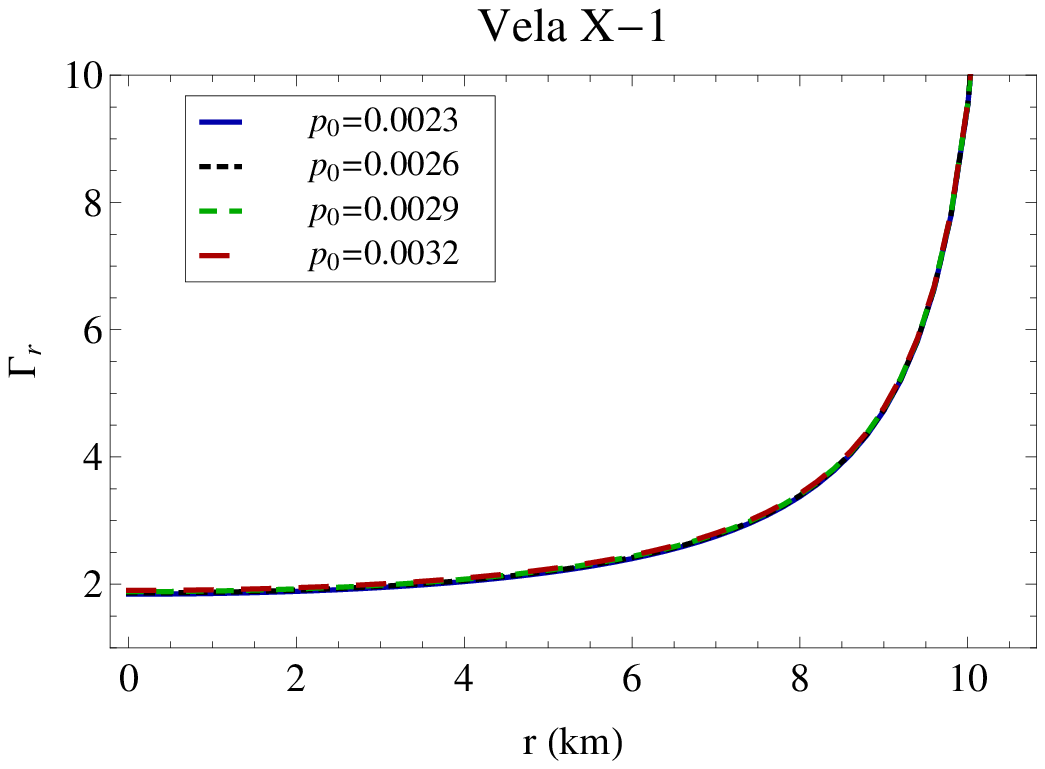}
        \includegraphics[scale=.45]{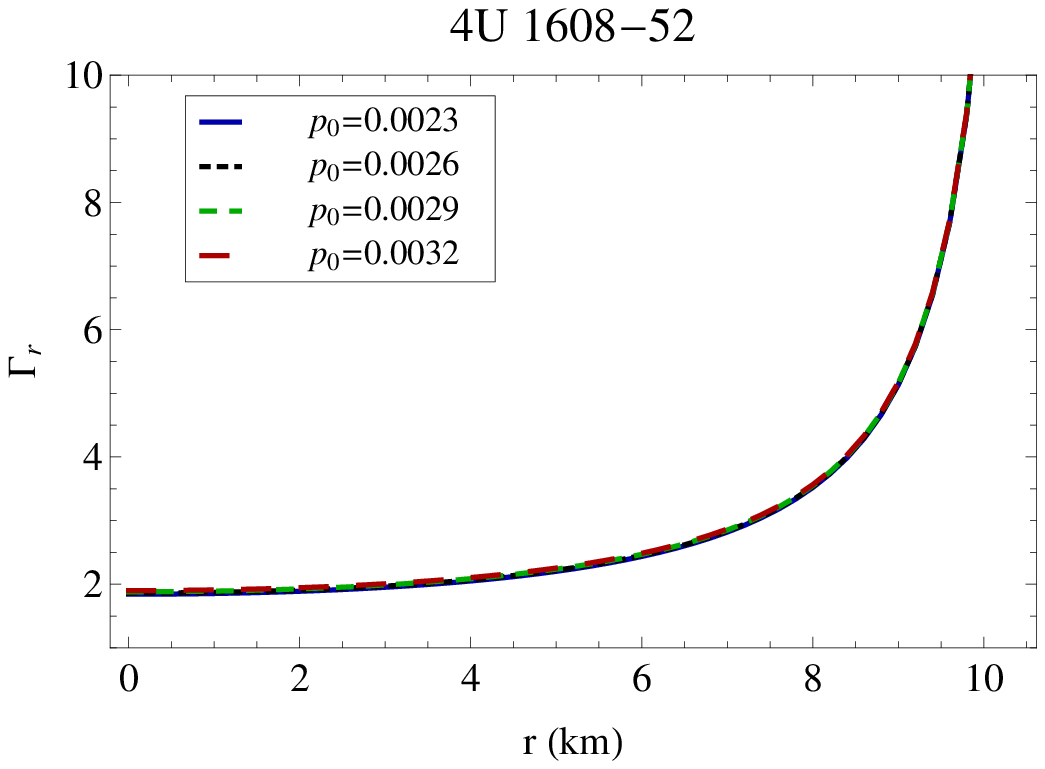}
       \caption{Adiabatic Index is plotted against $r$ inside the stellar interior for a possible modelling of the compact star Vela X-1 and 4U 1608-52 for different values of $p_0$ mentioned in the figures.\label{gamma}}
\end{figure}
For our model we have
\begin{eqnarray*}
\Gamma_r&=&\frac{3 a + p_0 + (a^2 + 5 b) r^2 + D_1 r^4 +
D_2 r^6}{p_0 (1- a r^2) \Psi}V_r^2,\\
\end{eqnarray*}
where $D_1$ and $D_2$ are constants given by,
$$D_1=2 a b - a^2 p_0 + b p_0~;~~D_2= b (b - a p_0).$$
Any stellar configuration will maintain its stability if adiabatic  index $\Gamma_r > 4/3 $ \cite{Heintzmann}. For our solution, the adiabatic index $\Gamma_r$ takes the value more than $4/3$ throughout the interior of the compact star, as evident from Fig.~\ref{gamma}.

\begin{table*}\label{table3}
\caption{The numerical values of central pressure ($p_c$) in dyne$\cdot$cm$^{-2}$ unit, surface anisotropy in dyne$\cdot$cm$^{-2}$ unit, central radial velocity ($V_r(0)$), surface radial velocity ($V_r(R)$), central transverse velocity ($V_t(0)$), surface transverse velocity ($V_t(R)$) for a possible modelling of the compact star 4U 1608-52 \cite{4u}.}
\begin{tabular*}{\textwidth}{@{\extracolsep{\fill}}lrrrrrrrrl@{}}
\hline
$p_0$ &\multicolumn{1}{c}{$p_c$}&\multicolumn{1}{c}{$\Delta_s=p_t(R)$}&\multicolumn{1}{c}{$V_{r}(0)$}&\multicolumn{1}{c}{$V_{r}(R)$}&\multicolumn{1}{c}{$V_t(0)$}&\multicolumn{1}{c}{$V_t(R)$}&\multicolumn{1}{c}{$\Gamma_{r}(0)$}\\
\hline
$0.0023$ &$1.11134\times10^{35}$ &$6.64583\times10^{34}$&$0.382269$&$0.305075$&$0.314703$&$0.185608$&$1.84249$\\
$0.0026$&$1.2563\times10^{35}$ &$6.26024\times10^{34}$&$0.406435$&$0.324362$&$0.360459$&$0.212585$&$1.86155$\\
$0.0029$&$1.40126\times10^{35}$ &$5.87465\times10^{34}$&$0.429243$&$0.342564$&$0.400895$&$0.236505$&$1.88061$\\
$0.0032$&$1.54621\times10^{35}$&$5.48905\times10^{34}$&$0.450899$&$0.359847$&$0.437487$&$0.258218$&$1.89967$\\
\hline
\end{tabular*}
\end{table*}

\subsection{Tov equation}
Now we are ready to check the stability of our present model under three different forces $viz$ gravitational force $F_g$, hydrostatics force $F_h$ and anisotropic force $F_a$. The effect of the above three forces can be described by the conservation equation given by
\begin{equation}\label{tov1}
\nabla^{\mu}T_{\mu\nu}=0,
\end{equation}
known as TOV equation.
Now using the expression given in (\ref{tmu}) into (\ref{tov1}) one can obtain the following equation:
\begin{equation}\label{tov3}
-\frac{\nu'}{2}(\rho+p_r)+\frac{2}{r}(p_t-p_r)=p_r'.
\end{equation}
The eqn.~\eqref{tov3} can be written as,
\begin{equation}
F_g+F_h+F_a=0,
\end{equation}

\begin{figure}[htbp]
        \includegraphics[scale=.3]{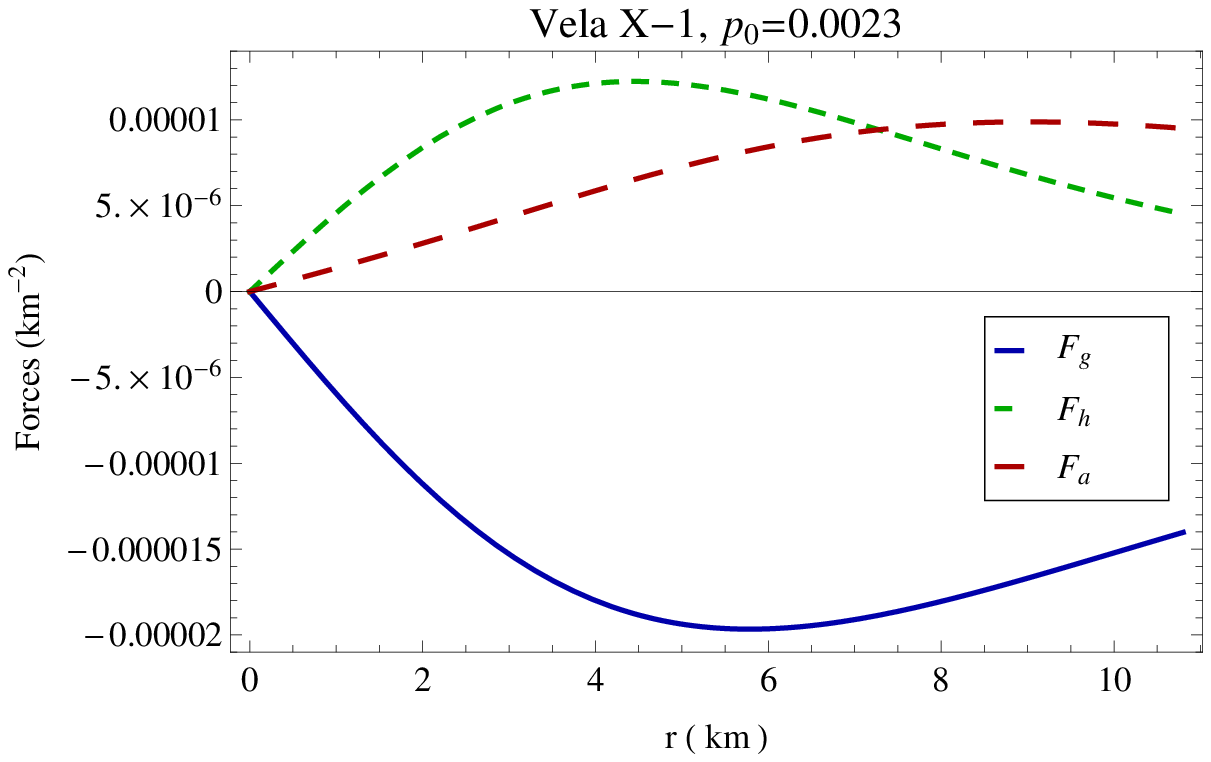}
        \includegraphics[scale=.3]{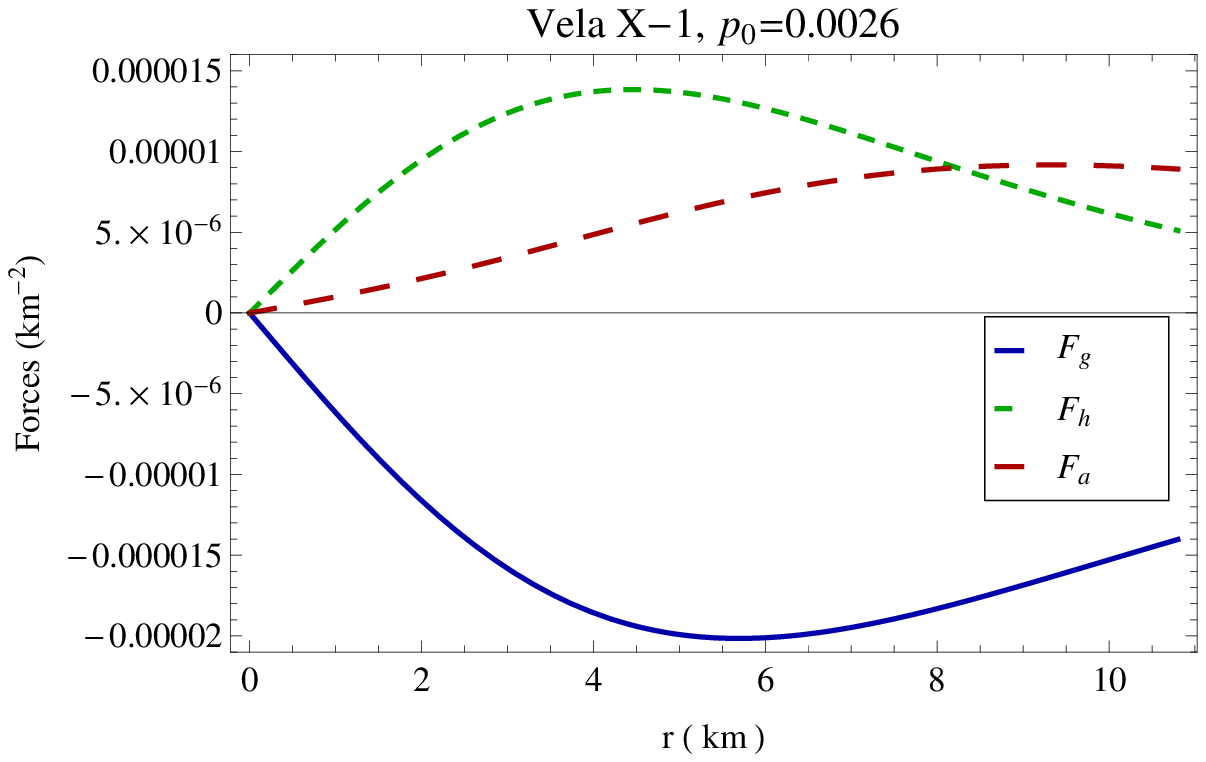}
        \includegraphics[scale=.3]{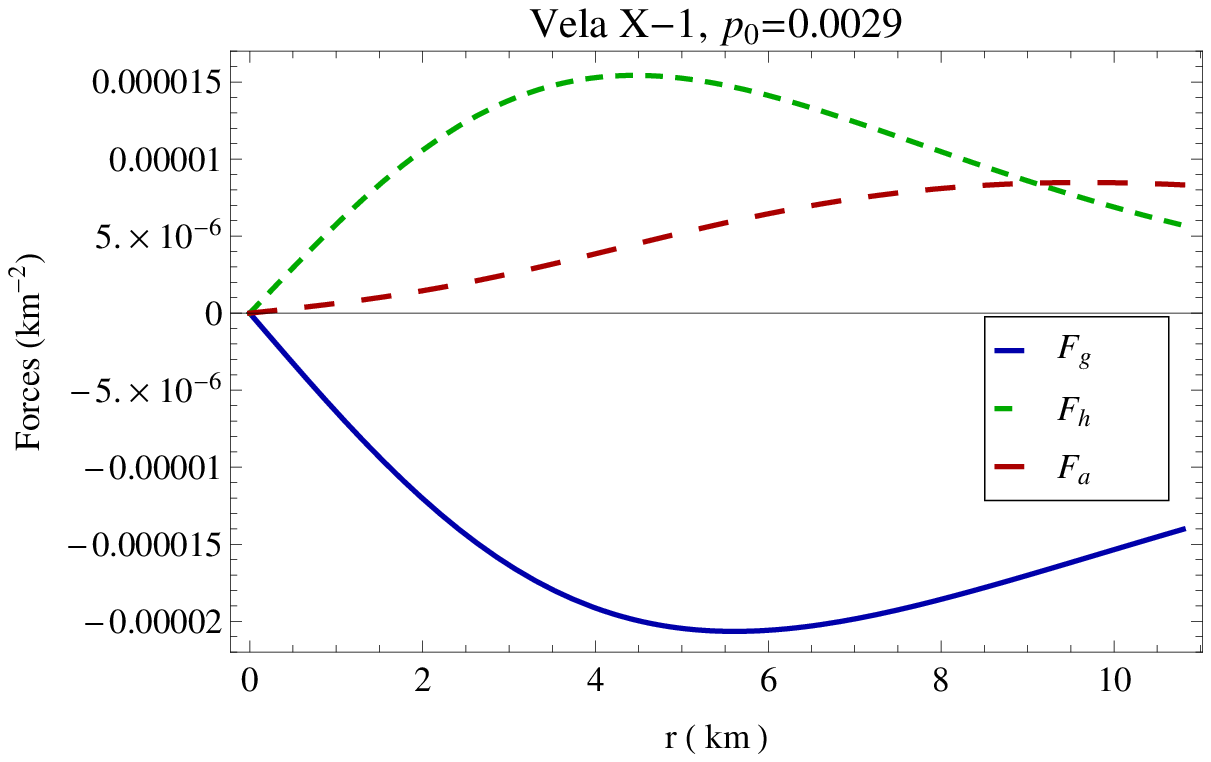}
        \includegraphics[scale=.3]{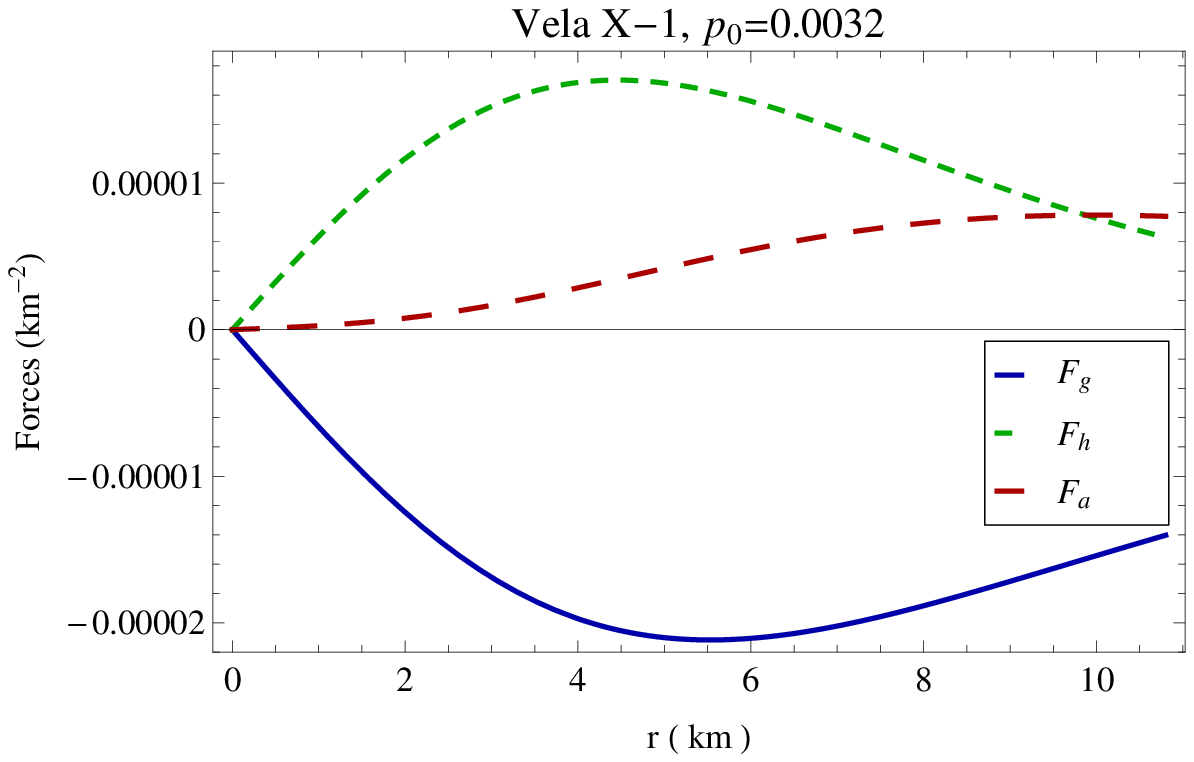}
        \includegraphics[scale=.3]{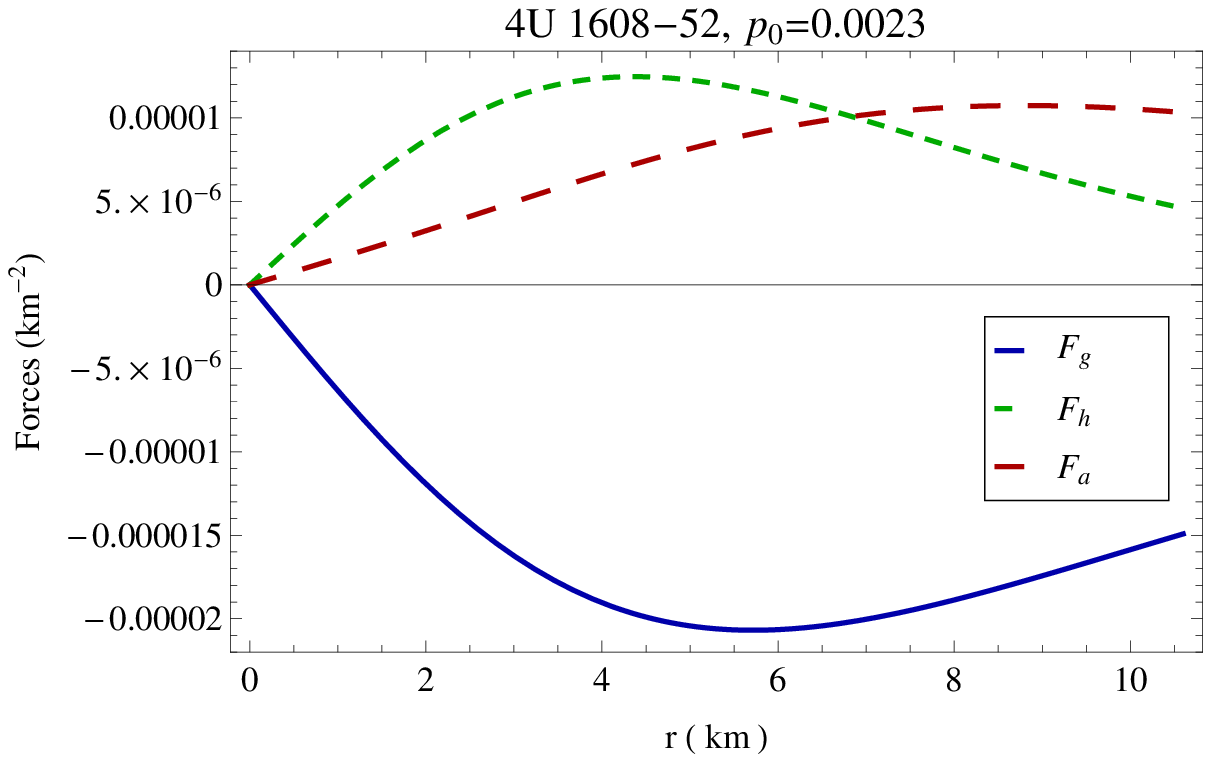}
        \includegraphics[scale=.3]{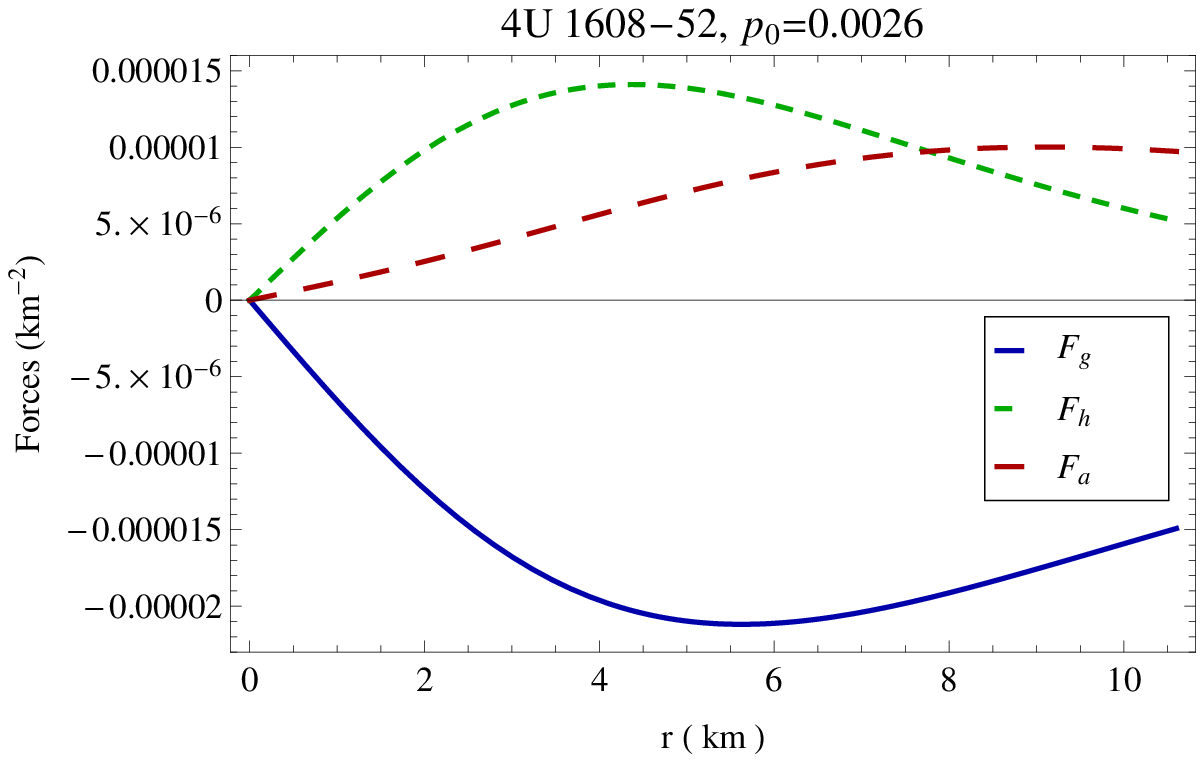}
        \includegraphics[scale=.3]{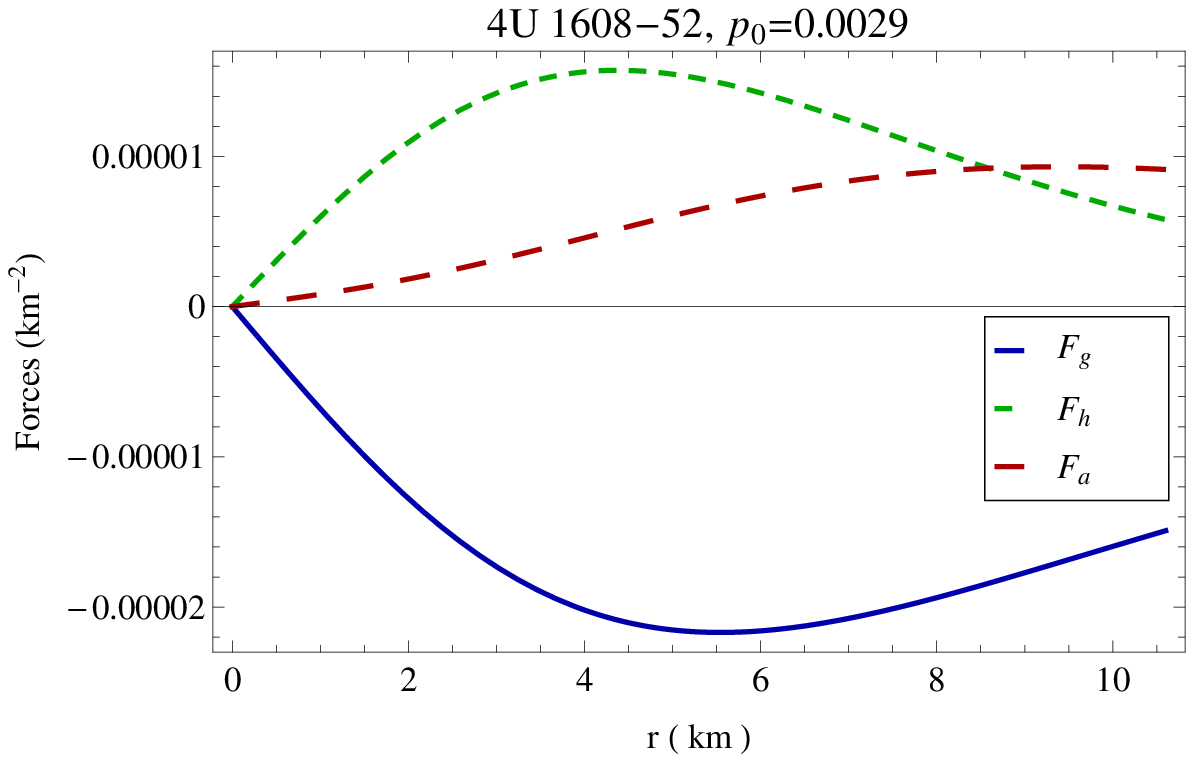}
        \includegraphics[scale=.3]{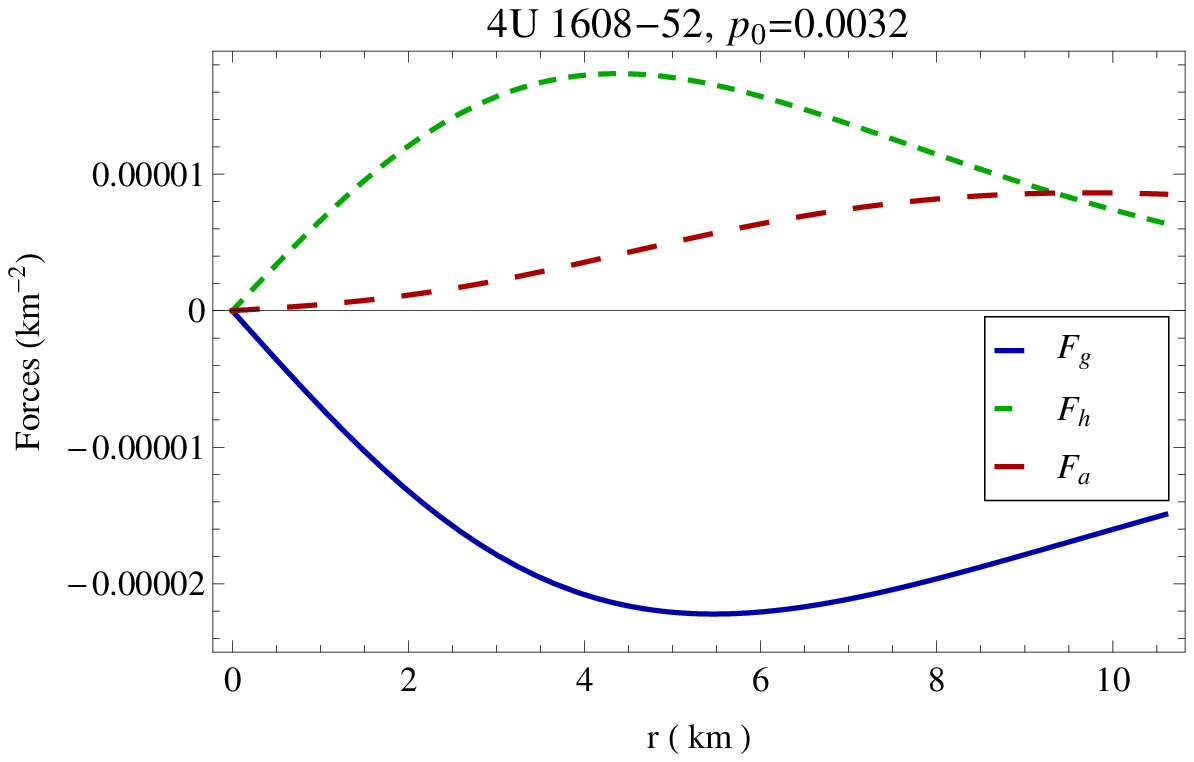}
       \caption{Different forces are plotted against radial parameter $r$ for a possible modelling of the compact star Vela X-1 and 4U 1608-52 for different values of $p_0$ mentioned in the figures.\label{tov}}
\end{figure}

where the expression for $F_g,\,F_h$ and $F_a$ are obtained as:
\begin{eqnarray}
F_g & = & -\frac{\nu'}{2}(\rho+p_r)\nonumber\\
&=&\frac{r}{2 \kappa \Psi^3} \Big\{\Big(a + p_0 + (a^2 + b - a p_0) r^2 + 2 a b r^4 \nonumber\\&& + b^2 r^6\Big)\times\Big(3 a +
   p_0 + (a^2 + 5 b - a p_0) r^2 \nonumber\\&&+ 2 a b r^4 + b^2 r^6\Big)\Big\},\\
F_h &=& - \frac{dp_r}{dr}=\frac{2 p_0}{\kappa\Psi^2} \Big(2 b r^3 + a r (2-b r^4)\Big),\\
F_a &=& \frac{2\Delta}{r}=\frac{r}{8\kappa \Psi^3}\big[A_1 +
    A_2 r^2 + A_3 r^4 +
   A_4 r^6 +
   A_5 r^8\nonumber\\&&+ 4 a b^3 r^{10} + b^4 r^{12}\big].
\end{eqnarray}
The three different forces acting on the system are shown in fig.~\ref{tov} for the compact stars Vela X-1 and 4U 1608-52  for different values of $p_0$.

\subsection{Harrison-Zeldovich-Novikov's stability condition}

Depending on the mass and central density of the star, Harrison et al. \cite{harri} and Zeldovich-Novikov \cite{novi} proposed the stability condition for the model of compact star. From their investigation they suggested that for stable configuration $\frac{\partial M}{\partial \rho_c}>0$, where $M,\,\rho_c$ denotes the mass and central density of the compact star.

\begin{figure}[htbp]
        \includegraphics[scale=.3]{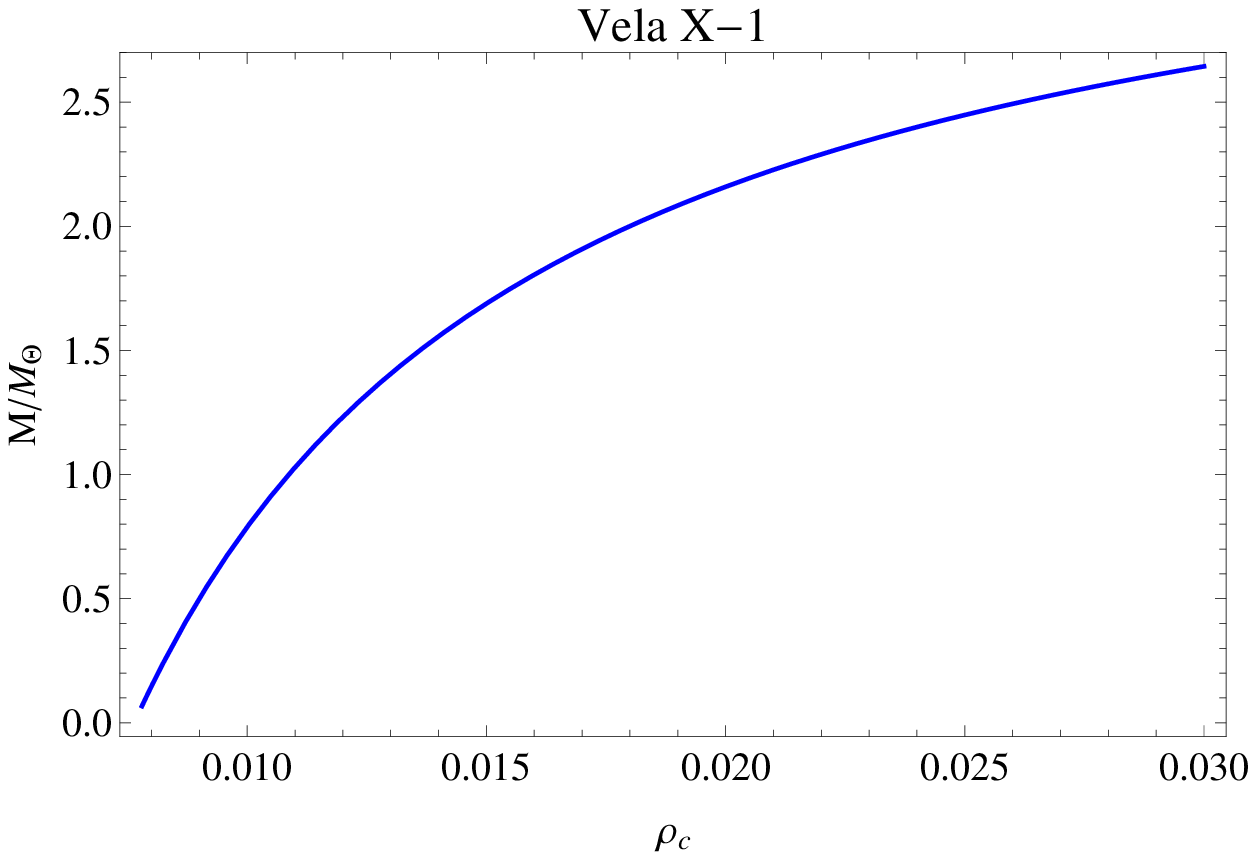}
        \includegraphics[scale=.3]{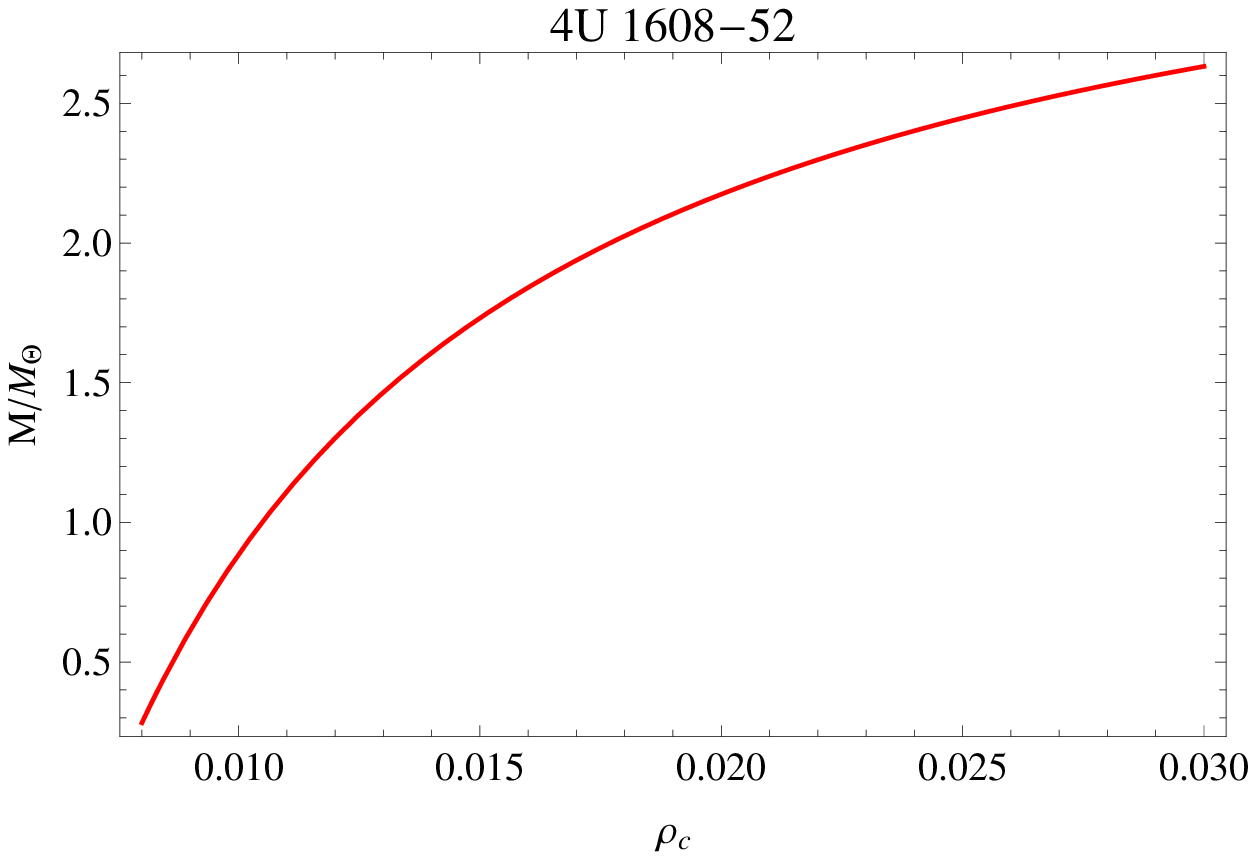}
        \includegraphics[scale=.3]{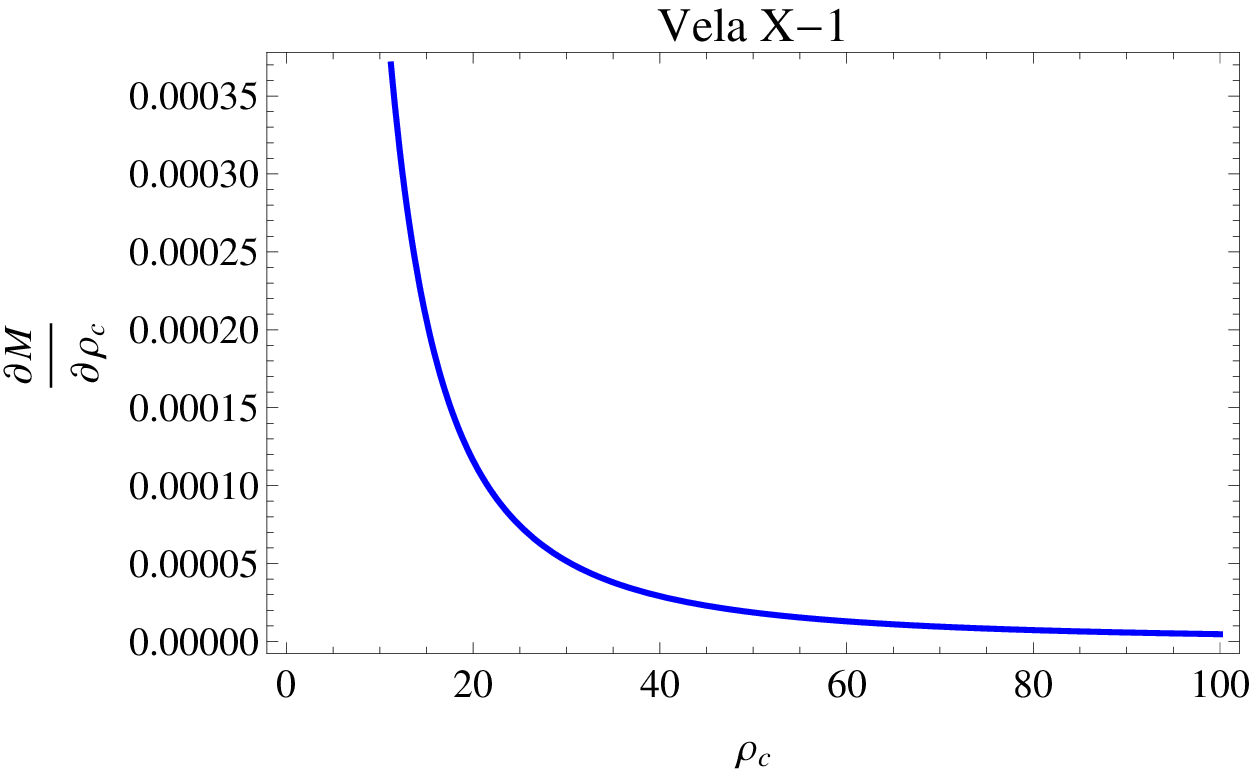}
        \includegraphics[scale=.3]{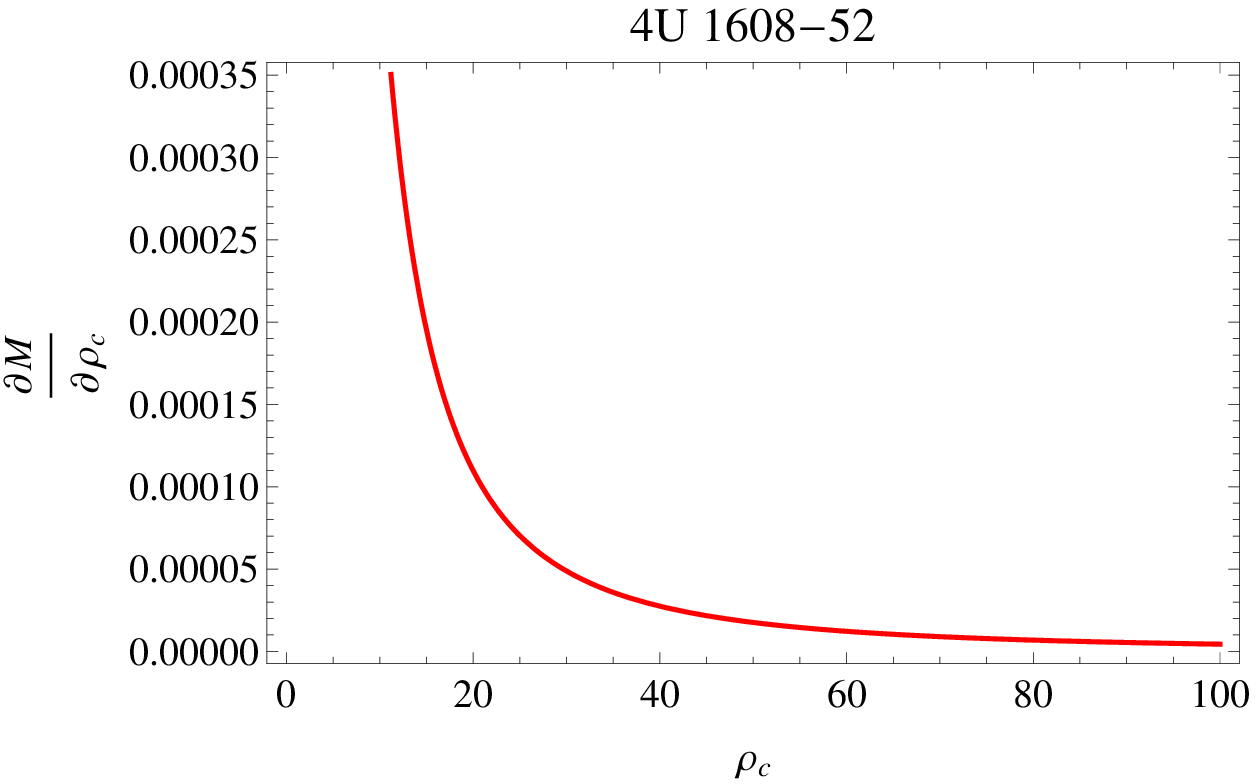}
       \caption{Variation of mass (M/$M_{\odot}$) and $\frac{\partial M}{\partial \rho_c}$ with respect to the central density $\rho_c$ for a possible modelling of the compact star Vela X-1 and 4U 1608-52. \label{tt1}}
\end{figure}

For our present model,
\begin{eqnarray}
\frac{\partial M}{\partial \rho_c}=\frac{12 \pi R^3 \left(1 + b (R^4 - R^6)\right)}{(3 + 3 b R^4 + 8 \pi R^2 \rho_c)^2}.
\end{eqnarray}
Above expression of $\frac{\partial M}{\partial \rho_c}$ is positive and hence the stability condition is well satisfied. The variation of the mass function as well as $\frac{\partial M}{\partial \rho_c}$ with respect to the central density is depicted in fig.~\ref{tt1}.

\begin{table*}\label{table4}
\caption{The numerical values of central pressure ($p_c$) in dyne$\cdot$cm$^{-2}$ unit, surface anisotropy in dyne$\cdot$cm$^{-2}$ unit, central radial velocity ($V_r(0)$), surface radial velocity ($V_r(R)$), central transverse velocity ($V_t(0)$), surface transverse velocity ($V_t(R)$) for a possible modelling of the compact star Vela X-1 \cite{vela}.}
\begin{tabular*}{\textwidth}{@{\extracolsep{\fill}}lrrrrrrrrl@{}}
\hline
$p_0$ &\multicolumn{1}{c}{$p_c$}&\multicolumn{1}{c}{$\Delta_s=p_t(R)$}&\multicolumn{1}{c}{$V_{r}(0)$}&\multicolumn{1}{c}{$V_{r}(R)$}&\multicolumn{1}{c}{$V_t(0)$}&\multicolumn{1}{c}{$V_t(R)$}&\multicolumn{1}{c}{$\Gamma_{r}(0)$}\\
\hline
$0.0023$ &$1.11134\times10^{35}$ &$6.2321\times10^{34}$&$0.38895$ &$0.311366$&$0.328293$&$0.200689$&$1.84302$\\
$0.0026$&$1.2563\times10^{35}$ &$5.84535\times10^{34}$&$0.413539$&$0.33105$ &$0.373797$&$0.227046$&$1.86275$ \\
$0.0029$&$1.40126\times10^{35}$ &$5.45861\times10^{34}$&$0.436746$&$0.349628$&$0.414194$&$0.250647$&$1.88248$  \\
$0.0032$&$1.54621\times10^{35}$&$5.07186\times10^{34}$&$0.45878$ &$0.367267$ &$0.450859$&$0.272209$&$1.90222$ \\
\hline
\end{tabular*}
\end{table*}

\subsection{Linearized stability analysis}

We have already matched the
interior space-time continuously to an exterior schwarzschild
vacuum solution with $p_r = 0$ at the junction interface $\Sigma$,
with junction radius R.\\
Now by using the standard Darmois-Israel formalism \cite{israel1,israel2} and with the help of Lanczos equations the surface stresses can be obtained,
as follows :
\begin{eqnarray}\label{sigma1}
\sigma &=&-\frac{1}{4\pi R}\Big[\sqrt{1-\frac{2M}{R}+\dot{R}^2}\nonumber\\&&-\sqrt{\frac{1}{(1+aR^2+bR^4)}+\dot{R}^2}~\Big],
\end{eqnarray}
and
\begin{eqnarray}
\mathcal{P}&=&\frac{1}{8\pi R}\Big[\frac{1-\frac{M}{R}+\dot{R}+R\ddot{R}}{\sqrt{1-\frac{2M}{R}+\dot{R}^{2}}}\nonumber\\&&
-\frac{(1+aR^2+bR^4)^{-1}
+\dot{R}^{2}\frac{1-bR^4}{1+aR^2+bR^4}+R\ddot{R}}{\sqrt{(1+aR^2+bR^4)^{-1}+\dot{R}^{2}}}\Big]
\end{eqnarray}

Where $\sigma$ and $\mathcal{P}$ represent the surface density and tangential surface pressure of the internal forces on the junction surface.\par

 We now proceed with the
surface mass of the thin shell is given by
\begin{equation}\label{ms1}
m_s=4\pi R^2\sigma.
\end{equation}
When one substitutes Eq. (\ref{sigma1}) into Eq. (\ref{ms1}), one can get
\begin{equation}
m_s=-R\left[\sqrt{1-\frac{2M}{R}+\dot{R}^2}-\sqrt{\dot{R}^2+(1+aR^2+bR^4)^{-1}}\right],
\end{equation}
Now, differentiating twice the expression in (\ref{ms1}), and taking into account the radial
derivative of $\sigma^\prime$, we can obtain,
\begin{equation}
\left(\frac{m_s}{2R}\right)^{\prime\prime}=\Upsilon-4\pi~\sigma^\prime\eta,
\end{equation}
where the parameters $\eta$ and $\Upsilon$ are given by,
\begin{equation}
\eta=\frac{\mathcal{P}^\prime}{\sigma^\prime},~~~~\Upsilon=\frac{4\pi}{R}(\sigma+\mathcal{P})+2\pi~R\Xi^\prime.
\end{equation}
The expression found above will play an important role in stability analysis of static solutions.
Here the parameter $\eta$ is used to determine the stability of the
system.  $\eta$ can be interpreted as the square of velocity of
sound on the shell and it lies in the range $0 < \eta \leq 1$. For our present model, \[\Xi=\frac{a R + 2 b R^3}{4 \pi + 4 a \pi R^2 + 4 b \pi R^4} \sqrt{\dot{R}^2+(1+aR^2+bR^4)^{-1}}.\]

Now we are going to check for the stability analysis of the solution.
In order to study the dynamical stability of the stellar model
we consider a linear perturbation around those static solutions.\\
Rearranging (\ref{sigma1}), we get,
\begin{equation}\label{p7}
\dot{R}^2+V(R)=0,
\end{equation}
with $V(R)$ given by
\begin{equation}
V(R)=1-\frac{m(R)+M}{R}-\left(\frac{m_s(R)}{2R}\right)^2-\left(\frac{M-m(R)}{m_s(R)}\right)^2.
\end{equation}
Note that the potential function $V(R)$ helps us to determine the stability region for
the thin shell under the linear perturbation. By considering the
Taylor series expansion around the static solution $R_0$, up to second order we get,
\begin{eqnarray}\label{p8}
V(R) &=& V(R_0)+(R-R_0)V^\prime(R_0)+\frac{(R-R_0)^2}{2}V^{\prime\prime}(R_0)\nonumber\\
&&+\mathcal{O}[(R-R_0)^3],
\end{eqnarray}
where prime corresponding to a derivative with respect to R. According to the standard method
we are linearizing around the static radius $R = R_0$, we must have $V(R_0)= 0$ and  $V^\prime(R_0)=0$.
Now, $V^\prime(R_0)=0$ gives the following relation

\begin{eqnarray}
\left(\frac{m_s(R_0)}{2R_0}\right)^\prime= \Phi &= &\frac{R_0}{m_s(R_0)}
\left[-\left(\frac{m(R_0)+M}{R_0}\right)^\prime\right.\nonumber\\
&\;&\left.-2\left(\frac{M-m(R_0)}{m_s(R_0)}\right)\left(\frac{M-m(R_0)}{m_s(R_0)}\right)^\prime~\right]\nonumber\\.
\end{eqnarray}
With this definition the second derivative $V^{\prime\prime}(R_0)$ can be written as
\begin{eqnarray}
V^{\prime\prime}(R_0)&=&-\left(\frac{m+M}{R_0}\right)^{\prime\prime}-2\left[\left(\frac{m_s}{2R_0}\right)^\prime\right]^2\nonumber\\&&
-2\left(\frac{m_s}{2R_0}\right)\left(\frac{m_s}{2R_0}\right)^{\prime\prime}-2\left[\left(\frac{M-m}{m_s}\right)^\prime~\right]^2\nonumber\\&&-2\left(\frac{M-m}{m_s}\right)
\left(\frac{M-m}{m_s}\right)^{\prime\prime}.
\end{eqnarray}
Thus, for a static configuration $V(R_0)=0,\,V'(R_0)=0$. Now from (\ref{p7}) and (\ref{p8}), we get,
\begin{eqnarray}\label{p9}
\dot{R}^2+\frac{1}{2}V^{\prime\prime}(R_0)\left(R-R_0\right)^2+\mathcal{O}[(R-R_0)^3] = 0,
\end{eqnarray}
To ensure the stability of static configuration at R = $R_0$, we must have, $V^{\prime\prime}(R_0) > 0$. For the sake of
simplicity we rearrange the Eq. (\ref{p9}), which turns out
\begin{equation}\label{p10}
\Rightarrow~ V^{\prime\prime} = \Pi-2\Phi^2-\frac{m_s}{R_0}\left( \Upsilon-4\pi~\sigma^\prime\eta \right) \Big|_{R_0}.
\end{equation}
where,
\begin{eqnarray}
\Pi &=&-\left(\frac{m+M}{R_0}\right)^{\prime\prime}-2\left[\left(\frac{M-m}{m_s}\right)^\prime\right]^2\nonumber\\
&&-2\left(\frac{M-m}{m_s}\right)\left(\frac{M-m}{m_s}\right)^{\prime\prime}.
\end{eqnarray}
(For details calculations please refer ref \cite{pb15}.)\par
\begin{figure*}[htbp]
    \centering
        \includegraphics[scale=.3]{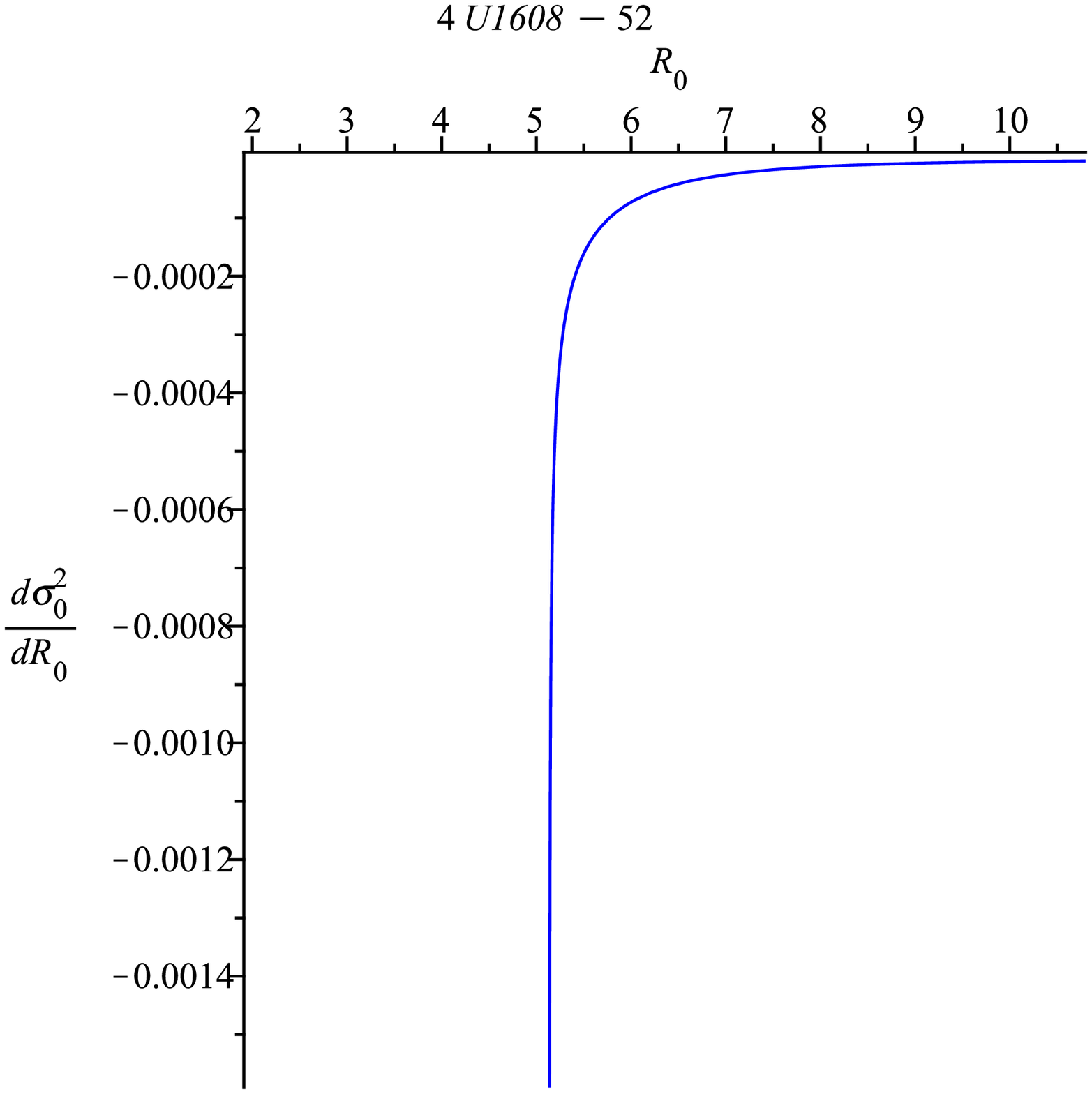}
        \includegraphics[scale=.3]{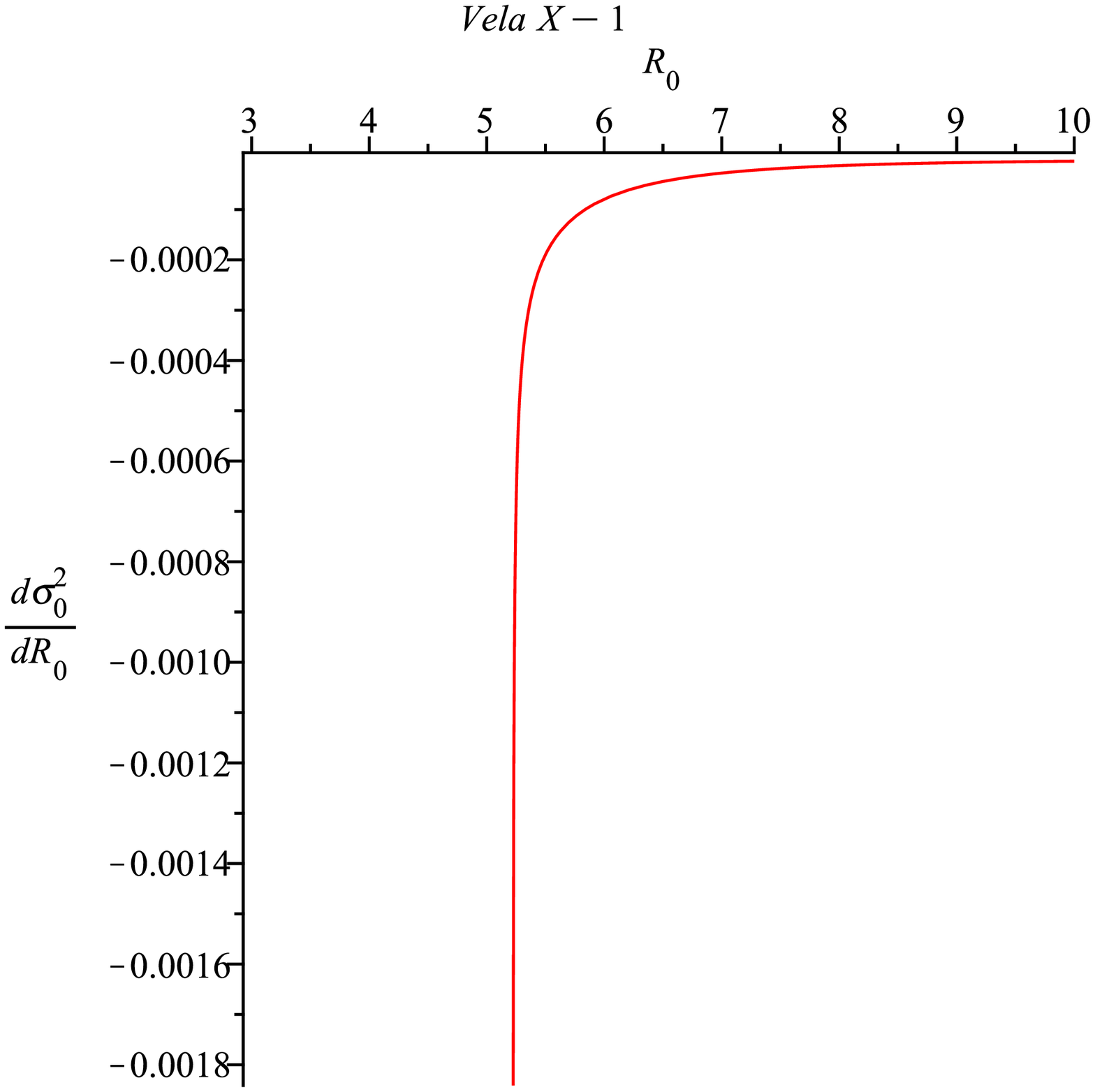}
       \caption{$\frac{d \sigma_0^2}{dR_0}$ has been shown against $R_0$ \label{stabb}}
\end{figure*}

Now by assuming $\eta(R_0)=\eta_0$
and using Eq.~(\ref{p10}) for $V^{\prime\prime}(R_0)>0$ we have
\begin{equation}
\eta_0\frac{d\sigma^2}{dR}\Big|_{R_0}>\frac{\sigma}{2\pi}\left[\Upsilon+\frac{R_0}{m_s}(2 \Phi^2-\Pi)\right] = \Omega ~(\mathrm{say}).
\end{equation}
From the above inequality we get the following two cases,
\begin{equation}\label{st1}
\eta_0>\Omega\left(\frac{d\sigma^2}{dR}\Big|_{R_0}\right)^{-1}, ~~~ \mathrm{if}~~~ \frac{d\sigma^2}{dR}\Big|_{R_0} > 0,
\end{equation}
\begin{equation}\label{eq:less}
\eta_0<\Omega\left(\frac{d\sigma^2}{dR}\Big|_{R_0}\right)^{-1}, ~~~ \mathrm{if}~~~ \frac{d\sigma^2}{dR}\Big|_{R_0} < 0.
\end{equation}
Now to make a comment regarding the stability region of the stellar configuration, the profile of $\frac{d\sigma_0^2}{dR_0}$ is plotted in fig~\ref{stabb}. So, we can conclude that the stability region is given by eq.~(\ref{eq:less}) since $\frac{d\sigma_0^2}{dR_0}<0$ for our present paper.

\section{Calculation of tidal love number of compact neutron star with anisotropic pressure}

Gravitational perturbation are very much important, though very much difficult to handle, since they convey information
about the structure of the gravitational field equations.  So let's start by trying to answer these question: Given a metric for instance Schwarzschild geometry  or  metric of a compact neutron star, how does a small perturburtation in the metric evolve? Here small perturbation means anything- a wave, a celestial body, a falling particle - that disturb the background metric slightly. Mathematically consider a background metric $^{(0)}g_{\mu \nu}(x^{\nu})$- metric of a neutron star. With small perturbation $h_{\mu \nu}(x^{\nu})$, the modified new metric can be written as

\begin{align}
g_{\mu \nu}\left(x^{\nu}\right)=^{(0)} g_{\mu \nu}\left(x^{\nu}\right)+h_{\mu \nu}\left(x^{\nu}\right), \label{48}
\end{align}

where the background geometry of spacetime of a spherical static star can be written as

\begin{align}
^{\left( 0\right)} ds^{2} &=^{\left( 0\right)} g_{\mu \nu }dx^{\mu }dx^{\nu } \nonumber\\ &=-e^{\nu(r) }dt^{2}+e^{\lambda(r) }dr^{2}+r^{2}\left( d\theta ^{2}+\sin ^{2}\theta d\phi ^{2}\right). \label{49}
\end{align}

For the linearized metric perturbation $h_{\alpha \beta }$, using the method as in \cite{Regge1957},  \cite{Biswas2019}, we restrict ourselves to static $l = 2, \, m=0$ even parity perturbation. With these assumptions the perturbed metric becomes

\begin{align}
h_{\alpha \beta}=\operatorname{diag}\left[H_{0}(r) e^{\nu}, H_{2}(r) e^{\lambda}, r^{2} K(r), r^{2} \sin ^{2} \theta K(r)\right] Y_{2 m}(\theta, \phi), \label{50}
\end{align}

Because of these external perturbation the equalibrium configuration of a star gets deformed tidally. As a cosequence spherically symmetric star develpes a quadrapole moment $Q_{ij}$. And $Q_{ij}$ can be related with the linear order external tidal field $\varepsilon_{ij}$ as \cite{Hinderer2008}

\begin{align}
Q_{ij} = - \Lambda \, \varepsilon_{ij}, \label{51}
\end{align}

where $\Lambda$ is the tidal deformability of the neutron star and it is related to the tidal love number $k_{2}$ as $ \Lambda = \dfrac{2}{3}k_{2}\, R^{5} $ \cite{Hinderer2008}.

Now for the spherically static metic \eqref{49}, the stress-energy tensor is given as

\begin{align}
^{(0)}T_{\chi}^{\xi}=\left(\rho+p_{t}\right) u^{\xi} u_{\chi}+p_{t} g_{\chi}^{\xi}+\left(p_{r}-p_{t}\right) \eta^{\xi} \eta_{\chi}. \label{52}
\end{align}

Furthermore the energy momentum tensor is perturbed by a perturbation tensor $\delta T_{\chi}^{\xi}$. The perturbed tensor is defined by

\begin{align}
T_{\chi}^{\xi} = ^{(0)}T_{\chi}^{\xi} + \delta T_{\chi}^{\xi}, \label{53}
\end{align}

where the non-zero components of $T_{\chi}^{\xi}$ are

\begin{align}
T_{t}^{t} &= - \frac{d\rho }{dp_{r}}  \delta p_{r}(r) Y(\theta ,\phi )- \rho (r), \label{54}\\
T_{r}^{r} &=     \delta p_{r}(r) Y(\theta ,\phi )+ p_{r}(r),\\
T_{\theta}^{\theta} &=   \frac{dp_{t}}{dp_{r}}  \delta p_{r}(r) Y(\theta ,\phi )+  p_{t}(r),\\
T_{\phi}^{\phi} &=    \frac{dp_{t}}{dp_{r}}  \delta p_{r}(r) Y(\theta ,\phi )+  p_{t}(r). \label{57}
\end{align}

Note that in Eq. \eqref{54}-\eqref{57}, and hereafter, we use $p \equiv p_{r}$  to denote
the radial pressure. With these
perturbed quantities we can write down the perturbed Einstein Field Equations

\begin{align}
G_{\chi}^{\xi} = 8 \pi T_{\chi}^{\xi}, \label{58}
\end{align}

where the Einstein tensor $G_{\chi}^{\xi} $ is calculated using the metric $g_{\chi \xi} $.

From the various components of  background Einstein field equation $^{(0)}G_{\chi}^{\xi} = 8 \pi ^{(0)}T_{\chi}^{\xi}$, we can have the following things (which will be used later on):

\begin{align}
^{(0)}G_{t}^{t} &= 8 \pi ^{(0)}T_{t}^{t}\nonumber\\
\Rightarrow \lambda'(r) &= \frac{8 \pi  r^2 e^{\lambda (r)} \rho (r)-e^{\lambda (r)}+1}{r}, \label{59}\\
^{(0)}G_{r}^{r} &= 8 \pi ^{(0)}T_{r}^{r}\nonumber\\
\Rightarrow \nu'(r) &= \frac{8 \pi  r^2 p_{r}(r) e^{\lambda (r)}+e^{\lambda (r)}-1}{r}. \label{60}
\end{align}

Also we know that $\nabla_{\xi}^{(0)}T_{\chi}^{\xi} = 0 $. Choosing $\xi = r$, by expanding and solving the equation, we can find the expression as

\begin{align}
p_{r}'(r) = \frac{1}{2 r}\left[-4 p_{r}(r)+4 p_{t}(r)-r p_{r}(r) \nu^{\prime}(r)-r \rho(r) \nu^{\prime}(r)\right]. \label{61}
\end{align}

Now from the various components of  perturbed Einstein equation  \eqref{58}, we get the following relations

\begin{align}
&G_{\theta}^{\theta} - G_{\phi}^{\phi} = 0 \Rightarrow H_{0}(r) = H_{2}(r) = H(r), \label{62}\\
&G_{r}^{\theta} =0 \Rightarrow K' = H' + H \nu', \label{63}\\
&G_{\theta}^{\theta} + G_{\phi}^{\phi} = 8 \pi (T_{\theta}^{\theta} + T_{\phi}^{\phi}) \Rightarrow \delta p = \frac{H(r) e^{-\lambda (r)} \left(\lambda '(r)+\nu '(r)\right)}{16 \pi r \frac{dp_{t}}{dp_{r}} }. \label{64}
\end{align}

Using the identity $$\dfrac{\partial^{2} Y(\theta,\phi)}{\partial \theta^{2}} + cot(\theta) \dfrac{\partial Y(\theta,\phi)}{\partial \theta}+ \csc ^2(\theta ) \dfrac{\partial^{2} Y(\theta,\phi)}{\partial \phi^{2}} = -6 Y(\theta ,\phi )$$, eqn \eqref{59},~\eqref{60}~,\eqref{61}~,\eqref{63},~\eqref{64}  we have the master equation for $H(r)$ as

\begin{align}
&- \frac{1}{e^{-\lambda (r)} Y(\theta ,\phi )} \left[ G_{t}^{t} - G_{r}^{r}\right] = - \frac{8 \pi}{e^{-\lambda (r)} Y(\theta ,\phi )} \left[ T_{t}^{t} - T_{r}^{r}\right]\nonumber\\
&\Rightarrow  H''(r) + \mathcal{R} H'(r) + \mathcal{S} H(r) = 0 \label{65}
\end{align}

Where,

\begin{align}
  \mathcal{R} = - \left[ \frac{-e^{\lambda (r)}-1}{r}-4 \pi  r e^{\lambda (r)} (p_{r}-\rho (r))\right] \label{83}
\end{align}

\begin{widetext}
\begin{align}
  	\mathcal{S} =- \left[16 \pi  e^{\lambda (r)} \left(p_{r} \left(e^{\lambda (r)}-2\right)-p_{t}(r)-\rho (r)\right)+64 \pi ^2 r^2 p_{r}^2 e^{2 \lambda (r)}+\frac{4 e^{\lambda (r)}+e^{2 \lambda (r)}+1}{r^2}\right. \nonumber\\
	 + \left.\frac{-4 \pi  \frac{d\rho}{dp_{r}} e^{\lambda (r)} (p_{r}+\rho (r))-4 \pi  e^{\lambda (r)} (p_{r}+\rho (r))}{\frac{dp_{t}}{dp_{r}}}\right] \label{84}
\end{align}
\end{widetext}

$e^{\lambda (r)} = 1 + a r^2 + b r^4, \, a = 1/R^2 ,\, b = \frac{1}{R^4} \left[ \left( 1 -\frac{2 M}{R}\right) ^{-1} -2\right] $. $M$ \& $R$ are mass and radius of the star respectively. And the expression for $p_{r}, \, p_{t}$ \& $\rho$ are given in the equation \eqref{Rhoo},\eqref{pr},\eqref{Pt} respectively.\\

Outside the star, considering Schwarzschild metric and by setting $ \rho =0, p_{r} =0, p_{t} =0, \, e^{\lambda} = 1/(1- 2 M/r)$, equation \eqref{65} becomes:
\begin{widetext}
\begin{align}
  -H''(r)+\left(\frac{1}{2 M-r}-\frac{1}{r}\right) H'(r) +\frac{2 H(r) \left(2 M^2-6 M r+3 r^2\right)}{r^2 (r-2 M)^2} = 0 \label{85}
\end{align}
\end{widetext}

Solution to the equation \eqref{85} is

\begin{widetext}
\begin{align}
    H(r) = 	\frac{3 c_1 r (2 M-r)}{M^2}+\frac{c_2 \left(-2 M \left(2 M^3+4 M^2 r-9 M r^2+3 r^3\right)-3 r^2 (r-2 M)^2 \log \left(\frac{r}{M}-2\right)+3 r^2 (r-2 M)^2 \log \left(\frac{r}{M}\right)\right)}{2 M^2 r (2 M-r)} \label{86}
\end{align}
\end{widetext}

Where $c_1$ \& $c_2$ are integration constant which yet to be determined. In order to get these constants, let's expand the equation \eqref{86}.

\begin{align}
   H(r) =  -\frac{3 c_1 r^2}{M^2}+\frac{6 c_1 r}{M}-\frac{c_2 \left(8 M^3\right)}{5 r^3}+\mathcal{O}\left(\left(\frac{1}{r}\right)^4\right) \label{87}
\end{align}

Now in the star's local
asymptotic rest frame (asymptotically mass-centered Cartesian coordinates) at large $r$ the metric coefficient $g_{tt}$ is given by \cite{Thorne1998}, \cite{Hinderer2008}

\begin{eqnarray}
\frac{\left(1-g_{t t}\right)}{2}&=&-\frac{M}{r}-\frac{3 Q_{i j}}{2 r^{3}}\left(n^{i} n^{j}-\frac{1}{3} \delta^{i j}\right)+\mathcal{O}(\frac{1}{r^{3}} )\nonumber\\&&
+\frac{1}{2} \mathcal{E}_{i j} x^{i} x^{j}+ \mathcal{O}(r^{3}), \label{88}
\end{eqnarray}

where $n^{i} = x^{i}/r$. Matching the asymptotic solution from equation \eqref{87} to the expansion from equation \eqref{88} and using the equation \eqref{51} we have

\begin{align}
    c_1 = - \frac{M^2  \mathcal{E}}{3}, \quad
    c_2 = \frac{15 \mathcal{Q}}{8 M^3} \label{89}
\end{align}

We now solve for $k_{2}$ in terms of $H$ and its derivative at the star's surface $r=R$ using equations \eqref{89} and \eqref{86},  and use equation $k_{2} = \frac{3}{2} \Lambda R^{-5}$ to obtain the expression for tidal love number

\begin{align}
    k_2 = [8 (1-2 \mathcal{C})^2 \mathcal{C}^5 (2 \mathcal{C} (\mathit{y}-1)-\mathit{y}+2)]/X \label{90}
\end{align}

Where, 
\begin{widetext}
\begin{align}
    X &= 5(2 \mathcal{C} (\mathcal{C} (2 \mathcal{C} (\mathcal{C} (2 \mathcal{C} (\mathit{y}+1)+3 \mathit{y}-2)-11 \mathit{y}+13)+3 (5 \mathit{y}-8))-3 \mathit{y}+6) \nonumber\\
&\left.+3 (1-2 \mathcal{C})^2 (2 \mathcal{C} (\mathit{y}-1)-\mathit{y}+2) \log \left(\frac{1}{\mathcal{C}}-2\right)-3 (1-2 \mathcal{C})^2 (2 \mathcal{C} (\mathit{y}-1)-\mathit{y}+2) \log \left(\frac{1}{\mathcal{C}}\right)\right) \label{91}
\end{align}
\end{widetext}

Here $\mathcal{C} = \frac{M}{R}$ and $ \mathit{y}$ depends on $r,\, H$ and it's derivatives
\begin{eqnarray}
\mathit{y} =\dfrac{r H'(r)}{H(r)}_R.  \label{92}
\end{eqnarray}

To get the numerical value of $k_2$ for a particular star, first of all we have to find the numerical value of $\mathit{y}$. To do this let's modify  master differential equation \eqref{65}  using the equation \eqref{92} as \cite{Rahmansyah2020}

\begin{align}
    r \mathit{y}' + \mathit{y}^2 + (r \mathcal{R} -1) \mathit{y} + r^2 \mathcal{S} = 0 \label{93}
\end{align}

At $\mathcal{C} \rightarrow 0$ we expect $k_2 \rightarrow 0$ and hence $\mathit{y}(0) = 2$. Also one can check numerically that at $\mathcal{C} = 1/2$ the tidal love number vanishes for all value of $\mathit{y}$. It is because the tidal deformability of any order $l$, both electric and magnetic, vanish in the Black hole limit. In
other words, the multipolar structure of a black hole is not affected by the tidal field. This can be view as a corollary of the no-hair theorem, and recently it has been proved beyond the perturbative level \cite{Gurlebeck2015}.\\

After solving the the differential equation \eqref{93} using initial value $\mathit{y}(0) = 2$, equation \eqref{83} \& eqn \eqref{84}, we can calculate $k_2$ from the equation \eqref{90}.\\

\begin{figure}[htbp]
    \includegraphics[scale =.3]{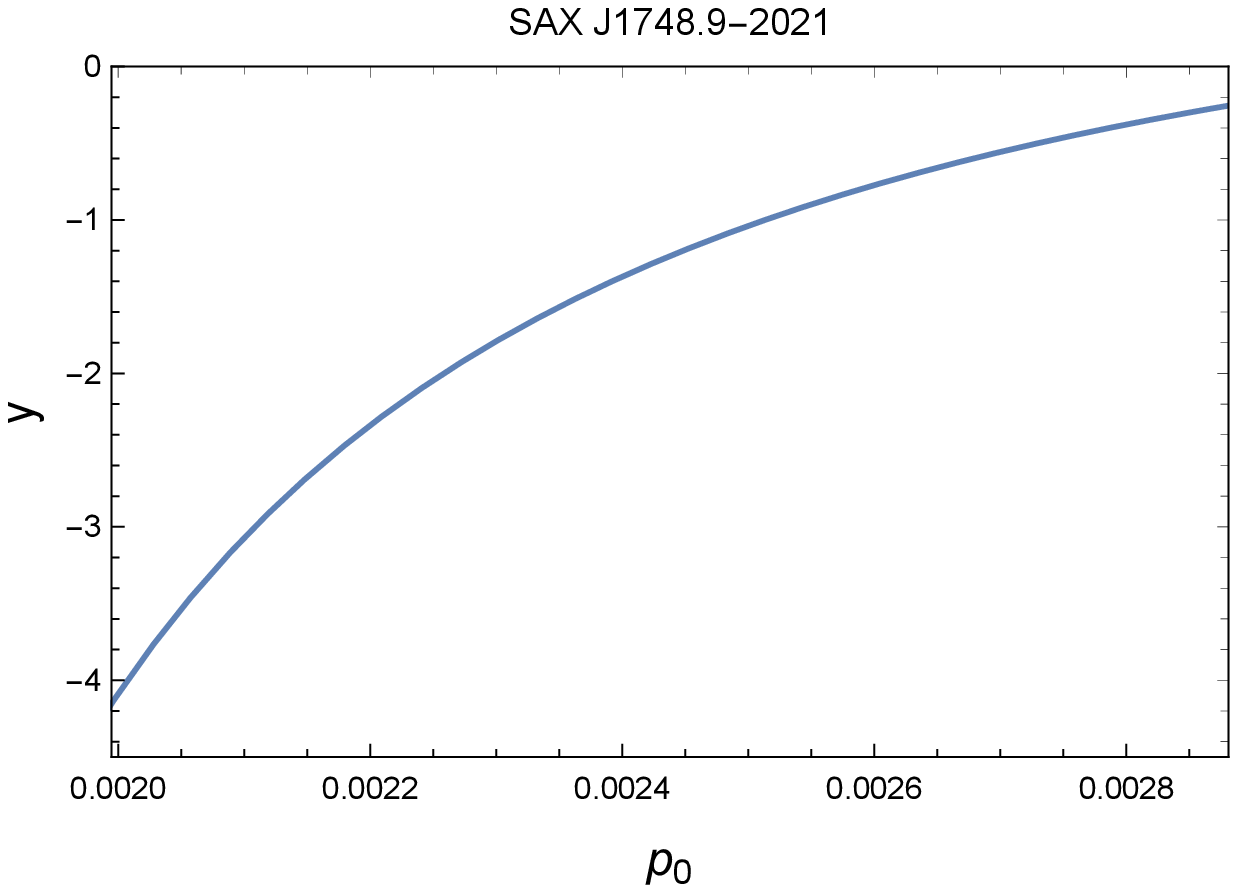}
    \includegraphics[scale =.3]{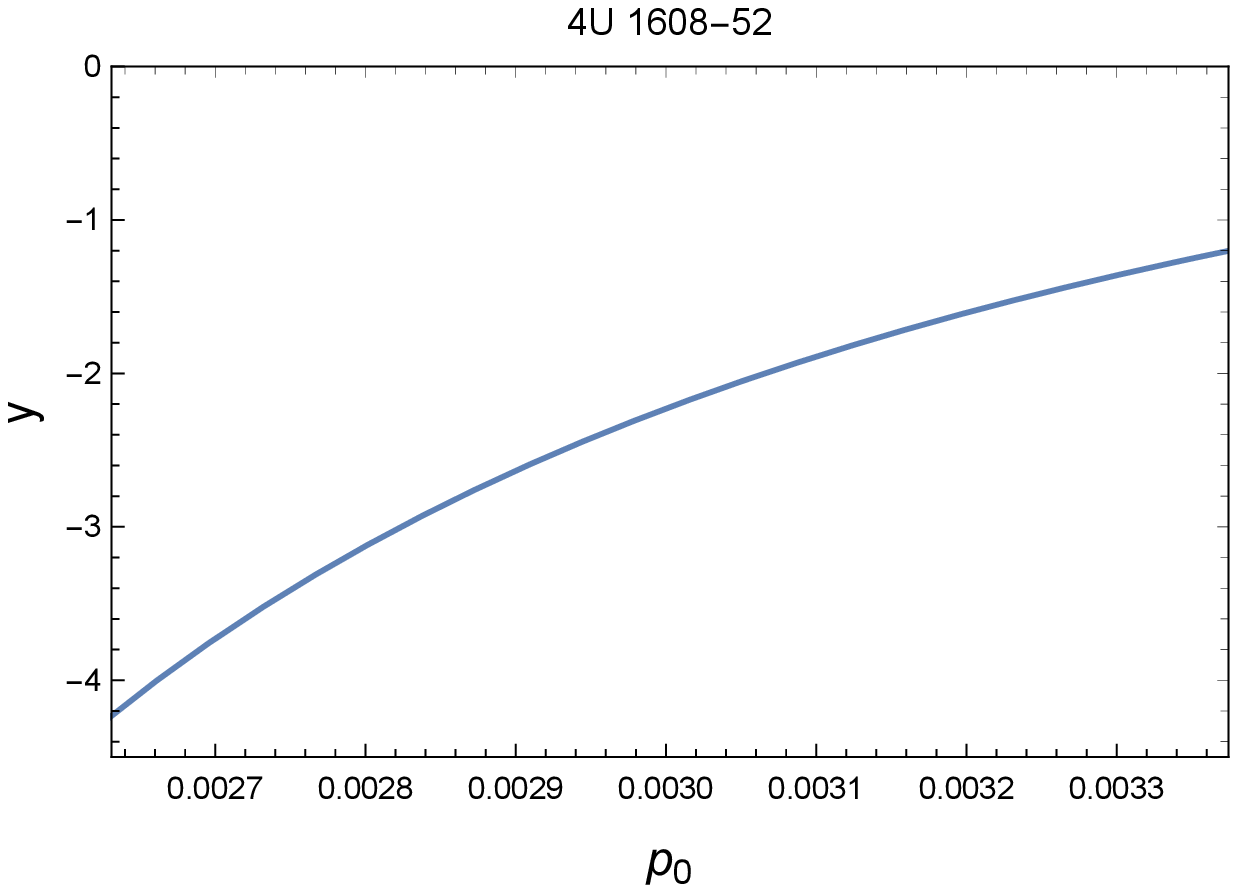}
    \includegraphics[scale =.3]{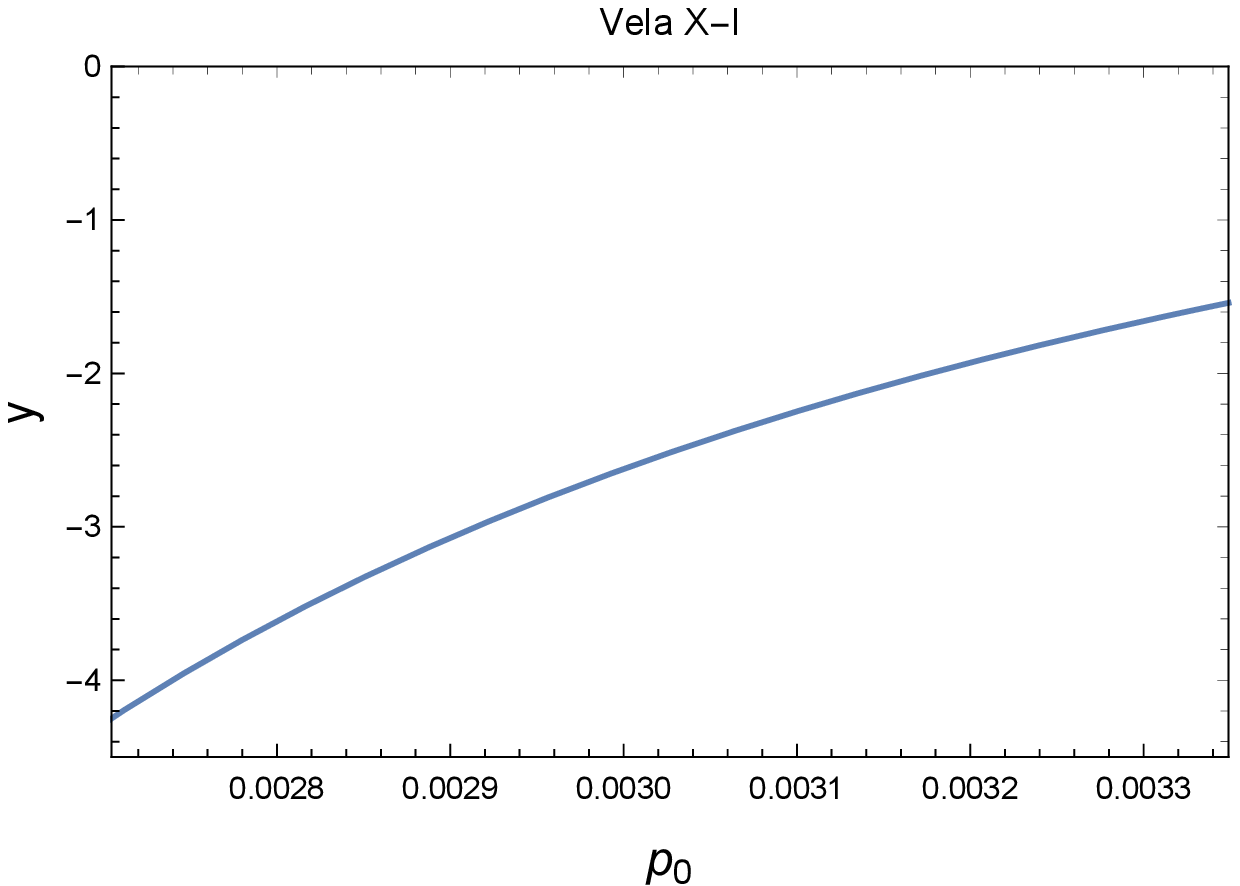}
    \includegraphics[scale =.3]{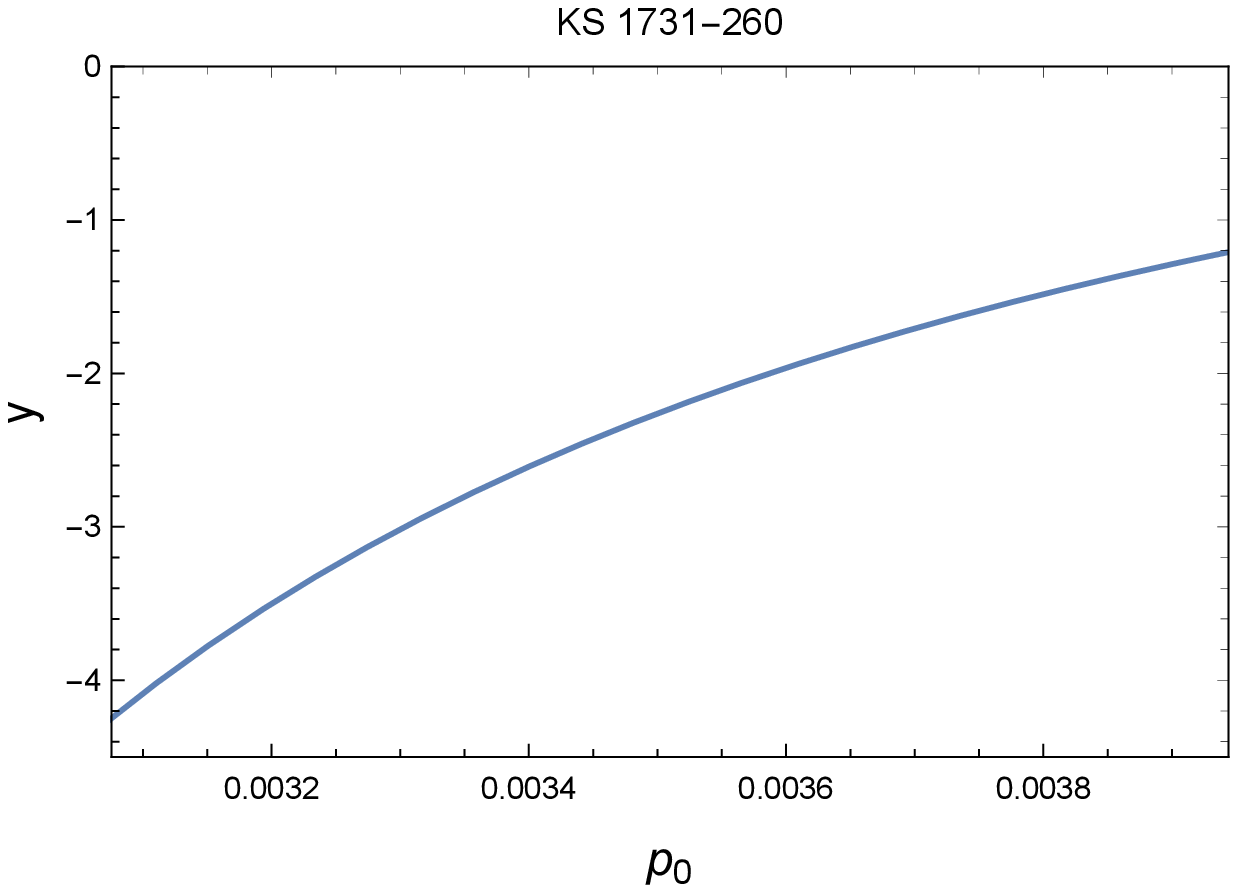}
    \caption{$y$ is plotted over   $p_{0}$ for four different compact star $SAX J1748 .9-2021$, $4U 1608-52$, $Vela X-I$, $KS 1731-260$. The value of $M$ \& $R$ of these stars are take from the paper \cite{Roupas2020}. Only allowed range of $p_0$ are considered. }
    \label{fig:13}
\end{figure}

\begin{figure}
    \includegraphics[scale=0.3]{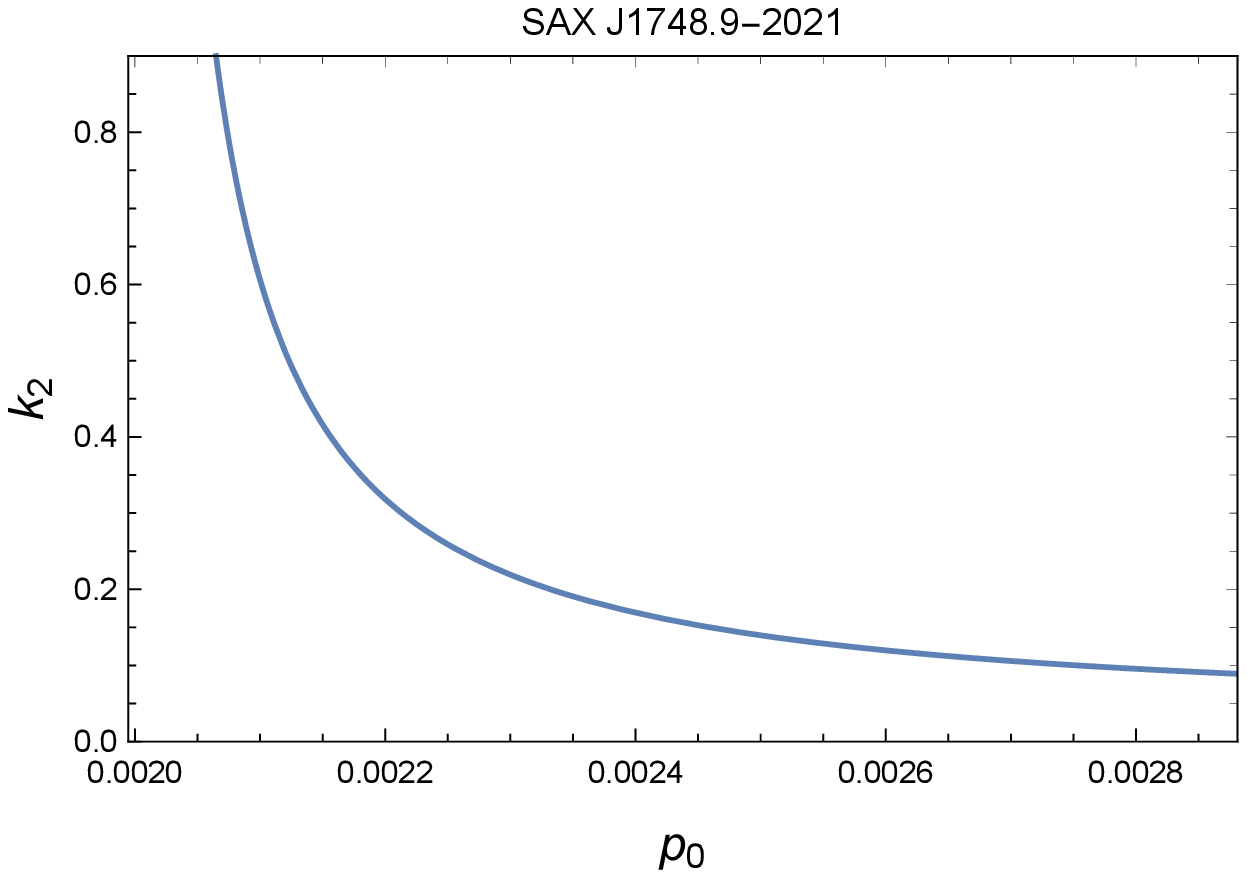}
    \includegraphics[scale=0.3]{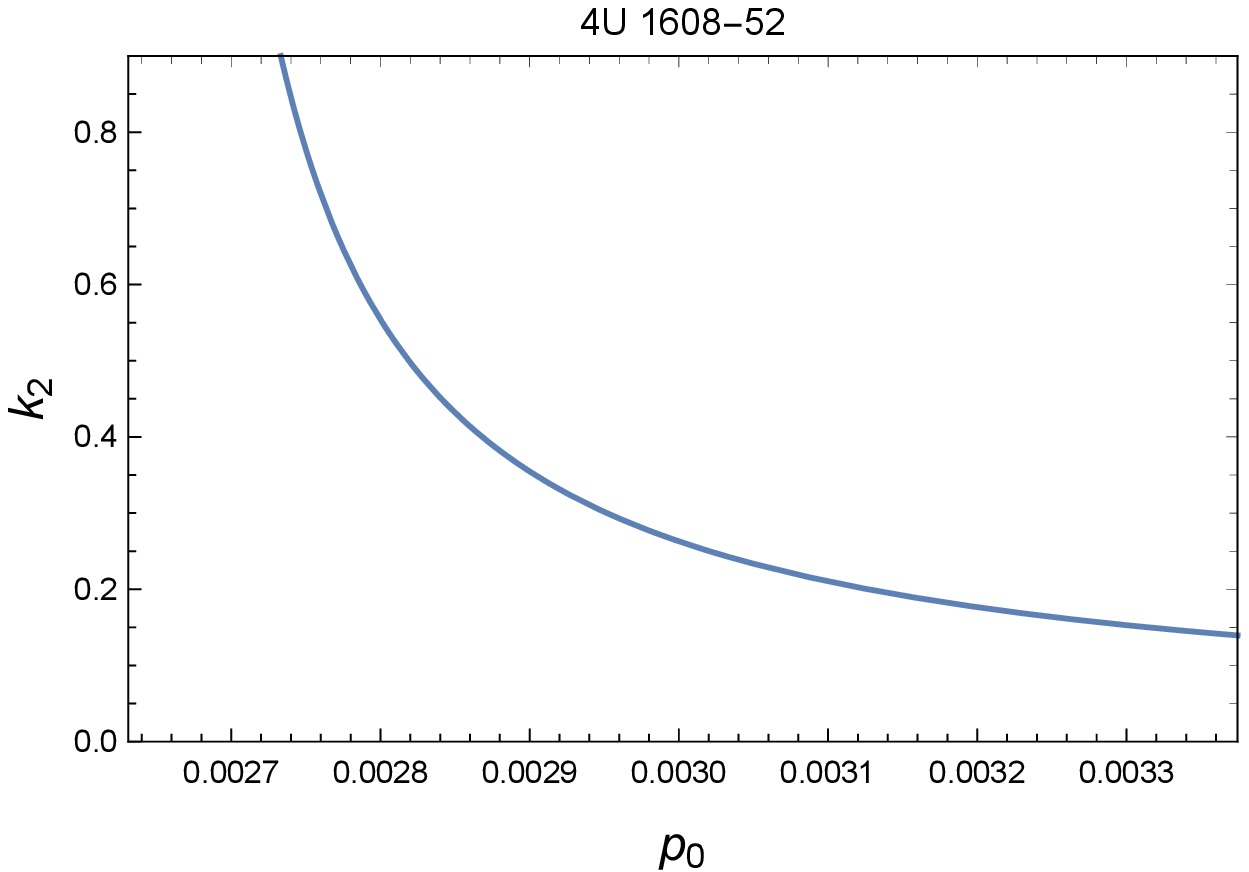}
    \includegraphics[scale=0.3]{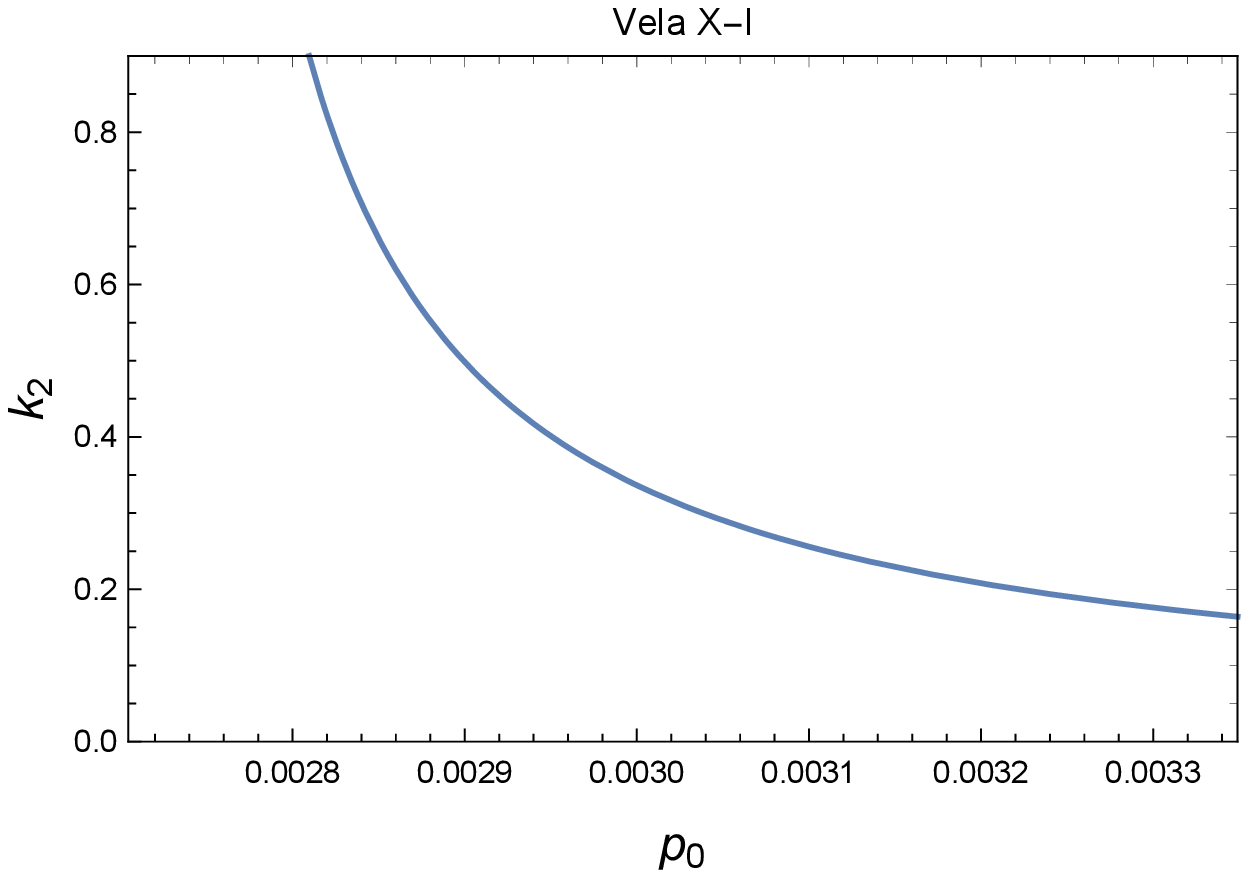}
    \includegraphics[scale=0.3]{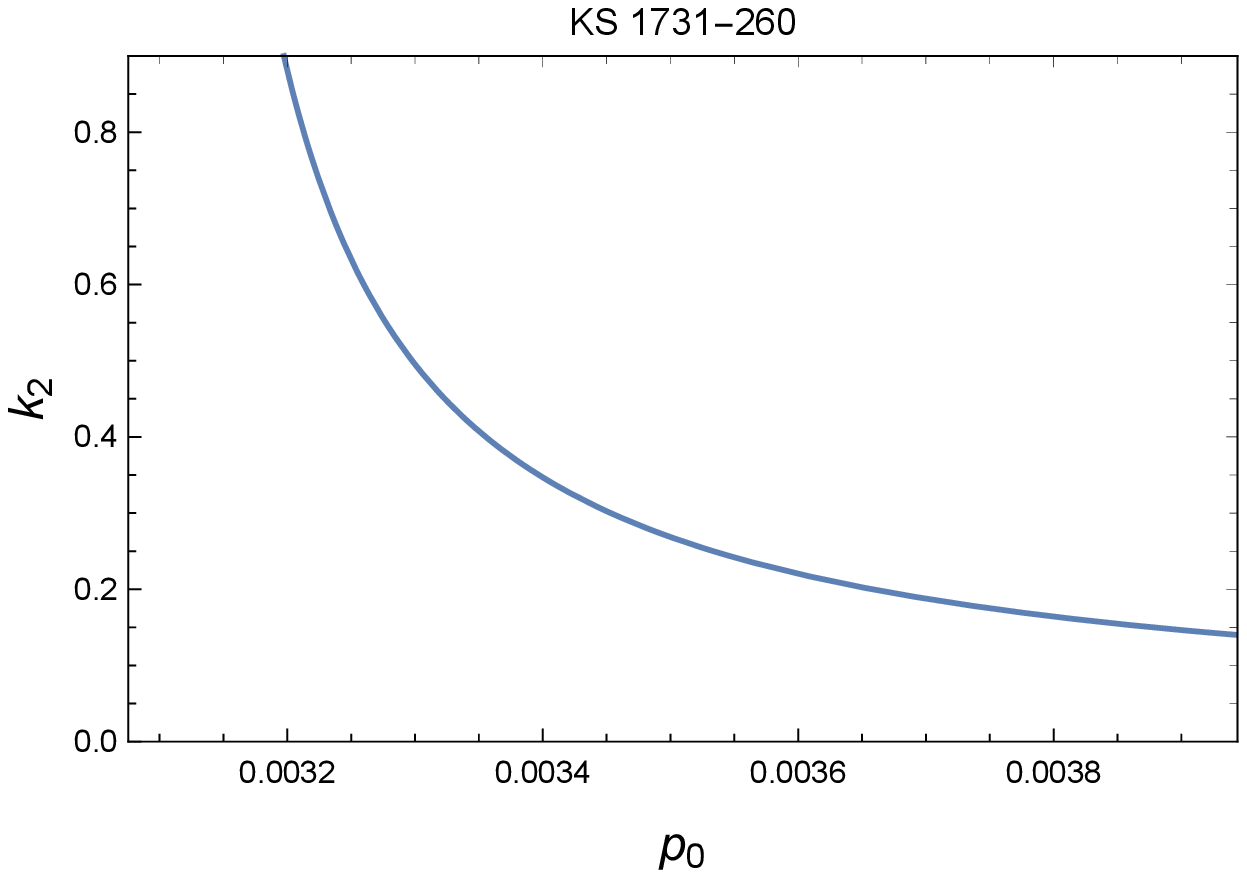}
    \caption{Tidal love number $k_2$ is plotted against  $p_{0}$ for compact star $SAX J1748 .9-2021$, $4U 1608-52$, $Vela X-I$, $KS 1731-260$. The value of $M$ \& $R$ of these stars are take from the paper \cite{Roupas2020}. Only allowed range of $p_0$ are considered.}
    \label{fig:14}
\end{figure}

In the fig. \ref{fig:13} the variation of $\mathit{y}$ w.r.t.  $p_{0}$ are shown for four different compact stars. In the figure \ref{fig:14} the variation of tidal love number $k_2$ against the central pressure $p_{0}$ are shown for four different compact objects. We found that tidal love number $k_2$ decreases monotonically with increasing $p_0$. Here the maximum range of $p_0$ is $0.3944\, a$ as in the equation \eqref{centp}. Where as the minimum range of $p_0$ varies for different compact stars. The minimum possible value of $p_0$ for some compact star is given in the table \ref{V}.

\begin{table}
\caption{The numerical value of Mass, Radius, minimum and maxium range of $p_0$ for different compact star. $a = 1/R^2$. Numerical values of mass and radius are taken from \cite{Roupas2020}}
 \begin{tabular}{|c| c| c| c | c|} 
 \hline
  Name & min. $p_0$ & max. $p_0$ & M($M_{\odot}$) &  $R$ km \\ [0.5ex] 
 \hline\hline
 SAX J1748.9-2021 & $0.27305a$ & $0.3944a$ & $1.81\pm0.25$ & $11.7\pm1.7$ \\ 
 \hline
 4U 1608-52 & $0.30750a$ & $0.3944a$ & $1.74\pm0.14$ & $10.811\pm0.19$ \\
 \hline
 Vela X-I & $0.318500a$ & $0.3944a$ & $1.77\pm0.08$ & $10.852\pm0.1$ \\
 \hline
 KS 1731-260 & $0.30755a$ & $0.3944a$ &  $1.61\pm0.35$ & $10\pm2.2$\\[1ex] 
 \hline
\end{tabular}\label{V}
\end{table}

\section{Discussion}\label{7x}
In our present paper we proposed a new model of compact star in the background of general theory of relativity. In this work we have focused on the compact object 4U 1608-52 and Vela X-1 whose estimated masses and radii are obtained in Table~I. Our present model satisfies the following conditions:
\begin{itemize}
\item Regularity of metric potential: Both the metric potentials $e^{\nu},\,e^{\lambda}$ are regular i.e., free from all kinds of singularity inside the stellar interior. $e^{\lambda}|_{r=0}=1,\,e^{\nu}|_{r=0}=\exp\bigg[\frac{2 p_0(a^2 + 2 b)}{b \sqrt{4 b-a^2}} \tan^{-1}\Big(\frac{a}{\sqrt{4b-a^2}}\Big)+B\bigg]$.
\item The matter density, both radial and transverse pressures take maximum value at the center of the star and gradually decreases towards the boundary, i.e., all are monotonic decreasing functions of r and at the same time they do not suffer from any kinds of singularity. Both the pressures depend on dimensionless quantity $p_0$ and the central pressure increases for the increasing values of $p_0$.
\item The anisotropic factor vanishes at the center of the star and $\Delta>0$ for $0\leq r\leq R$. For positive anisotropic factor makes the system more stable and it helps to construct more compact object \cite{gm}. The equation of state parameters are plotted against `r' inside the stellar interior in fig.~\ref{e8} and one can note that $0~<\omega_r,\,\omega_t~<1$.
\item We have verified that all the energy conditions are satisfied by our model with the help of graphical representation and the model is stable under the effect of the three different forces. From the fig. \ref{tov} we see that the gravitational force is attractive but both the anisotropic and hydrostatic forces are repulsive and the combine effect of these two forces is counterbalanced by the gravitational force which makes the system in static equilibrium. One interesting thing we can also notice that $F_h$ and $F_a$ intersects at some point inside the fluid sphere. The causality conditions and the the Herrera's cracking conditions are also satisfied as well. Our model is potentially stable as well since $V_t^2-V_r^2<0$ everywhere inside the fluid sphere.
\item We have obtained a reasonable bound for the dimensionless quantity $p_0$ from the physical analysis and we have shown that it depends on 'a'.
\item We have successfully demonstrated the possibility that the anisotropic stars could be ultra compact objects due to the additional support from anisotropy and could be sources of gravitational wave echoes. In this model, we have calculated tidal deformability of the anisotropic stars and show how the tidal love number varies with the compactness factor of a star. As an interesting consequence, we can see that in the black hole limit the love number vanishes. Also, the existence of the non zero value of tidal love number at zero compactness factor agrees with some of the recent findings.  
\item Generating function: Lake \cite{lake1} proposed an algorithm based on the choice of a single monotone function subject to boundary conditions which generates all regular static spherically symmetric perfect-fluid solutions of Einstein equations. Herrera  et al. \cite{h1} extended this work to the case of locally anisotropic fluids and proved that two functions instead of one is required to generate all possible solutions for anisotropic fluid. To find these two generating function we first consider pressure anisotropy for our present model given as,
\begin{eqnarray}\label{tr}
  p_t-p_r&=&\frac{e^{-\lambda}}{\kappa}
    \bigg[\frac{-\lambda'\nu'}{4}+\frac{\nu'^2}{4}+\frac{\nu''}{2}-\frac{\lambda'-\nu'}{2r}\nonumber\\&&-\frac{\nu'}{r} + \frac{e^{\lambda} - 1}{r^2}\bigg].
\end{eqnarray}
Now by introducing DB transformation \[x=r^2,~~~ V(x)=e^{-\lambda},~~\text{ and}~~~ y(x)=e^{\nu},\]
the Eqn.~(\ref{tr}), transform to,
\begin{eqnarray}
  \frac{dV}{dx}\Big(1+x\frac{\dot{y}}{y}\Big)+V\Big[\Big(2\frac{\ddot{y}}{y}-\frac{\dot{y}^2}{y^2}\Big)x-\frac{1}{x} \Big]=\kappa \Delta-\frac{1}{x},
\end{eqnarray}
The above equation can be denoted as,
\begin{eqnarray}
\frac{dV}{dx}+F(x)V=S(x),
\end{eqnarray}
which is linear equation of $x$.
The integrating factor of the above equation is, \[e^{\int F(x) dx}\]
and the solution of the above equation is,
\[V(x)=e^{-\int F(x) dx}\int S(x)e^{\int F(x) dx} dx +k,\]
where $k$ is the constant of integration and
\begin{eqnarray}
F(x)&=&\frac{\Big(2\frac{\ddot{y}}{y}-\frac{\dot{y}^2}{y^2}\Big)x-\frac{1}{x}}{1+x\frac{\dot{y}}{y}},\label{p1}\\
S(x)&=&\frac{\kappa \Delta-\frac{1}{x}}{1+x\frac{\dot{y}}{y}}\label{p2}.
\end{eqnarray}
From the above discussion it is clear that the model of the compact star stands on two functions $F(x)$ and $S(x)$, where as $F(x)$ and $S(x)$ depends on $y(x)$ and $\Delta$. Therefore for our model the two generating functions are therefore,
\begin{eqnarray}
y(x)&=& \exp\Big[\frac{1}{4}\Big\{2 ax + bx^2 +\frac{
   2 p_0(a^2 + 2 b)}{b \sqrt{a^2-4b}} \times  \nonumber\\&&\tan^{-1}\Big(\frac{a + 2 b x}{\sqrt{a^2-4b}}\Big)
   -\frac{ap_0}{b} \log(1 + ax + bx^2)\Big\}\nonumber\\&&+B\Big] ,\\
\Delta &=& \frac{x}{4 \Psi^3}\big[A_1 +
    A_2 x + A_3 x^2 +
   A_4 x^3 +
   A_5 x^4\nonumber\\&&+ 4 a b^3 x^5 + b^4 x^6\big].
\end{eqnarray}
\end{itemize}

\bibliographystyle{plain}
\bibliography{./ref}

\section*{Acknowledgement}
PB is thankful to IUCAA , Pune, Govt of India for providing visiting associateship. SD  gratefully acknowledges support from the Inter-university Centre for Astronomy and Astrophysics (IUCAA), Pune, India, where a part of this work was carried out under its Visiting Research Associateship Programme.

\end{document}